\documentclass[acmsmall,screen,nonacm]{acmart}

\usepackage{listings}
\usepackage{booktabs}
\usepackage{multirow}
\usepackage{xspace}
\usepackage{amsmath}
\usepackage[strings]{underscore}
\usepackage{tikz}
\usepackage{placeins}
\usetikzlibrary{positioning,fit,arrows.meta,calc,decorations.pathreplacing}

\newif\ifdraft\drafttrue

\lstdefinestyle{halide}{
  language=C++,
  basicstyle=\ttfamily\footnotesize,
  keywordstyle=\bfseries,
  commentstyle=\itshape\color{black!70},
  columns=fullflexible,
  keepspaces=true,
  showstringspaces=false,
  morekeywords={Func,Var,RDom,Expr,Approximation,Compose,Parallel,Derive,Pointwise,TrustedInverse},
  xleftmargin=1em,
  aboveskip=0.5em,
  belowskip=0.5em,
}
\newcommand{\code}[1]{\texttt{#1}}
\newcommand{\cc}[1]{\texttt{#1}}

\newcommand{\enc}{\mathsf{enc}}
\newcommand{\dec}{\mathsf{dec}}
\newcommand{\Vals}{V}
\newcommand{\Encs}{E}

\newcommand{\fpsixteen}{\mathrm{fp16}}
\newcommand{\ulp}{\mathrm{ulp}}

\newcommand{\GGML}{GGML\xspace}
\newcommand{\Kleidi}{KleidiAI\xspace}
\newcommand{\GH}{G/H\xspace}

\newcommand{\nFormats}{31\xspace}
\newcommand{\nQuantFormats}{29\xspace}
\newcommand{\nGGMLTypes}{27\xspace}
\newcommand{\nGGMLTypesAll}{27\xspace}
\newcommand{\nFormatsLayouts}{53\xspace}
\newcommand{\nQuantizers}{16\xspace}
\newcommand{\nLossless}{24\xspace}

\newcommand{\nCoreUnits}{7\xspace}
\newcommand{\nKernels}{347\xspace}
\newcommand{\nKernelsCPU}{265\xspace}
\newcommand{\nKernelsMetal}{82\xspace}

\newcommand{\nFormatRows}{87\xspace}
\newcommand{\nMetalFormats}{16\xspace}
\newcommand{\nKnobRows}{38\xspace}
\newcommand{\nChecks}{3118\xspace}

\newcommand{\locOurs}{4304\xspace}
\newcommand{\locGGML}{16111\xspace}

\newcommand{\locGGMLMetal}{1341\xspace}

\newcommand{\locRatio}{3.7\xspace}
\newcommand{\locSchedule}{1879\xspace}
\newcommand{\locAlgorithm}{216\xspace}
\newcommand{\locSchemes}{1880\xspace}
\newcommand{\locApproxLib}{3939\xspace}
\newcommand{\locSever}{144\xspace}
\newcommand{\locApproxSpecific}{4083\xspace}
\newcommand{\locDirectives}{5300\xspace}
\newcommand{\locBackends}{1737\xspace}

\newcommand{\locPropTesting}{1161\xspace}
\newcommand{\locGeneral}{7037\xspace}
\newcommand{\locOursApprox}{8387\xspace}
\newcommand{\locOursFramework}{15424\xspace}
\newcommand{\locRatioApprox}{1.9\xspace}
\newcommand{\locRatioApproxNoKleidi}{1.3\xspace}
\newcommand{\locRatioFramework}{1.04\xspace}     %

\newcommand{\nChecksOutputs}{7.0M\xspace}       %
\newcommand{\nChecksCodec}{780\xspace}
\newcommand{\nChecksKernel}{2338\xspace}
\newcommand{\nChecksKernelOurs}{1952\xspace}
\newcommand{\nChecksKernelOursVecDot}{428\xspace}
\newcommand{\nChecksKernelOursMulMat}{1524\xspace}
\newcommand{\nChecksKernelGGML}{386\xspace}
\newcommand{\nPropChecks}{568\xspace}
\newcommand{\nPropTrials}{3846\xspace}
\newcommand{\nExhaustive}{41\xspace}

\newcommand{\nExhFormats}{11\xspace}
\newcommand{\nExhScales}{65{,}536\xspace}
\newcommand{\nExhBlocksQEight}{66.8M\xspace}     %
\newcommand{\nExhPairsFull}{2{,}015{,}299{,}584\xspace}
\newcommand{\nExhBlocksQFourOne}{32.2G\xspace}
\newcommand{\nExhBlocksQFiveOne}{64.5G\xspace}

\newcommand{\nExhUlpQFourOne}{128\xspace}       %
\newcommand{\nExhUlpQFiveOne}{64\xspace}
\newcommand{\nExhMaxFailUlpQFourOne}{127.5\xspace}
\newcommand{\nExhMaxFailUlpQFiveOne}{63.5\xspace}

\newcommand{\nMutFamilies}{7\xspace}
\newcommand{\nMutKinds}{9\xspace}
\newcommand{\nMutRuns}{621\xspace}         %
\newcommand{\nMutEquiv}{20\xspace}         %
\newcommand{\nMutNonEquiv}{601\xspace}
\newcommand{\nMutTRejects}{563\xspace}     %
\newcommand{\nMutMissedSampled}{14\xspace} %
\newcommand{\nMutAnyRejects}{601\xspace}

\newcommand{\nMutZeroRatioSOne}{0.032\xspace}
\newcommand{\nChecksBitExact}{800\xspace}
\newcommand{\nChecksGGMLDot}{20\xspace}
\newcommand{\nChecksTolOnly}{2318\xspace}

\newcommand{\nExactFormats}{20\xspace}
\newcommand{\nExactBlocks}{50.8M\xspace}
\newcommand{\nLegacyBlocks}{29.4M\xspace}
\newcommand{\nLegacyUBDiff}{218{,}533\xspace}
\newcommand{\nLegacyNonStrict}{139{,}260\xspace}  %

\newcommand{\nExactGridBlocks}{4.08M\xspace}

\newcommand{\genTimeCPU}{356\xspace}
\newcommand{\genTimeMetal}{421\xspace}
\newcommand{\genTimeMax}{17\xspace}
\newcommand{\sizeCPU}{4.8\xspace}
\newcommand{\sizeMetal}{14\xspace}

\newcommand{\nKnobs}{7\xspace}
\newcommand{\nKnobSettings}{46\xspace}
\newcommand{\nKnobFormats}{14\xspace}

\newcommand{\nEqExact}{14\xspace}
\newcommand{\nEqUlp}{2\xspace}

\newcommand{\nEqRefit}{7\xspace}
\newcommand{\nEqNone}{7\xspace}

\newcommand{\firstLoadSrcLo}{0.44\xspace}
\newcommand{\firstLoadSrcHi}{2.25\xspace}
\newcommand{\firstLoadLibLo}{54\xspace}
\newcommand{\firstLoadLibHi}{88\xspace}

\newcommand{\sizeMetalLibsSrc}{332\xspace}
\newcommand{\sizeMetalLibsLib}{29\xspace}

\newcommand{\nCases}{2041\xspace}               %
\newcommand{\nCasesCPU}{1649\xspace}
\newcommand{\medCPU}{1.81\xspace}               %
\newcommand{\nCasesMetal}{392\xspace}
\newcommand{\pctMatchMetal}{86.5\%\xspace}      %
\newcommand{\medMetal}{1.13\xspace}             %
\newcommand{\nBelowCPU}{14\xspace}              %
\newcommand{\nBelowMetal}{53\xspace}            %
\newcommand{\nBelowMetalSmall}{22\xspace}       %
\newcommand{\nBelowMetalSmallGEMV}{16\xspace}   %
\newcommand{\metalMeightLo}{0.68\xspace}        %
\newcommand{\metalMeightHi}{0.90\xspace}
\newcommand{\pctMatchMetalPerSample}{84.9\%\xspace} %
\newcommand{\medMetalPerSample}{1.06\xspace}    %
\newcommand{\nBelowMetalPerSample}{59\xspace}
\newcommand{\pctMatchAllPerSample}{96.4\%\xspace} %

\newcommand{\noiseCPU}{5\%\xspace}              %

\newcommand{\nHH}{703\xspace}                   %
\newcommand{\pctHH}{91.5\%\xspace}              %
\newcommand{\nHHCPU}{311\xspace}
\newcommand{\nHHMetal}{392\xspace}              %
\newcommand{\pctHHMetal}{86.5\%\xspace}         %
\newcommand{\medHHMetal}{1.13\xspace}           %
\newcommand{\nLoopCPU}{1338\xspace}
\newcommand{\pctLoopCPU}{99.5\%\xspace}         %
\newcommand{\medLoopGEMM}{2.36\xspace}          %
\newcommand{\medLoopGEMV}{1.18\xspace}          %
\newcommand{\nPairsCPU}{54\xspace}
\newcommand{\nPairsTiled}{8\xspace}

\newcommand{\nPairsNoGGML}{38\xspace}
\newcommand{\nLawBlocks}{100{,}000\xspace}
\newcommand{\pctTruncRecipBlocks}{30.3\%\xspace} %
\newcommand{\pctTruncRecipCodes}{3.8\%\xspace}
\newcommand{\nLawQEightK}{200{,}001\xspace}
\newcommand{\nLawQai}{260{,}010\xspace}
\newcommand{\pctLawQEightKFail}{13\%\xspace}
\newcommand{\pctLawQaiFail}{11.7\%\xspace}  %
\newcommand{\pctTruncDivBlocks}{50.1\%\xspace}   %

\newcommand{\oracleRatioLo}{1.39\xspace}    %
\newcommand{\oracleRatioHi}{1.49}

\newcommand{\oracleQuantLo}{0.7\%\xspace}   %
\newcommand{\oracleQuantHi}{3.2\%\xspace}
\newcommand{\pctQEightOneDiffer}{25--27\%\xspace}   %
\newcommand{\pctQEightOneStrict}{24--27\%\xspace}   %
\newcommand{\pctQEightOneTie}{0.2--1.7\%\xspace}    %

\newcommand{\nKnobCases}{126\xspace}
\newcommand{\nKnobDefaultMatch}{103\xspace}
\newcommand{\pctKnobDefault}{82\%\xspace}
\newcommand{\medKnobGain}{1.02\xspace}

\newcommand{\blasCores}{1.6\xspace}             %
\newcommand{\nBlasCases}{432\xspace}            %
\newcommand{\blasMedMfivetwelveTone}{0.09\xspace} %
\newcommand{\nBlasApplies}{351\xspace}          %
\newcommand{\nBlasFaster}{254\xspace}           %
\newcommand{\pctMatchCPUBlas}{87.2\%\xspace}    %
\newcommand{\medCPUBlas}{1.53\xspace}           %
\newcommand{\pctMatchAllBlas}{87.1\%\xspace}    %
\newcommand{\medAllBlas}{1.31\xspace}           %
\newcommand{\pctMatchBlasEligible}{43.9\%\xspace} %
\newcommand{\medBlasEligible}{0.79\xspace}      %
\newcommand{\nBlasRouted}{19\xspace}            %
\newcommand{\pctMatchCPUBlasRouted}{89.1\%\xspace} %
\newcommand{\pctMatchAllBlasRouted}{88.6\%\xspace} %

\newcommand{\pctMatchAllFixed}{96.5\%\xspace}   %
\newcommand{\pctExceedAllFixed}{93.1\%\xspace}  %
\newcommand{\medAllFixed}{1.53\xspace}          %
\newcommand{\pctHHFixed}{91.0\%\xspace}         %
\newcommand{\medHHFixed}{1.13\xspace}           %
\newcommand{\pctHHCPUFixed}{96.8\%\xspace}      %
\newcommand{\medHHCPUFixed}{1.10\xspace}        %
\newcommand{\pctLoopCPUFixed}{99.3\%\xspace}    %
\newcommand{\medLoopCPUFixed}{1.94\xspace}      %
\newcommand{\pctMatchAllBlasFixed}{86.8\%\xspace} %
\newcommand{\pctMatchCPUBlasFixed}{86.9\%\xspace} %
\newcommand{\medCPUFixed}{1.77\xspace}          %
\newcommand{\medCPUBlasFixed}{1.49\xspace}      %
\newcommand{\pctHHCPUBlasFixed}{86.5\%\xspace}  %
\newcommand{\medHHCPUBlasFixed}{1.07\xspace}    %
\newcommand{\pctLoopCPUBlasFixed}{87.0\%\xspace} %
\newcommand{\medLoopCPUBlasFixed}{1.73\xspace}  %
\newcommand{\medLoopGEMMFixed}{2.36\xspace}     %
\newcommand{\medLoopGEMVFixed}{1.17\xspace}     %
\newcommand{\medLoopGEMMBlasFixed}{1.96\xspace} %
\newcommand{\pctMatchAllRule}{96.2\%\xspace}    %
\newcommand{\pctHHCPURule}{94.9\%\xspace}       %
\newcommand{\nFixedLost}{5\xspace}              %
\newcommand{\nBelowCPUFixed}{19\xspace}         %
\newcommand{\nLayoutSwitch}{69\xspace}          %
\newcommand{\nLayoutGroups}{108\xspace}         %
\newcommand{\pctHHHind}{91.5\%\xspace}          %
\newcommand{\pctMatchAllHind}{96.6\%\xspace}    %
\newcommand{\nGGMLOffRule}{25\xspace}           %
\newcommand{\pctLoopGEMV}{98.1\%\xspace}        %
\newcommand{\pctLoopBlas}{87.1\%\xspace}        %
\newcommand{\medLoopBlas}{1.75\xspace}          %
\newcommand{\pctLoopGEMMBlas}{82.9\%\xspace}    %
\newcommand{\medLoopGEMMBlas}{2.00\xspace}      %
\newcommand{\nLoopBlasElig}{247\xspace}         %
\newcommand{\pctLoopBlasElig}{33.2\%\xspace}    %
\newcommand{\medLoopBlasElig}{0.64\xspace}      %
\newcommand{\bwSpec}{400\xspace}
\newcommand{\bwPeakCPU}{128\xspace}             %
\newcommand{\bwPeakMetal}{389\xspace}           %
\newcommand{\bwCPUOurs}{177\xspace}             %
\newcommand{\bwCPUGGML}{150\xspace}
\newcommand{\bwMetalOurs}{368\xspace}           %
\newcommand{\bwMetalGGML}{364\xspace}
\newcommand{\bwFracCPUOurs}{87\%\xspace}        %
\newcommand{\bwFracCPUGGML}{72\%\xspace}
\newcommand{\bwFracMetalOurs}{89\%\xspace}      %
\newcommand{\bwFracMetalGGML}{89\%\xspace}
\newcommand{\bwFracAllCPUOurs}{19\%\xspace}     %
\newcommand{\bwFracAllCPUGGML}{16\%\xspace}
\newcommand{\bwFracAllMetalOurs}{47\%\xspace}   %
\newcommand{\bwFracAllMetalGGML}{44\%\xspace}
\newcommand{\bwDenCPU}{149--224\xspace}         %
\newcommand{\bwDenMetal}{407--470\xspace}       %

\newcommand{\ggmlVersion}{0.26.0\xspace}
\newcommand{\ggmlCompiler}{Apple clang 21.0.0\xspace}  %
\newcommand{\ggmlFlags}{\texttt{-O3 -mcpu=native+dotprod+i8mm+nosve+nosme}\xspace} %
\newcommand{\medFHVecDot}{1.19\xspace}          %
\newcommand{\medFHGEMM}{1.9\xspace}             %
\newcommand{\medFHGEMV}{1.05\xspace}            %
\newcommand{\medBFVecDot}{13.5\xspace}          %
\newcommand{\rangeBFGEMV}{9.5--19\xspace}       %
\newcommand{\rangeBFGEMM}{23--27\xspace}        %
\newcommand{\pctMatchAllFixedNoBF}{96.6\%\xspace} %
\newcommand{\medAllFixedNoBF}{1.51\xspace}      %
\newcommand{\pctHHFixedNoBF}{91.4\%\xspace}     %
\newcommand{\medHHFixedNoBF}{1.13\xspace}       %
\newcommand{\tokDecMed}{1.15\xspace}            %
\newcommand{\tokDecHHMed}{1.18\xspace}          %
\newcommand{\tokDecFbMed}{1.13\xspace}          %
\newcommand{\tokDecToneMed}{1.08\xspace}        %
\newcommand{\tokDecToneHHMed}{1.05\xspace}      %
\newcommand{\tokPreMed}{1.88\xspace}            %
\newcommand{\tokPreHHMed}{1.17\xspace}          %
\newcommand{\tokPreFbMed}{2.44\xspace}          %
\newcommand{\tokPreBlasMed}{0.70\xspace}        %
\newcommand{\nTokPreBlasMatch}{8\xspace}        %
\newcommand{\tokPreBlasHHMed}{1.06\xspace}      %
\newcommand{\nTokPreBlasHHMatch}{7\xspace}      %
\newcommand{\tokPreBlasFbMed}{0.53\xspace}      %
\newcommand{\nTokPreBlasFbMatch}{1\xspace}      %
\newcommand{\tokMetalDecMed}{1.03\xspace}       %
\newcommand{\nTokMetalDecMatch}{11\xspace}      %
\newcommand{\tokMetalPreMed}{1.08\xspace}       %
\newcommand{\nTokMetalPreMatch}{15\xspace}      %

\newcommand{\pctMatchAllFixedEstFirst}{96.2\%\xspace} %
\newcommand{\pctMatchAllFixedEstMedian}{96.5\%\xspace} %
\newcommand{\nBelowAllFixed}{72\xspace}         %
\newcommand{\nBelowAllFixedEstFirst}{77\xspace} %
\newcommand{\nBelowAllFixedEstMedian}{72\xspace} %
\newcommand{\pctMatchCPUFixed}{98.8\%\xspace}   %
\newcommand{\pctMatchCPUFixedEstFirst}{98.5\%\xspace} %
\newcommand{\pctMatchCPUFixedEstMedian}{98.8\%\xspace} %
\newcommand{\pctMatchMetalEstFirst}{86.5\%\xspace} %
\newcommand{\pctMatchMetalEstMedian}{86.7\%\xspace} %
\newcommand{\pctHHFixedEstFirst}{90.9\%\xspace} %
\newcommand{\pctHHFixedEstMedian}{91.3\%\xspace} %
\newcommand{\pctLoopCPUFixedEstFirst}{99.0\%\xspace} %
\newcommand{\pctLoopCPUFixedEstMedian}{99.2\%\xspace} %
\newcommand{\pctHHEstFirst}{91.3\%\xspace}      %
\newcommand{\pctHHEstMedian}{91.7\%\xspace}     %
\newcommand{\pctLoopCPUEstFirst}{99.2\%\xspace} %
\newcommand{\pctLoopCPUEstMedian}{99.3\%\xspace} %

\newcommand{\xCases}{1620\xspace}                %
\newcommand{\xPctMatch}{83.5\%\xspace}           %

\newcommand{\xMed}{1.23\xspace}                  %
\newcommand{\xFSTwelveN}{412\xspace}             %
\newcommand{\xFSTwelveMed}{1.16\xspace}
\newcommand{\xFSTwelvePct}{78.6\%\xspace}
\newcommand{\xFSOneN}{323\xspace}                %
\newcommand{\xFSOneMed}{1.07\xspace}
\newcommand{\xFSOnePct}{67.8\%\xspace}           %
\newcommand{\xFSOnePctPerCase}{73.1\%\xspace}
\newcommand{\xFSN}{789\xspace}                  %
\newcommand{\xFSMed}{1.11\xspace}
\newcommand{\xFSPct}{73.9\%\xspace}
\newcommand{\xBDN}{831\xspace}                  %
\newcommand{\xBDMed}{1.64\xspace}
\newcommand{\xBDPct}{92.7\%\xspace}
\newcommand{\xVdN}{54\xspace}                    %
\newcommand{\xVdMed}{1.00\xspace}
\newcommand{\xVdPct}{74.1\%\xspace}
\newcommand{\xBDTwelveN}{452\xspace}             %
\newcommand{\xBDTwelveMed}{1.71\xspace}
\newcommand{\xBDTwelvePct}{92.9\%\xspace}
\newcommand{\xBDOneN}{379\xspace}
\newcommand{\xBDOneMed}{1.60\xspace}
\newcommand{\xBDOnePct}{92.3\%\xspace}
\newcommand{\xTwelveN}{864\xspace}               %
\newcommand{\xTwelveMed}{1.32\xspace}
\newcommand{\xTwelvePct}{86.1\%\xspace}
\newcommand{\xOneN}{702\xspace}                  %
\newcommand{\xOneMed}{1.20\xspace}
\newcommand{\xOnePct}{81.1\%\xspace}
\newcommand{\xAmxTwelveN}{101\xspace}
\newcommand{\xAmxTwelveMed}{1.21\xspace}
\newcommand{\xAmxTwelvePct}{80.2\%\xspace}
\newcommand{\xAmxOneN}{91\xspace}
\newcommand{\xAmxOneMed}{0.95\xspace}
\newcommand{\xAmxOnePct}{49.5\%\xspace}
\newcommand{\xRpTwelveN}{38\xspace}
\newcommand{\xRpTwelveMed}{1.80\xspace}
\newcommand{\xRpTwelvePct}{71.1\%\xspace}
\newcommand{\xRpOneN}{30\xspace}
\newcommand{\xRpOneMed}{1.77\xspace}
\newcommand{\xRpOnePct}{66.7\%\xspace}
\newcommand{\xTlTwelveN}{273\xspace}
\newcommand{\xTlTwelveMed}{1.14\xspace}
\newcommand{\xTlTwelvePct}{79.1\%\xspace}
\newcommand{\xTlOneN}{202\xspace}
\newcommand{\xTlOneMed}{1.12\xspace}
\newcommand{\xTlOnePct}{76.2\%\xspace}
\newcommand{\xSmallFSN}{256\xspace}
\newcommand{\xSmallFSMed}{1.28\xspace}           %
\newcommand{\xSmallFSPct}{93.0\%\xspace}         %
\newcommand{\xSmallTlMed}{1.31\xspace}           %
\newcommand{\xSmallTlPct}{98.3\%\xspace}
\newcommand{\xSmallBDMed}{2.74\xspace}           %

\newcommand{\xPreAmxMed}{1.27\xspace}

\newcommand{\xPreAmxBigMed}{1.39\xspace}         %

\newcommand{\xPreAmxFiveMed}{1.35\xspace}        %
\newcommand{\xPreRpMed}{2.18\xspace}             %
\newcommand{\xPreBDMed}{4.35\xspace}             %
\newcommand{\xFiveTlMed}{0.93\xspace}

\newcommand{\xFiveTlMedPerCase}{0.97\xspace}
\newcommand{\xGemvRpLo}{0.54\xspace}             %
\newcommand{\xGemvRpHi}{0.84\xspace}
\newcommand{\xBelow}{267\xspace}                 %
\newcommand{\xLocSched}{261\xspace}
\newcommand{\xLocTable}{14\xspace}

\newcommand{\xLocBackend}{518\xspace}
\newcommand{\xLocOurs}{793\xspace}               %
\newcommand{\xLocOursMax}{1478\xspace}
\newcommand{\xLocGGMLDots}{7229\xspace}
\newcommand{\xLocGGMLAmx}{2244\xspace}
\newcommand{\xLocGGMLTiled}{1634\xspace}
\newcommand{\xLocGGML}{11107\xspace}
\newcommand{\xLocRatio}{14\xspace}               %
\newcommand{\xLocRatioMin}{7.5\xspace}           %

\begin{document}
\makeatletter\def\usc@dischyph{}\makeatother

\title{User-Schedulable Approximations for Quantized Tensor Programs}

\author{Alex Reinking}
\affiliation{\institution{Adobe Research}\country{USA}}
\email{areinking@adobe.com}
\authorsaddresses{}

\begin{abstract}
Quantization lets language models run on edge devices, but it sharply increases
the number of tensor kernels a library must maintain: every weight format must
be paired with every activation format on every instruction set and GPU, with
an encoder and a decoder that agree on the stored bytes. In \GGML, the tensor
library used by llama.cpp, these kernels are written by hand. Only
\nPairsTiled of the \nPairsCPU weight--activation pairs it runs on our Arm CPU
have an optimized, tiled matrix-multiplication kernel; the rest compute one
block dot product at a time.

We extend Halide's separation of algorithm and schedule to the representation
of values. A programmer specifies an algorithm, selects approximations for its
values, and schedules the resulting computations. Each approximation is a
codec, an encoder paired with a decoder, composed from reusable units:
quantizers, integer representations, bit packers and layouts. A new directive,
\code{approximate\_by}, applies a codec to one stage of a program and returns
the encoder and decoder as ordinary stages that Halide can schedule: the
weights are encoded offline, the activations inside the kernel, and both
decoders are fused into the register tiles of a matrix product. Optional
contracts, such as a round-trip law (re-encoding decoded values reproduces
their encoding), are checked by property tests; the law exposed an
inconsistent block sum in \GGML's q8\_1.

One matrix-product algorithm and one schedule library generate \nKernels
kernels for \nFormats of \GGML's and \Kleidi's formats on Arm CPUs and Apple
GPUs. They cover \nPairsNoGGML weight--activation pairs for which \GGML has
no kernel, with no additional schedule code. Our application uses
\locRatio{}$\times$ less code than \GGML and \Kleidi. Our decoders and
encoders match \GGML's bit for bit on every input where \GGML's behavior is
defined in C, except q8\_1's corrected block sum and mxfp4's encoder on
non-finite inputs. On an Apple M3 Max, our kernels reach at least 95\% of the
speed of \GGML's format-specific kernels in \pctHHFixed of the cases that
have one, with a median speedup of \medHHFixed{}$\times$, and are faster
still where \GGML falls back to blocked dot products. For larger f32 matrix
products, \GGML can instead dequantize the weights and call Accelerate,
Apple's BLAS library, which is usually faster than our kernels; counting
that path as part of \GGML, our kernels still reach the 95\% threshold in
\pctMatchCPUBlasFixed of CPU cases. A port to x86 with Intel's AMX adds
\xLocOurs lines of schedule and compiler code against \GGML's \xLocGGML, and
matches \GGML in \xPctMatch of its cases.
\end{abstract}

\maketitle

\section{Introduction}\label{sec:intro}

Weight quantization lets language models run on edge devices such as laptops
and phones: each weight is stored in a few bits, in small blocks that share a
scale. For kernel writers, quantization adds new dimensions to a space that
was already combinatorial. A tensor library writes a kernel per operation, matrix shape
and instruction set; now each weight format must also be paired with each
activation format, be packed for each CPU's dot-product and matrix instructions, and be
decoded inside each GPU kernel, with an encoder and a decoder that agree on
the stored bytes. In \GGML~\cite{ggml}, the tensor library used by
llama.cpp~\cite{llamacpp}, these combinations are written by hand: for the
formats we study, they take \locGGML nonblank, noncomment lines, including
\Kleidi's kernels.
Even so, only \nPairsTiled of its \nPairsCPU CPU weight--activation
pairs have tiled kernels, which reuse each loaded block across several rows
and columns; the rest compute one block dot product at a time or, where
applicable, call Apple's BLAS library on dequantized weights
(\S\ref{sec:eval:matrix}). The gaps are the cost of hand-writing a
combinatorial space.

Halide~\cite{halide-pldi13} already factors part of that space. It separates
an \emph{algorithm}, which specifies what to compute, from a \emph{schedule},
which says how, so that one algorithm serves many loop structures and targets.
But a schedule preserves values, while quantization changes them deliberately:
it rounds them, packs two integers into a byte and splits an integer into bit
planes. What Halide lacks is a third, separate choice, the representation of the algorithm's values. With
this separation, a programmer specifies an algorithm, selects approximations for
its values, and schedules the resulting computations. Selecting an
approximation may change values over the reals; scheduling, by contrast, preserves them.

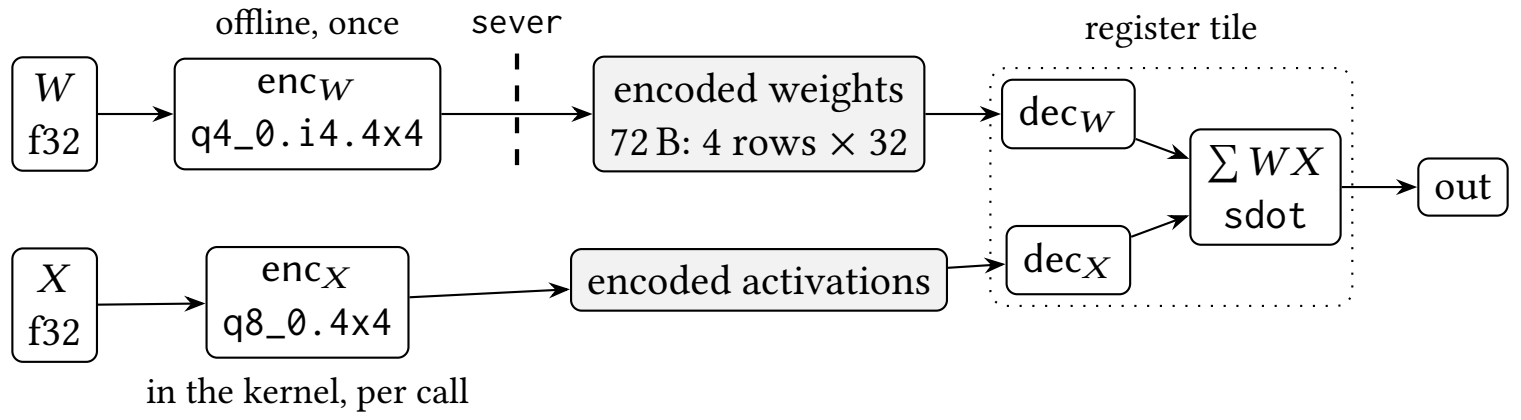
\begin{figure}[t]
\centering
\begin{tikzpicture}[
  font=\footnotesize,
  box/.style={draw, rounded corners=2pt, inner sep=3pt, align=center},
  arr/.style={-{Stealth[length=4pt]}},
  node distance=4mm and 5mm]
\node[box] (W) {$W$\\f32};
\node[box, right=of W] (ew) {$\enc_W$\\\code{q4_0.i4.4x4}};
\node[box, right=10mm of ew, fill=gray!10] (rw) {encoded weights\\72\,B: 4 rows $\times$ 32};
\node[box, right=of rw] (dw) {$\dec_W$};
\node[box, below=5mm of W] (X) {$X$\\f32};
\node[box, below=5mm of ew] (ex) {$\enc_X$\\\code{q8_0.4x4}};
\node[box, below=5mm of rw, fill=gray!10] (rx) {encoded activations};
\node[box, below=5mm of dw] (dx) {$\dec_X$};
\node[box, right=8mm of $(dw)!0.5!(dx)$] (dot) {$\sum W X$\\\cc{sdot}};
\node[box, right=of dot] (out) {out};
\draw[arr] (W) -- (ew); \draw[arr] (ew) -- (rw); \draw[arr] (rw) -- (dw);
\draw[arr] (X) -- (ex); \draw[arr] (ex) -- (rx); \draw[arr] (rx) -- (dx);
\draw[arr] (dw) -- (dot); \draw[arr] (dx) -- (dot); \draw[arr] (dot) -- (out);
\draw[dashed, thick] ($(ew.east)!0.5!(rw.west) + (0,4mm)$) -- ++(0,-8mm)
  node[pos=0, above, font=\scriptsize] {\code{sever}};
\node[font=\scriptsize, above=0mm of ew] {offline, once};
\node[draw, dotted, rounded corners=3pt, inner sep=2pt, fit=(dw)(dx)(dot),
  label={[font=\scriptsize]above:register tile}] {};
\node[font=\scriptsize, below=0mm of ex] {in the kernel, per call};
\end{tikzpicture}
\caption{The running example, q4\_0 weights and q8\_0 activations, after
scheduling. Dashed: the staging cut (\code{sever}) between the offline encoder
and the kernel. Dotted: the register tile, with both decoders inlined.}\label{fig:teaser}
\end{figure}

We represent an approximation by a \emph{codec}, an encoder paired with a
decoder. A codec is an executable definition of a storage format such as
\GGML's q4\_0. It is written once, apart from any algorithm, and every kernel
that reads the format uses it. The intent is an almost-identity round trip, in
which encoding then decoding changes values only as much as the application
accepts. Codecs compose from small reusable units, such as quantizers, bit
packers and layouts, and one composition yields both directions, so the
encoder and decoder need not be synchronized by hand. We extend Halide with
\code{approximate_by}, which inserts a codec in front of a Func (Halide's
unit of computation) and returns its encoder and decoder as ordinary Funcs.
From there it is just Halide: the schedule, not the format, decides when
values are encoded and where they are decoded. In \autoref{fig:teaser}, a
shared schedule encodes the weights once, offline, encodes the activations
inside the kernel, and fuses both decoders into the register tile of a matrix
product. Another placement changes only the schedule, where in \GGML it would
mean another kernel. Optional contracts make accuracy and stable re-encoding
testable. Chief among them is a round-trip law, which says that re-encoding
decoded values reproduces their encoding.

\paragraph{Contributions.}
\begin{itemize}
\item \emph{A representation layer for Halide}: codecs, composable
encoder--decoder pairs defined apart from the algorithms that use them,
with optional contracts (a round-trip law, losslessness, error bounds), and
\code{approximate_by}, which makes both halves schedulable
(\S\S\ref{sec:ident}--\ref{sec:example}).
\item \emph{Reusable formats}: \nFormats \GGML and \Kleidi formats
(\nQuantFormats quantized) from shared units
(\S\ref{sec:example}).
\item \emph{Shared execution recipes}: one matrix-product algorithm and one
schedule library for Arm CPUs and Apple GPUs, ported to x86 with AMX
(\S\ref{sec:eval:x86}); new general directives
rewrite dot products of decoded values into integer dot products, with
preconditions proved by Halide's interval analysis (\S\ref{sec:sched}).
\item \emph{Contract-based validation}: property tests of contracts,
exhaustive for the round-trip law of \nExhFormats formats' compiled codecs,
and differential checks against references; the round-trip law exposed a
block-sum inconsistency in \GGML's q8\_1
(\S\S\ref{sec:props}, \ref{sec:eval:correct}).
\end{itemize}

\paragraph{Results.}
On \GGML's widely deployed catalog of quantized formats, one algorithm and one
schedule library generate \nKernels kernels, including \nPairsNoGGML
weight--activation pairs for which \GGML has no kernel, from an application
\locRatio{}$\times$ smaller than the \GGML and \Kleidi code it replaces
(\S\ref{sec:eval:size}). On an Apple M3 Max, our kernels are on par with
\GGML's format-specific kernels, and faster wherever \GGML falls back to
blocked dot products, including when it calls Apple's BLAS
(\S\ref{sec:eval:perf}). A second CPU target, x86 with AMX, takes \xLocOurs
lines of schedules and back end; there our kernels match \GGML in most
cases and run a median \xPreAmxMed{}$\times$ faster on multithreaded prefill (\S\ref{sec:eval:x86}).

\section{Background}\label{sec:background}

\paragraph{Halide.}
A Halide~\cite{halide-pldi13} program is a graph of \emph{Funcs}. A Func is a
pure function from integer coordinates to values, defined by an expression
over other Funcs. It may also have update definitions, which reduce over a
reduction domain (\code{RDom}). For example, the product of a $K\times N$
weight matrix $W$ and a $K\times M$ matrix $X$ of activations, the inputs that
arrive at run time, is:
\begin{lstlisting}
Func out; Var n, m; RDom k(0, K);
out(n, m) += W(k, n) * X(k, m);
\end{lstlisting}
\noindent This definition is the \emph{algorithm}: it says what value each element of
\code{out} has, but not how to compute it. A separate \emph{schedule} decides
the rest. It chooses where each Func is computed and stored: inlined into its
consumers, computed once in full (\code{compute_root}), or computed piece by
piece inside a loop of a consumer (\code{compute_at}). It splits and reorders
loops and maps them to hardware (\code{vectorize}, \code{parallel}, GPU
threads). \code{rfactor}~\cite{halide-rfactor} splits an associative reduction into
partial reductions that can run in parallel, and \code{specialize} branches on
run-time conditions. Bounds inference then computes the region of each Func
that each loop needs. A schedule never changes the values the algorithm
specifies, up to float rounding (\S\ref{sec:ident:round}). Halide compiles
\emph{generators} ahead of time; each generator
declares typed input buffers (\code{ImageParam}s) and outputs.

\paragraph{Quantized formats.}
\GGML stores quantized tensors in \emph{block formats}. Each row of a weight
matrix is cut into blocks of 32 or 256 values. A block is stored as small
integers, its \emph{quantized integers}, together with one or a few
floating-point fields, such as a scale, that map the integers back to real
values (we write f32, f16 and bf16 for IEEE binary32, binary16 and
bfloat16). A \emph{format} fixes everything about how a block becomes bytes: the
block size; the \emph{quantizer}, which computes the scale and the integers
from the values; how each integer is represented (for example, offset binary
or two's complement); how the integers are packed into bits; and where each
field lies within the block's bytes. A \emph{layout} may then rearrange the
bytes of many blocks, as repacking does below. \GGML names a format by the bits per integer and its
kind of scaling. A \code{_0} format such as q4\_0 decodes an integer $q$ to
$d\,q$ with one scale $d$ per block; a \code{_1} format adds a per-block
minimum, $d\,q+m$; a \code{_K} format (a ``K-quant'') groups 256 values into
sub-blocks whose own scales are themselves quantized; and the \code{iq}
formats decode groups of integers through lookup tables. Activations use the
8-bit formats q8\_0, q8\_1 and q8\_K, which pair with the weight formats.

\paragraph{Running example.}
\autoref{fig:teaser} shows our running example: q4\_0 weights multiplied by
q8\_0 activations. A q4\_0 block is 18 bytes: an f16 scale $d$ and 32
unsigned four-bit integers $u$, packed two per byte, with positions 0--15 of
the block in the low nibbles and 16--31 in the high nibbles. A value decodes
to $d\,(u-8)$; \S\ref{sec:example:parts} gives the encoder. The activations
arrive in f32 and are quantized at run time to q8\_0, which stores, per block,
a scale $d=\max|x|/127$ and 32 integers $q=\mathrm{round}(x/d)$ in int8. The
product of a q4\_0 block with a q8\_0 block is then
$d_w d_x\sum_j (u_{w,j}-8)\,q_{x,j}$: an integer dot product, scaled once per
block, which Arm's \cc{sdot} computes four byte products per 32-bit lane.
When it loads a model, \GGML can \emph{repack} the weights for its fastest Arm
kernels: it interleaves the blocks of 4 rows in pieces of 4 or 8 bytes
(\cc{q4\_0\_4x4}, \cc{q4\_0\_4x8}), so that one vector load serves 4 output
rows, and XORs every byte with \cc{0x88}, turning each nibble $u$ into the
two's-complement nibble of $u-8$; activations are quantized into matching
interleaved records. On Apple's CPUs, \GGML prefers \Kleidi's~\cite{kleidiai}
kernel, with a layout of its own. Each of these paths is a separate
hand-written kernel, and so is each GPU kernel. We use the repacked
$4{\times}4$ layout. The \code{approximate_by} directive attaches a codec to
each operand of the f32 product; after scheduling, both decoders are inlined
into the register tile, where the product of 8-bit integers becomes \cc{sdot}
instructions. \autoref{lst:matmul} gives the user's three choices: the
algorithm, an approximation of each operand, and the schedule.

\section{From Identities to Approximations}\label{sec:ident}

The codec, the basis of our design, comes from identities on values. A
Halide schedule applies identities to the \emph{structure} of an algorithm:
splitting and reordering loops, or computing an intermediate early, leaves
every value unchanged (\S\ref{sec:background}). Compilers also apply identities to
\emph{values}: strength reduction replaces $2x$ with $x\ll 1$, and Halide's
simplifier is a large set of exact rewrite rules. This section separates
approximation from scheduling (\S\ref{sec:ident:round}), defines codecs and
their composition (\S\ref{sec:ident:split}), exposes their halves to the
schedule (\S\ref{sec:ident:handles}) and states optional contracts for
validating them (\S\S\ref{sec:ident:contracts}--\ref{sec:ident:consequences});
\autoref{app:laws} proves its propositions. Halide already provides Funcs,
schedules, \code{strict_float}, \code{rfactor} and the storage directives.
This paper adds codecs, the \code{approximate_by} directive that inserts
them, \code{sever}, which splits off offline work, and a few
general directives and types (\autoref{tab:core}).

\subsection{Rounding and permutation}\label{sec:ident:round}

Kernels also change values on purpose, for example storing f16 instead of f32 when
accurate enough. But the f32 program is itself an approximation: the
algorithm describes real numbers, and every float operation rounds its exact
result to a nearby float. Write $\mathit{round}(x)$ for the float nearest to
a real number $x$, and $\mathbb{F}\subset\mathbb{R}$ for the set of floats.
Rounding is not the identity, but it is close to it: within the range of the
floats, it moves $x$ by at most half an \emph{ulp}, the spacing of the floats
near $x$, so $|\mathit{round}(x)-x|\le\frac12\ulp(x)$. Rounding is also
idempotent: a float rounds to itself, so
$\mathit{round}(\mathit{round}(x))=\mathit{round}(x)$.

Kernels also permute data for storage: a matrix-multiply (GEMM) library pre-transposes a matrix,
\GGML interleaves the blocks of 4 rows, a format packs two 4-bit integers into one
byte. The reader undoes the permutation in its index arithmetic. A permutation
$\varphi$ followed by its inverse is exactly the identity,
$\varphi^{-1}\circ\varphi=\mathrm{id}$. Halide's existing
\code{reorder_storage} and \code{split_storage} directives move the elements
of a buffer, but never the bits inside an element. Packing two integers into
a byte, flipping the top bit of each nibble, or splitting 5-bit integers into
a 4-bit and a 1-bit plane are exact too, but they rearrange bits and change
the stored type, so no existing directive expresses them.

We distinguish the \emph{specification}, the algorithm read over
$\mathbb{R}$, from the \emph{f32 program} that implements it. Scheduling
preserves the specification's value over $\mathbb{R}$, though it may change
float behavior; an approximation may change the real value and must be
selected explicitly. This is a design rule that this paper's directives
follow; we test it but have not proved that Halide's compiler
upholds it (\S\ref{sec:props:laws}).

A schedule may thus reassociate or factor a float sum, or evaluate an
exact-over-$\mathbb{R}$ sum in integers (\S\ref{sec:sched:micro}). Accumulating in f16 or reading a stored block sum
(\S\ref{sec:props:derived}) can change the value over $\mathbb{R}$, so each
is an approximation: the programmer selects it, it may declare an error bound
against the specification, and the schedule chooses only where it applies.

Every scheduling directive may change only the f32 program's rounding,
overflow, underflow and NaN behavior, on supported inputs (finite data,
normal-range scales, no overflowing partial sum). An encoder's
roundings belong to the codec: encoders use \code{strict_float}, and Halide
rejects directives that would reorder their updates (\autoref{app:directives}).
With integers fixed, the
schedule may reorder an integer dot if interval analysis
shows no partial sum wraps (\S\ref{sec:sched:micro}).

\subsection{Factoring the round trip}\label{sec:ident:split}

Rounding between reals and floats factors into two halves:
$\pi:\mathbb{R}\to\mathbb{F}$ rounds a real to a float, and
$\iota:\mathbb{F}\to\mathbb{R}$ casts it back, so
$\mathit{round}=\iota\circ\pi$ and $\pi\circ\iota=\mathrm{id}_{\mathbb{F}}$.
A quantized format splits the same way: its encoder quantizes and permutes
(packs and interleaves); its decoder un-permutes and dequantizes.

\paragraph{Codecs.}
Let $\Vals$ be the set of possible values of a Func, taken as a whole. For a
weight matrix, $\Vals=\mathbb{R}^{K\times N}$; the program stores each element
as an f32 and the matrix as a dense array. Let $\Encs$ be a set of encodings:
tuples of named integer or floating-point arrays, which we call
\emph{components}; operationally, each component is a Func, and the array is
its value. A q4\_0 encoding,
for example, has two components: the quantized integers and the per-block
scales. A \emph{codec} is a pair of maps $\enc:\Vals\to\Encs$ and
$\dec:\Encs\to\Vals$; its \emph{round trip} is $\rho=\dec\circ\enc$.
Selecting the codec as an approximation of a Func replaces its value $x$ by
$\rho(x)$ for the Func's consumers. Both maps are Halide Funcs:
the encoder computes one Func per component, and the decoder computes values
back from them.

\paragraph{Composition.}
Codecs combine in two ways. \emph{Sequential} composition,
$\mathrm{Compose}(c_1,c_2)$, encodes with $c_1$, then encodes $c_1$'s
encoding with $c_2$, and decodes in reverse: $\enc=\enc_2\circ\enc_1$ and
$\dec=\dec_1\circ\dec_2$. Its round trip is therefore
$\rho=\dec_1\circ\rho_2\circ\enc_1$, not $\rho_1\circ\rho_2$: the inner codec
$c_2$ approximates $c_1$'s encoding rather than the values, and its error
reaches the values only through $\dec_1$. For example, a quantizer computes
an f32 scale for each block, and an inner \code{fp16()} codec rounds that scale
to f16. \emph{Componentwise} composition, $\mathrm{Parallel}(c_a,c_b,\ldots)$,
applies a separate codec to each component of an encoding, such as one codec
to the integers and another to the scales. (The name refers to the
components lying side by side; it does not mean parallel execution.) A
\emph{unit} is a small reusable codec, such
as a quantizer, an integer representation, a bit packer or a layout, and a
\emph{codec stage} is one occurrence of a unit in a composition
(\S\ref{sec:example}).

\subsection{Exposing both halves}\label{sec:ident:handles}

Inserted in front of a Func, a codec on its own only adds work: the program
now quantizes and dequantizes the weights on every call. Exposing each half
lets the schedule avoid repeating that work. Because both halves are ordinary
Funcs, a schedule can place them like any other Func, with Halide's
existing directives and the general ones we add. The encoder of a weight
matrix can run once, offline. The encoder of the activations can run inside
the kernel, on just the activations that one tile of the output needs. The
decoder can be inlined into the innermost loop, where the compiler merges its
arithmetic into the dot product, so that the loop multiplies the integers
directly.

\paragraph{The three choices.}
\autoref{lst:matmul} shows the three choices a user makes for the running
example. The algorithm is the plain f32 product, written over blocks of 32
values. Two calls to \code{approximate_by} select an approximation of each
operand's values. The schedule then places the halves; its first decision,
\code{sever}, moves the weights' encoder offline, so the kernel reads stored
records (the \code{ImageParam} \code{w\_records}); \S\ref{sec:sched} gives
the rest.

\begin{lstlisting}[float=h,caption={The three choices for the running example, with simplified names: an algorithm, an approximation of each operand, and a schedule, here the cut that moves the weights' encoder offline.},label=lst:matmul]
RDom r(0, 32, 0, K / 32);                      // r.x: value in a block, r.y: block
W(j, b, n) = Wf(b * 32 + j, n);                // f32 weights, in blocks of 32
X(j, b, m) = Xf(b * 32 + j, m);                // f32 activations, in blocks of 32
dot(n, m) += W(r.x, r.y, n) * X(r.x, r.y, m);  // 1. algorithm: the f32 product
ApproximationResult aw = W.approximate_by(q4_0_i4_4x4, {dot}); // 2. approximations,
ApproximationResult ax = X.approximate_by(q8_0_4x4, {dot});    //    one per operand
Pipeline(dot).sever(aw.encoded, {w_records});  // 3. schedule: encode W offline
\end{lstlisting}

\paragraph{Approximation.}
\code{W.approximate\_by(codec, consumers)}
constructs the encoder of \code{W} and the matching decoder, checks that it
produces one Func with \code{W}'s arguments and type, and redirects every
call to \code{W} in the listed consumers to the decoded Func. The result
exposes every Func a schedule may need: \code{aw.encoded} holds the final encoded components, which a
kernel stores or reads, and \code{aw.decode_funcs()} the decoders; each
stage's components and intermediates are reachable too. Each Func carries a role annotation (a \emph{tag}), such as \code{codes}
for quantized integers or \code{scale}, so a schedule can ask for ``the
decoders of the integers'' without naming a format. Like \code{rfactor},
\code{approximate_by} rewrites the graph at once: later directives,
\code{sever} included, see the result.

\paragraph{Staging.}
\code{Pipeline::sever(fs, params)} introduces an explicit offline/online
staging boundary. It replaces every call to each Func in
\code{fs} with a same-shaped \code{ImageParam}, the \emph{online} input, and
returns the \emph{offline} pipeline that computes them. Severing at
\code{aw.encoded} turns ``quantize the weights'' into a separate artifact and
leaves a kernel that reads encoded records; severing a codec's round trip on
one row builds standalone quantize and dequantize functions. The result is
unchanged provided the caller passes in what the offline pipeline computes.

\paragraph{Why not plain Funcs?}
A C++ library could build a q4\_0 encoder and decoder as Funcs from one
description, and a float algorithm parameterized by its operands could
receive them decoded. But the user wires each use by hand, and no role
annotations let a schedule identify encoders, decoders or components. A library
alone also cannot perform the compiler transformations needed here: derived-component matching (\S\ref{sec:props:derived})
checks a term against a component's definition, and \code{change\_type}'s interval
proof and \code{hoist\_invariants} (\S\ref{sec:sched:micro}) rewrite encoded
reductions.

\subsection{Contracts}\label{sec:ident:contracts}

A codec needs only its two maps to be inserted, scheduled and tested. It may
declare two independent, optional contracts. A
\emph{round-trip law} makes conversion stable: quantizing an
already-quantized tensor should change nothing. An \emph{error bound} says how
far the decoded values may lie from the unencoded input. These contracts support
validation and reasoning about composition; no compiler rewrite assumes the law.

\begin{definition}[Round-trip law]\label{def:law}
A codec $(\enc,\dec)$ satisfies the \emph{round-trip law} on
$D\subseteq\Vals$ if $\enc\circ\dec\circ\enc=\enc$ on $D$: re-encoding a
decoded value reproduces its encoding. Here encodings are compared
numerically, written $e\approx e'$: float components that are equal as numbers
count as equal even if their bits differ, so we do not distinguish signed
zeros or NaN payloads. A decoder must give the same result on encodings that
compare equal ($e\approx e'\Rightarrow\dec(e)=\dec(e')$). Ours do, because
they only compute with float components and never inspect their sign bits.
For example, rounding satisfies the law, with $\enc=\pi$ and $\dec=\iota$:
rounding a float again returns the same float. So does any permutation, whose
round trip is the identity.
\end{definition}

\noindent Narcissus~\cite{narcissus} states such laws for lossless formats;
$\dec$ is a generalized inverse~\cite{penrose-ginverse} of $\enc$ on $D$.

\paragraph{Error bounds.}
A codec is \emph{lossless} on a declared input set $D\subseteq\Vals$ if
$\rho(x)=x$ for all $x\in D$. Otherwise it may declare a pointwise \emph{error
bound} $\beta$, with $|x_i-\rho(x)_i|\le\beta(x,\enc(x))_i$ for every element
$i$. The law and the bound are independent: $\enc(x)=0$, $\dec(e)=0$
satisfies the law, and so, with exact division, does a truncating encoder,
which only the bound and comparisons with \GGML's bytes catch
(\autoref{app:checks}). Bounds compose through the decoders: an inner stage's
error is scaled by the outer decoder's \emph{sensitivity} to that component
(rounding a scale by $\varepsilon$ moves $q\,d$ by at most $|q|\varepsilon$),
and lossless stages such as packers and layouts add none. Adding a stage thus
means stating its bound, not re-deriving the format's. The compiler neither
infers nor checks bounds; ours are derived by hand and tested
(\autoref{app:laws}).

\subsection{Consequences of the law}\label{sec:ident:consequences}

The law buys four things: conversion can be repeated, decoding loses nothing
an encoding holds, an encoding can be re-laid out without changing the round
trip, and an encoding stores nothing that its decoded values do not determine.

\begin{proposition}\label{prop:split}
Let $(\enc,\dec)$ satisfy the law on $D$. (a) On $D$, the round trip
$\rho=\dec\circ\enc$ is idempotent, $\rho\circ\rho=\rho$, so $\rho^n=\rho$ and
$\enc\circ\rho^n=\enc$ for all $n\ge1$: repeated conversion is stable. (b)
$\enc\circ\dec$ is the identity on $\enc(D)\subseteq\Encs$. (c) For
\emph{any} codec $(\enc,\dec)$, law or not, if $\psi\circ\varphi$ is the
identity on $\enc(D)$, then $(\varphi\circ\enc,\dec\circ\psi)$ has the same
round trip $\rho$ on $D$. If, moreover, $(\enc,\dec)$ satisfies the law on
$D$, $\varphi$ respects $\approx$, and $\psi$ respects $\approx$ on
$\varphi(\enc(D))$, then the relaid-out codec also satisfies the law. (d) For $x,y\in D$, if
$\rho(x)=\rho(y)$, then $\enc(x)=\enc(y)$: every stored field $s$ of the encoding, such as a
\emph{derived component} (precomputed metadata such as a block sum), is a
function of the decoded values, $s(x)=s(\rho(x))$.
\end{proposition}

\noindent \GGML's q8\_1 block sum violates (d) (\S\ref{sec:props:q81}).
The law is optional. \GGML and \Kleidi deploy q8\_K and qai8dx, whose
encoders fail it: re-encoding their output can move a scale by up to 2 ulp
(\autoref{app:checks}). The cost is bounded: re-encoding a decoded block, as
a chain of identity layers would, stays within the encoder's error bound, as
encoding any input does, and every layer already pays that error.
Such formats are still built and scheduled like any other; composing stages
requires only that the later stages be lossless on what the earlier stages
produce, as (C1) and (C2) show.

\emph{(C1) Layouts keep the round trip.} Interleaving rows, packing nibbles
and changing an integer representation are bijections on encodings: they
rearrange or re-represent the stored data without losing any of it. By
Proposition~\ref{prop:split}(c), appending one after any encoder leaves the
round trip unchanged. A format's numerics, once validated, therefore carry
over to every layout of it: a kernel that needs a new layout swaps the layout
stage without re-testing accuracy. This holds even for the search-based
formats that fail the law, and when the encoder does satisfy the law, the
re-laid-out codec satisfies it too.

\emph{(C2) Ranges discharge preconditions.} A 4-bit packer is lossless only on
integers in $[0,15]$. Appending a stage $L$ after an approximation $A$
preserves $\rho$, and the law if $A$ satisfies it, when $\dec_L\circ\enc_L$ is
the identity on the image of $\enc_A$ (every encoding $A$ can produce); a
declared output range of $A$ inside $L$'s declared input range
discharges that without running anything. Ranges cannot express every
precondition: a unit of \GGML's repacked q4\_0 also needs inputs in
multiples of 16 (\S\ref{app:q40variants}).

\emph{(C3) Lossy stages do not compose for free.} Two lossy codecs that each
satisfy the law need not satisfy it in sequence; we test each
composition; \S\ref{sec:props:q81} shows a deployed format that fails.
\autoref{tab:contracts} lists the status of every format.

\section{Assembling q4\_0 from Parts}\label{sec:example}

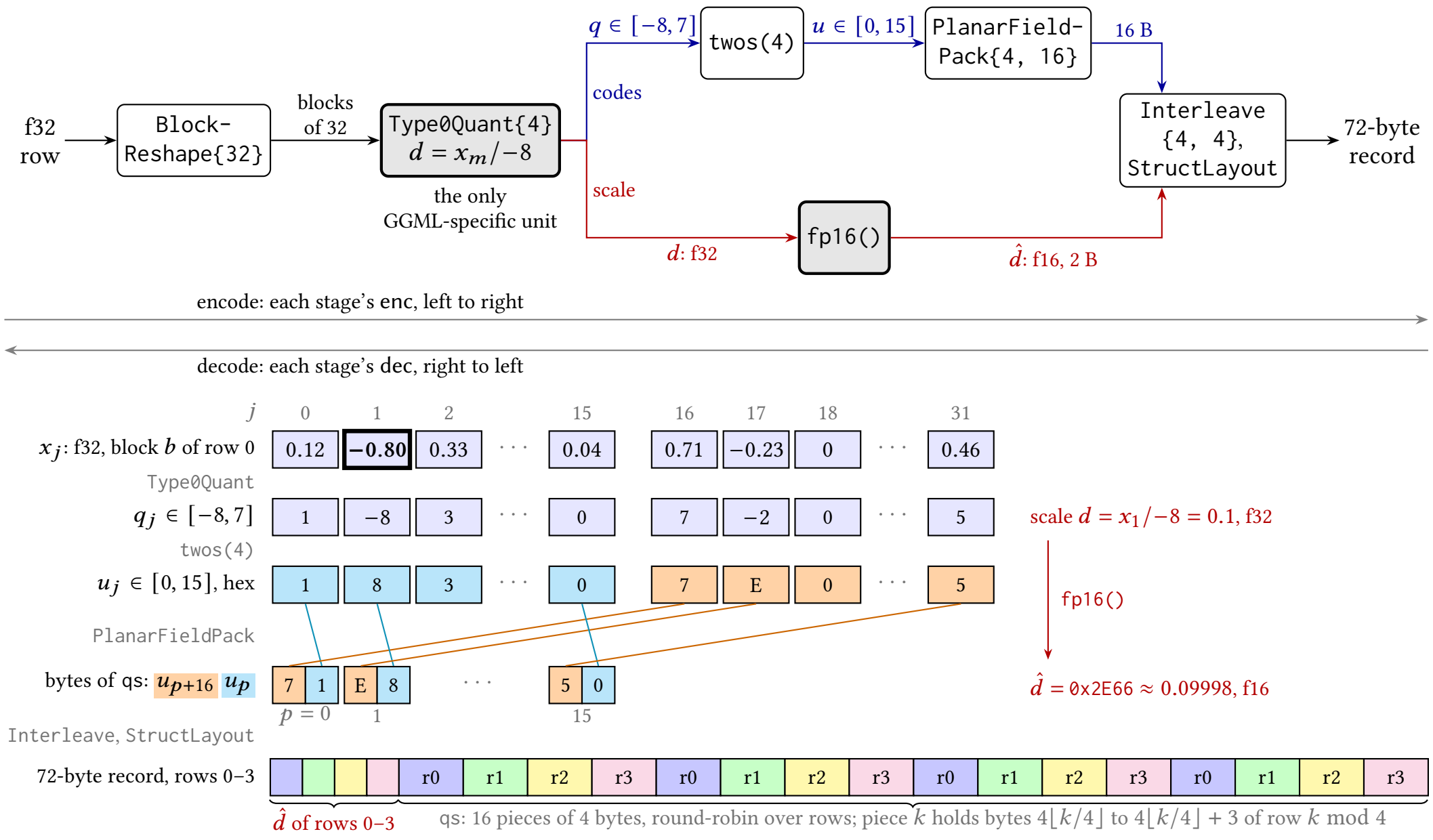
\begin{figure}[t]
\centering
\colorlet{lo}{cyan!25}\colorlet{hi}{orange!35}\colorlet{loL}{cyan!70!black}\colorlet{hiL}{orange!80!black}%
\colorlet{rA}{blue!22}\colorlet{rB}{green!22}\colorlet{rC}{yellow!40}\colorlet{rD}{magenta!18}%
\colorlet{sc}{red!70!black}\colorlet{cd}{blue!60!black}%
\begin{tikzpicture}[
  font=\scriptsize,
  st/.style={draw, rounded corners=2pt, inner sep=2pt, align=center, minimum height=7mm},
  lossy/.style={st, fill=gray!20, thick},
  dat/.style={align=center, inner sep=1pt},
  arr/.style={-{Stealth[length=3.5pt]}},
  lab/.style={font=\tiny, align=center, inner sep=1.5pt}]
\node[dat] (in) at (0.35,0) {f32\\row};
\node[st] (br) at (1.85,0) {\code{Block-}\\\code{Reshape\{32\}}};
\node[lossy] (tq) at (4.55,0) {\code{Type0Quant\{4\}}\\$d=x_m/{-8}$};
\node[lab, below=1pt of tq] {the only\\\GGML-specific unit};
\node[st] (tw) at (7.3,0.95) {\code{twos(4)}};
\node[st] (pk) at (9.8,0.95) {\code{PlanarField-}\\\code{Pack\{4, 16\}}};
\node[lossy] (fp) at (8.2,-0.95) {\code{fp16()}};
\node[st] (il) at (11.7,0) {\code{Interleave}\\\code{\{4, 4\}},\\\code{StructLayout}};
\node[dat] (out) at (13.45,0) {72-byte\\record};
\draw[arr] (in) -- (br);
\draw[arr] (br) -- node[lab, above] {blocks\\of 32} (tq);
\draw[arr, cd] (tq.east) -- ++(0.25,0) |- node[lab, pos=0.25, right] {codes} node[lab, pos=0.75, above] {$q\in[-8,7]$} (tw.west);
\draw[arr, sc] (tq.east) -- ++(0.25,0) |- node[lab, pos=0.25, right] {scale} node[lab, pos=0.75, below] {$d$: f32} (fp.west);
\draw[arr, cd] (tw) -- node[lab, above] {$u\in[0,15]$} (pk);
\draw[arr, cd] (pk.east) -| node[lab, pos=0.3, above] {16 B} ($(il.north)+(-0.4,0)$);
\draw[arr, sc] (fp.east) -| node[lab, pos=0.3, below] {$\hat d$: f16, 2 B} ($(il.south)+(-0.4,0)$);
\draw[arr] (il) -- (out);
\draw[arr, gray] (0,-1.75) -- node[lab, above, black, pos=0.25] {encode: each stage's $\enc$, left to right} (13.9,-1.75);
\draw[arr, gray] (13.9,-2.05) -- node[lab, below, black, pos=0.75] {decode: each stage's $\dec$, right to left} (0,-2.05);
\end{tikzpicture}

\vspace{4pt}
\begin{tikzpicture}[
  font=\scriptsize,
  cell/.style={draw, minimum width=6.4mm, minimum height=3.6mm, inner sep=0pt, font=\tiny},
  half/.style={draw, minimum width=3.2mm, minimum height=3.6mm, inner sep=0pt, font=\tiny},
  rl/.style={anchor=east, font=\tiny, align=right, inner sep=1pt},
  sl/.style={anchor=east, font=\tiny, inner sep=1pt, text=black!55},
  idx/.style={font=\tiny, text=black!55, inner sep=0.5pt},
  arr/.style={-{Stealth[length=3pt]}},
  rec/.style={draw, minimum height=3.6mm, inner sep=0pt, font=\tiny, anchor=west}]
\node[idx, anchor=east] at (2.45,0.36) {$j$};
\node[idx] at (2.9,0.36) {0};
\node[idx] at (3.6,0.36) {1};
\node[idx] at (4.3,0.36) {2};
\node[idx] at (5.6,0.36) {15};
\node[idx] at (6.6,0.36) {16};
\node[idx] at (7.3,0.36) {17};
\node[idx] at (8.0,0.36) {18};
\node[idx] at (9.3,0.36) {31};
\node[rl] at (2.45,0) {$x_j$: f32, block $b$ of row 0};
\node[cell, fill=rA!45] (c0) at (2.9,0) {$0.12$};
\node[cell, fill=rA!45,very thick] (c1) at (3.6,0) {$\boldsymbol{-0.80}$};
\node[cell, fill=rA!45] (c2) at (4.3,0) {$0.33$};
\node[idx] at (4.95,0) {$\cdots$};
\node[cell, fill=rA!45] (c4) at (5.6,0) {$0.04$};
\node[cell, fill=rA!45] (c5) at (6.6,0) {$0.71$};
\node[cell, fill=rA!45] (c6) at (7.3,0) {$-0.23$};
\node[cell, fill=rA!45] (c7) at (8.0,0) {$0$};
\node[idx] at (8.65,0) {$\cdots$};
\node[cell, fill=rA!45] (c9) at (9.3,0) {$0.46$};
\node[rl] at (2.45,-0.66) {$q_j\in[-8,7]$};
\node[cell, fill=rA!45] (c0) at (2.9,-0.66) {1};
\node[cell, fill=rA!45] (c1) at (3.6,-0.66) {$-8$};
\node[cell, fill=rA!45] (c2) at (4.3,-0.66) {3};
\node[idx] at (4.95,-0.66) {$\cdots$};
\node[cell, fill=rA!45] (c4) at (5.6,-0.66) {0};
\node[cell, fill=rA!45] (c5) at (6.6,-0.66) {7};
\node[cell, fill=rA!45] (c6) at (7.3,-0.66) {$-2$};
\node[cell, fill=rA!45] (c7) at (8.0,-0.66) {0};
\node[idx] at (8.65,-0.66) {$\cdots$};
\node[cell, fill=rA!45] (c9) at (9.3,-0.66) {5};
\node[rl] at (2.45,-1.32) {$u_j\in[0,15]$, hex};
\node[cell, fill=lo] (c0) at (2.9,-1.32) {1};
\node[cell, fill=lo] (c1) at (3.6,-1.32) {8};
\node[cell, fill=lo] (c2) at (4.3,-1.32) {3};
\node[idx] at (4.95,-1.32) {$\cdots$};
\node[cell, fill=lo] (c4) at (5.6,-1.32) {0};
\node[cell, fill=hi] (c5) at (6.6,-1.32) {7};
\node[cell, fill=hi] (c6) at (7.3,-1.32) {E};
\node[cell, fill=hi] (c7) at (8.0,-1.32) {0};
\node[idx] at (8.65,-1.32) {$\cdots$};
\node[cell, fill=hi] (c9) at (9.3,-1.32) {5};
\node[sl] at (2.45,-0.33) {\code{Type0Quant}};
\node[sl] at (2.45,-0.99) {\code{twos(4)}};
\node[sl] at (2.45,-1.81) {\code{PlanarFieldPack}};
\node[font=\tiny, anchor=west, text=sc] (dd) at (9.85,-0.66) {scale $d=x_1/{-8}=0.1$, f32};
\node[font=\tiny, anchor=west, text=sc] (dh) at (9.85,-2.3) {$\hat d=$ \texttt{0x2E66} $\approx0.09998$, f16};
\draw[arr, sc] ($(dd.south west)+(0.3,0)$) -- node[font=\tiny, right] {\code{fp16()}} ($(dh.north west)+(0.3,0)$);
\node[rl] at (2.45,-2.3) {bytes of \code{qs}: {\setlength{\fboxsep}{0.8pt}\colorbox{hi}{$u_{p+16}$}\,\colorbox{lo}{$u_p$}}};
\node[half, fill=hi] (h0) at (2.7399999999999998,-2.3) {7};
\node[half, fill=lo] (l0) at (3.06,-2.3) {1};
\node[half, fill=hi] (h1) at (3.44,-2.3) {E};
\node[half, fill=lo] (l1) at (3.7600000000000002,-2.3) {8};
\node[half, fill=hi] (h2) at (5.4399999999999995,-2.3) {5};
\node[half, fill=lo] (l2) at (5.76,-2.3) {0};
\node[idx] at (4.6,-2.3) {$\cdots$};
\node[idx] at (2.9,-2.5999999999999996) {$p=0$};
\node[idx] at (3.6,-2.5999999999999996) {1};
\node[idx] at (5.6,-2.5999999999999996) {15};
\draw[loL, thin] (c0.south) -- (l0.north);
\draw[hiL, thin] (c5.south) -- (h0.north);
\draw[loL, thin] (c1.south) -- (l1.north);
\draw[hiL, thin] (c6.south) -- (h1.north);
\draw[loL, thin] (c4.south) -- (l2.north);
\draw[hiL, thin] (c9.south) -- (h2.north);
\node[rl] at (2.45,-3.2) {72-byte record, rows 0--3};
\node[sl] at (2.45,-2.8) {\code{Interleave}, \code{StructLayout}};
\node[rec, minimum width=0.314cm, fill=rA] at (2.550,-3.2) {};
\node[rec, minimum width=0.314cm, fill=rB] at (2.864,-3.2) {};
\node[rec, minimum width=0.314cm, fill=rC] at (3.178,-3.2) {};
\node[rec, minimum width=0.314cm, fill=rD] at (3.492,-3.2) {};
\node[rec, minimum width=0.628cm, fill=rA] at (3.806,-3.2) {r0};
\node[rec, minimum width=0.628cm, fill=rB] at (4.434,-3.2) {r1};
\node[rec, minimum width=0.628cm, fill=rC] at (5.062,-3.2) {r2};
\node[rec, minimum width=0.628cm, fill=rD] at (5.690,-3.2) {r3};
\node[rec, minimum width=0.628cm, fill=rA] at (6.318,-3.2) {r0};
\node[rec, minimum width=0.628cm, fill=rB] at (6.946,-3.2) {r1};
\node[rec, minimum width=0.628cm, fill=rC] at (7.574,-3.2) {r2};
\node[rec, minimum width=0.628cm, fill=rD] at (8.202,-3.2) {r3};
\node[rec, minimum width=0.628cm, fill=rA] at (8.830,-3.2) {r0};
\node[rec, minimum width=0.628cm, fill=rB] at (9.458,-3.2) {r1};
\node[rec, minimum width=0.628cm, fill=rC] at (10.086,-3.2) {r2};
\node[rec, minimum width=0.628cm, fill=rD] at (10.714,-3.2) {r3};
\node[rec, minimum width=0.628cm, fill=rA] at (11.342,-3.2) {r0};
\node[rec, minimum width=0.628cm, fill=rB] at (11.970,-3.2) {r1};
\node[rec, minimum width=0.628cm, fill=rC] at (12.598,-3.2) {r2};
\node[rec, minimum width=0.628cm, fill=rD] at (13.226,-3.2) {r3};
\draw[decorate, decoration={brace, mirror, amplitude=2pt}] (2.550,-3.4200000000000004) -- node[idx, below=1.5pt, text=sc] {$\hat d$ of rows 0--3} (3.806,-3.4200000000000004);
\draw[decorate, decoration={brace, mirror, amplitude=2pt}] (3.806,-3.4200000000000004) -- node[idx, below=1.5pt] {\code{qs}: 16 pieces of 4 bytes, round-robin over rows; piece $k$ holds bytes $4\lfloor k/4\rfloor$ to $4\lfloor k/4\rfloor+3$ of row $k\bmod 4$} (13.854,-3.4200000000000004);
\end{tikzpicture}
\caption{q4\_0 in \GGML's repacked $4{\times}4$ layout, assembled from six
units (\autoref{lst:q40}). Only the quantizer is specific to \GGML; the
other five are generic. The two shaded units are lossy: the quantizer and
the f16 rounding of the scale. The other units only rearrange or
re-represent data, so they are lossless. The edges show the data that flows
between stages; each range lies inside the input range of the next stage, so
that stage is lossless on what it receives (\S\ref{sec:example:compose}).
Bottom: one block's values after each encoder stage, with $x_1$ (\textbf{bolded}) as $x_m$, the weight with greatest magnitude;
fields and bytes in hex. Byte $p$ packs $u_p$ and $u_{p+16}$. A record holds
block $b$ of four rows, one color per row.}\label{fig:q40parts}
\end{figure}

The running example is the weight format
\code{q4\_0.i4.4x4}, which is q4\_0 with two's-complement nibbles (\code{i4}) in \GGML's
4-row, 4-byte interleave. The encoder forms blocks of 32 values, produces
signed integers and a scale, converts the integers to two's-complement fields, packs
the fields while rounding the scale to f16, and groups four rows into 72-byte
records (\autoref{fig:q40parts}). The decoder runs the stages in reverse;
\autoref{lst:q40} defines the stages once for both directions. Every part satisfies the round-trip law (Definition~\ref{def:law}) on its own, on a
domain given in \S\ref{sec:example:compose}. Only the quantizer
carries \GGML's arithmetic.

\begin{lstlisting}[float=h,caption={The running example: one list of stages gives both its encoder and its decoder. Simplified from the application's one parameterized builder, given with the units' code in \autoref{app:formats}. \code{QK} is 32.},label=lst:q40]
Interleave il{4, 4,                                  // 4 rows, 4-byte pieces
  {{"d", Float(16)}, {"qs", UInt(8), QK / 2}},       // fields, memory order
  {"qs", "d"}};                                      // components codes, scale -> fields
Approximation q4_0_i4_4x4 = Compose{
  BlockReshape{QK},                                  // k -> (j, b): blocks of 32 values
  Type0Quant{4},                                     // -> integers q in [-8, 7], f32 scale d
  Parallel{{"codes", twos(4)}},                      // q -> 4-bit field in [0, 15]
  Parallel{{"codes", PlanarFieldPack{4, QK / 2}},    // two fields per byte,
           {"scale", fp16()}},                       // f16 scale: MiniFloat{5, 10, 15}
  il,                                                // block b of 4 rows -> one record
  StructLayout{il.record(), {"qs", "d"}}};           // {d[4], qs[64]}: 72 bytes
\end{lstlisting}

\subsection{The parts}\label{sec:example:parts}

\paragraph{\code{Type0Quant\{4\}}.}
The only stage specific to \GGML, it implements \GGML's reference quantizer. For the first value $x_m$ of largest magnitude, the
scale is $d=x_m/{-8}$, giving $x_m$ the integer $-8$, and each value's integer
is $q=\min(15,\lfloor x/d+8.5\rfloor)-8$: $x/d$ rounded half up, with the top
integer clamped (the cast truncates, which is the floor as $x/d+8.5>0$).
As in \GGML, $x/d$ is $x$ times the f32 reciprocal of $d$ (zero if $d=0$). Its
components are \code{codes} $q$ in $[-8,7]$ and an f32 \code{scale} $d$ per block.
Its decode is $d\,q$. Its declared error bound is one quantization step,
$|d|$, rather than the half step that rounding alone would give, because the
clamp can move the largest positive value by a full step.

\paragraph{\code{twos(4)}.}
Converts each signed integer in $[-8,7]$ to its 4-bit two's-complement field,
an unsigned value in $[0,15]$. The listing below is the whole unit, and it
shows what writing a unit takes: an encode expression, a decode expression,
and declarations of types, ranges and losslessness that Halide checks when
the unit is composed (\S\ref{sec:example:compose}).
\begin{lstlisting}
Approximation twos4 = Pointwise{"twos",
    [](Expr q) { return cast<uint8_t>(q) & cast<uint8_t>(15); },
    [](Expr u) { return cast<int8_t>(u << 4) >> 4; }}
  .with_types(Int(8), UInt(8))
  .with_ranges(ApproximationRange(-8, 7), ApproximationRange(0, 15))
  .with_lossless();
\end{lstlisting}
The maps are inverse on $[-8,7]$. \GGML's plain block instead uses
\code{offset(8)}; the repack's XOR with \cc{0x88} converts the fields.
\autoref{lst:q40gen} gives the parameterized version.

\paragraph{\code{fp16()}.}
Abbreviates \code{MiniFloat::binary16()}, the minifloat unit
\code{MiniFloat\{5, 10, 15\}}: 5 exponent bits, 10 mantissa bits and bias 15.
It rounds the scale to f16, to nearest even. It is the rounding codec
$(\pi,\iota)$ of \S\ref{sec:ident:split} for f16: lossy, with a declared bound
of half an f16 ulp, and idempotent.

\paragraph{Generic units.}
The other four units are core-library bijections. \code{BlockReshape\{32\}} maps index $k$ to
$(k\bmod 32,\lfloor k/32\rfloor)$ and passes further dimensions through, so it
applies to a row and to a matrix. \code{PlanarFieldPack\{4, 16\}} puts field $p$
in the low nibble of byte $p$ and field $p+16$ in the high, as \GGML does.
\code{Interleave\{4, 4\}} gathers block $b$ of four
rows into one record in 4-byte pieces, round-robin, and \code{StructLayout}
places the fields into \GGML's 72-byte \cc{block\_q4\_0x4}, scales first.

\subsection{Composition}\label{sec:example:compose}

\paragraph{Units.}
An \code{Approximation}, the API handle for a codec, lossy or lossless, is a
value-semantic, type-erased handle, like \code{std::function}. Any C++ type with \code{encode} and \code{decode} members
converts to one; there is no base class. Encode and decode take and return
vectors of Funcs, so a quantizer can produce integers and a scale and the next
stage can consume either. A unit may also declare a signature, an error bound
and whether it is lossless.

\paragraph{Combinators.}
\code{Compose\{s$_1$, \ldots, s$_n$\}} is the composition of
\S\ref{sec:ident:split} iterated, each stage encoding its predecessor's
encoding, so both halves come from one
list of stages. \code{Parallel} is a product. Its named form,
\code{Parallel\{\{"scale", fp16()\}\}}, routes one component to a child and passes
the others through. By C1 and C2, appending lossless stages preserves a codec's round trip
and law.

\paragraph{Components and ranges.}
Each component (\S\ref{sec:ident:split}) has a name, a type, a dimension count,
a range and tags. \code{codes\_and\_scale(-8, 7)}
declares the quantizer's integers as \code{Int(8)} in $[-8,7]$, tagged
\emph{codes}. That range lies inside \code{twos(4)}'s declared input range,
whose output range $[0,15]$ lies inside the range \code{PlanarFieldPack}
requires. This chain is C2: each lossless stage's precondition is
discharged by the stage before it. \code{check\_ranges} verifies the chain
statically, and reports a broken link as, for example,
\code{input 'value': [0, 15] not within [0, 7]}. Component names flow from
context and are checked, never silently substituted; decode's outputs carry
encode's input names.

\paragraph{Domain.}
The wiring above serves every codec; the guarantees that follow are
q4\_0's. It has two lossy stages, \code{Type0Quant} and \code{fp16()}. Composed, they
satisfy the law on blocks of finite values whose scale $d=x_m/{-8}$ is
zero or rounds to a nonzero finite f16 value: the decoded block's first extreme value,
$-8\,\fpsixteen(d)$, is exact and re-encodes to the same scale and integers.
Outside it, when the scale underflows to zero, blocks with different integers
decode to the same values; \GGML writes the same bytes there, and so do we.

\paragraph{Bounds.}
q4\_0's error is at most one quantization step plus the scale's rounding
error times the integer: rounding the scale to f16 moves a decoded value $d\,q$
by $|q|$ times the scale's error, and $|q|\le8$. Each lossy stage declares a
bound, which the tests check at that stage (\S\ref{sec:props}):
\code{Type0Quant} one step, $|x-d\,q|\le|d|$ in real arithmetic (its f32
decode adds a rounding margin), and \code{fp16()} half
an f16 ulp. Composed (Proposition~\ref{prop:bound}), they give
$|x-\hat d\,q|\le|d|(1+10^{-4}+2^{-8})$ for $\hat d=\fpsixteen(d)$ finite and the
unrounded $|d|\ge2^{-14}$; the $10^{-4}$ is a margin for the quantizer's f32
reciprocal. The bound holds in f32 too, as $\hat d\,q$ is exact there.
\autoref{app:checks} gives the domain's endpoints, the derivation, its
conditions and the bound for smaller scales.

\paragraph{Variants.}\label{sec:example:variants}
Format strings select quantizer, integer representation and layout independently
(\autoref{app:glossary}). The scaled variant keeps the bytes but uses integers
$16q$ and scale $d/16$: a shift or mask then feeds \cc{sdot}
(\autoref{lst:asm}). This is a format choice because the components change meaning;
\S\ref{app:q40variants} gives its construction and divisibility precondition.

\subsection{Other formats}\label{sec:example:formats}

The other formats are built the same way (\autoref{tab:formats}). Each has one quantizer,
transcribed from \GGML or \Kleidi, surrounded by generic lossless parts:
reshapes, integer representations, bit packers, codebooks, derived components and layouts.
\code{Tabulate} turns a unit's decode into a lookup table built by running that
decode on every possible byte, which is how \GGML's Arm code decodes q5's fifth
bits. \code{Derive} appends a component that is a function of other components, such as
q8\_1's block sum (\S\ref{sec:props}). \code{MiniFloat} is a generic
minifloat unit, which covers f16 (\code{fp16()}) and bf16. Bit-level layouts that no
storage directive could express are ordinary units: iq1\_m, for example,
spreads its f16 super-block scale over the top bits of the words that hold its
sub-block scales.

\paragraph{External encoders.}
\code{TrustedInverse\{e, d\}}
pairs an externally supplied encoder $e$ with a decoder $d$, trusting,
without checking, that $d$ decodes every encoding $e$ writes as the format's
definition says. We use it only for the seven grid formats (iq1--iq3), whose
encoder is \GGML's own quantizer, called as an extern stage, while the decoder
is an ordinary, native composition. Those quantizers are importance-weighted
searches over trial scales and grid neighbours, run once, offline.
Differential testing against \GGML's decoder checks the pairs
(\S\ref{sec:eval:correct}).

\section{Scheduling the Halves}\label{sec:sched}

A quantized matrix multiply is written as an ordinary f32 matrix product,
with a codec selected for each operand (\autoref{lst:matmul}); the stored
format is a generator parameter, fixed before scheduling. Everything else is
schedule, and none of it names a format. The schedule finds Funcs by the tags
their components carry and picks a recipe from the codec's structure: the
record shape, the component types and the form of the decoders. One recipe,
for integers and a per-block scale, serves most formats; about a quarter of
the schedule's lines handle structure that only some formats have, such as
float sub-block scales or high bits shared by several integers.

\subsection{Placing the halves}\label{sec:sched:place}

\paragraph{Offline and online.}
The weight encoder runs once, when the model is converted. \code{sever}
removes the encoded Funcs from the kernel and replaces them with the
\code{ImageParam}s that carry the stored records. A second generator, shared
by all formats, severs the same round trip at the same place and compiles its
encoder half, and its decoder half for testing. The bytes the kernel reads are
therefore the bytes the encoder wrote, and both are generated from one
\code{Compose}. Moving the cut changes the storage format, not what is
computed. Some formats also store a \emph{derived component}, a value
precomputed from the others, such as the sum of each activation block
(\S\ref{sec:props:derived}). The dot uses it only after a rewrite, so the
generator names its stored Func explicitly when it severs.
The severed inputs are indexed in records, $K/32$ blocks by $N/4$
row groups for the running example, so the alignment the micro-kernel needs
holds by construction. The generator constrains the other components and the
output to the first component's extents; Halide's buffer checks reject a mismatch
when called.

\paragraph{Activations.}
For f32 activations the kernel itself quantizes them, to \code{q8\_0.4x4}
(\code{f32:q8\_0.4x4} in our format strings). This is the second approximation
in \autoref{lst:matmul}. Its encode Funcs are computed at the root, once per
call, in parallel over row chunks, each stage at the record it fills, so the
record is written once, in its final layout, as \GGML's quantizer writes it.

\paragraph{Decoders.}
A decoder is ordinary Halide code. Once it is inlined into the dot, its
arithmetic is part of the dot's expression, so the directives below rewrite
and vectorize the two together. \code{Func::precompute} evaluates a Func, here \code{Tabulate}'s lookup tables, at
compile time into constant buffers, which are inlined when they fit in a
register, so that the decode uses a vector table lookup.

\paragraph{Layout is a format choice.}
The interleave is not a schedule directive: it changes the bytes, so it is part
of the format and must be chosen before the weights are converted. But it is
chosen for the schedule. The 4x4 interleave makes four rows' integers one 16-byte
load, which is what a by-element \cc{sdot} reads; 4x8 is what \cc{smmla} reads.
Here the approximation and the schedule depend on each other: the format is
fixed before the schedule runs, but the best format depends on which
instructions the schedule will use. The schedule
reads the record shape of the chosen format and selects a micro-kernel for it,
and it rejects combinations for which no execution recipe is available, such as affine weights
without the activation's block sums. The schedule library is otherwise
format-agnostic: format names appear only in the \nFormatRows-row format
table, keyed by weight format, operation and activation format, whose
\nKnobRows knob rows set tuning parameters, such as how many blocks a loop
iteration covers (\autoref{app:knobs}); at their defaults, most cases already
reach \GGML's speed (\S\ref{sec:sched:cost}). The dispatch rule and the table
together select a schedule of the one algorithm for each format combination.

\subsection{The micro- and macro-kernels}\label{sec:sched:micro}

\begin{figure}[t]
\small
\setlength{\tabcolsep}{4pt}
\begin{tabular}{@{}lll@{}}
\toprule
Directive & The per-block dot & Step (rule) \\
\midrule
\code{eager\_inline} & $\sum_j (d_w\,q_w(j))\cdot(d_x\,q_x(j))$ & the decoded product, in f32 \\
\code{hoist\_invariants} & $d_w d_x\cdot g$, \ $g=\sum_j q_w(j)\,q_x(j)$ & scales outside the reduction (\textsc{Hoist}) \\
\code{change\_type(Int(32))} & $g'=\sum_j\mathrm{widen}(q_w(j)\cdot q_x(j))$, in i32 & an integer dot (\textsc{Widen}, \textsc{Retype}) \\
 & $g=\mathrm{cast}_{\mathrm{f32}}(g')$ & cast back, exact for this block \\
instruction selection & $g'$ by \cc{sdot} (\cc{smmla} with i8mm) & 8-bit products into 32-bit lanes \\
\bottomrule
\end{tabular}
\caption{The running example's per-block dot under the directives of
\autoref{lst:blockdot}. $q_w(j)$ unpacks, de-interleaves and sign-extends a
weight integer. Small capitals name the general rewrite rules, stated with all
their side conditions in \autoref{fig:rules}.}\label{fig:chain}
\end{figure}

After an \code{rfactor} gives one partial sum per block, the dot's update is a
sum over the values $j$ of a block of $\dec_W(\ldots)\cdot\dec_X(\ldots)$.
Three directives turn it into an integer dot (\autoref{lst:blockdot});
\autoref{fig:chain} shows each step for the running example.

\begin{lstlisting}[float=h,caption={The per-block dot, abridged. \code{decoders(rs)} selects decode Funcs by tag, never by format.},label=lst:blockdot]
Func block_dot(Func blk, const vector<ApproximationResult> &rs, bool integer, ...) {
  blk.update().eager_inline(decoders(rs));              // the decode, inlined now
  Func codes = blk.update().hoist_invariants();         // scales out: a dot of integers
  return integer ? codes.change_type(Int(32)) : codes;  // exact: an integer dot
}
\end{lstlisting}

\code{eager\_inline} substitutes the decoders into the update now, not
during lowering, so the next directive sees the decoded
expression: $\sum_j (d_w\,q_w(j))\cdot(d_x\,q_x(j))$, where $q_w(j)$ unpacks,
de-interleaves and sign-extends. Like \code{rfactor}~\cite{halide-rfactor},
it is eager: later directives see, and schedule, its result.

\code{hoist\_invariants} factors loop-invariant scales out of the reduction
by distributivity. Because the scales depend on block and row but not on $j$,
the sum becomes $d_w d_x\sum_j q_w q_x$. In general, every factor that does
not depend on the update's reduction variables moves out, and the inner
reduction is returned as a new Func; hoisting nothing is an error. Invariance is decided by Halide's simplifier, given
the bounds of the reduction variables and of the pure variables that
\code{rfactor} created: after a split by 32, the scale read $s((32u+j)/32)$
is $s(u)$. The K-quants' 6-bit sub-block scales vary inside a 256-value block,
so the schedule splits its reduction into sub-blocks and hoists twice.

\code{change\_type(Int(32))} retypes the inner sum (\textsc{Widen},
\textsc{Retype}): it returns a Func that accumulates in 32-bit integers, with
widening multiplies, and makes the old Func a cast of it back to f32. Its
preconditions are proved by Halide's interval analysis, not taken from
declared component ranges; a failed proof is an error.
The integer sum is the exact sum over $\mathbb{R}$, so the rewrite is a
schedule (\S\ref{sec:ident:round}). Converting it back to f32 rounds at most
once. For a q4\_0 by q8\_0 block it never rounds: with scaled integers, every
partial sum has magnitude at most $32\cdot128\cdot128=2^{19}<2^{24}$. A K-quant's super-block sum, which still includes the
integer sub-block scales, may round: its magnitude is bounded by
$256\cdot128\cdot32\cdot127<2^{27}$ for q6\_K, and q4\_K's and q5\_K's
bounds also exceed $2^{24}$ (q2\_K's and q3\_K's do not). That rounding is the permitted f32 change,
not a change of value over $\mathbb{R}$. Then the scales multiply.
Instruction selection does the rest: the widening multiply-add of
8-bit lanes into 32-bit lanes is \cc{sdot}, and with the matrix-multiply
extension a $2\times 8$ by $8\times 2$ product is \cc{smmla}.

\autoref{fig:rules} states each rewrite in
general, with its preconditions. One edge case is an empty reduction domain,
where multiplying the hoisted sum by its factor can change the result; the
write-back is then guarded by a nonemptiness test, which folds away for
constant extents.

\paragraph{Floats.}
Retyping a sum is a schedule only when it is exact, as it is for integers.
When an operand stays in floating point, the hoist still applies, but the
sum stays in f32. Accumulating in a narrower float type, as \GGML's f16 dot
does, changes the result, so we declare it as an approximation:
\code{Accumulate}, a unit whose encode retypes the float sum and whose decode
widens the result back to f32. Its contract bounds the rounding error for
every summation order on supported inputs (\S\ref{app:accumulate}), and the
schedule fixes one order within that bound.

\paragraph{By-element dot products.}
For the running example, \code{blocked\_rows} \code{rfactor}s the in-block
reduction into groups of four integers, stages each group of activation
integers as one 32-bit register lane, and multiplies it by element against a
16-byte load of four weight rows: \Kleidi's \cc{dotprod} kernel. The weight
decode is one shift or mask (\S\ref{app:byelem}).

\paragraph{Other formats.}
Tags select every format's decoders (\autoref{tab:formats}).
An affine format's minimum term ($d\,q+m$) is split off by grouped
\code{hoist\_invariants} and read from activation block sums
(\S\ref{sec:props:derived}).

\paragraph{The macro-kernel.}
\code{tile\_out} tiles the output into register tiles, computes the dot at the
tile, and parallelizes over tasks of balanced cost. Several tile shapes are
kept, each under a \code{specialize} on the output's size: the dispatch on
matrix shape that a kernel library writes by hand. 4x8 records select
\code{blocked\_mmla} (\cc{smmla}) instead of \code{blocked\_rows}.

\subsection{Metal: a compute format}\label{sec:sched:metal}

Apple's GPUs multiply $8\times 8$ matrices of f16 values in hardware
(\cc{simdgroup\_matrix}), but not 4-bit integers. So on Metal the kernel computes
in a second format. The weight format \code{q4\_0:f16} names a \emph{storage}
format, q4\_0, and a \emph{compute} format, f16. The decoded weight is
approximated once more:
\begin{lstlisting}
ApproximationResult aw16 = aw.replacement.approximate_by(f16, {dot});
\end{lstlisting}
where \code{replacement} is the decoded Func that stands for $W$ in the dot.
The activations are approximated by f16 inside the kernel. The second
approximation is lossy, since a decoded weight, the f16 scale times an
integer, need not be an f16 value: the
compute format is an approximation choice, like the stored format, not part of
the schedule. \GGML's Metal matrix kernel makes the same choice, rounding each
decoded weight to f16, so its CPU and GPU paths compute different
approximations and its result depends on the device. In \GGML, only a
template argument of the Metal kernel records this choice; in our code, the
second \code{approximate\_by} states it explicitly. The schedule multiplies the two f16 encodings.
Per step of $K$, it decodes a tile of weight records into threadgroup memory,
where the second encoder runs, and stages a tile of f16 activations beside it.
It multiplies the two in $8\times 8$ fragments, accumulating in f32 in Funcs
stored in \code{MemoryType::Tile}. This matches the structure of \GGML's Metal
kernel, without a hand-written dequantization routine for each format. We added a Metal backend for Halide's tile-memory
matrix fragments, and \code{gpu\_lanes} reductions. Tile shapes are specialized on $M$
as on the CPU, with a matrix--vector schedule for small $M$.

\subsection{What the schedule costs}\label{sec:sched:cost}

One schedule library serves all formats and is a fraction of the application
(\autoref{tab:loc}). With all \nKnobRows rows of per-format knobs
at their defaults, \nKnobDefaultMatch of the
\nKnobCases cases measured for the \nKnobFormats formats that have knobs
(\pctKnobDefault) still reach at least 95\% of \GGML's speed; tuned, all do. The directives we added to Halide, \code{eager\_inline},
\code{hoist\_invariants}, \code{change\_type}, \code{Func::precompute}, Arm
\cc{smmla}/\cc{usdot} patterns and the Metal fragment backend, are general,
not quantization-specific (\autoref{tab:core}).

\section{Validating Codec Contracts}\label{sec:props}

Contracts are declared by the programmer; the compiler checks some of them
when a pipeline is built, and tests check the rest.

\subsection{Properties from laws}\label{sec:props:laws}

A small
testing library runs encode and decode as JIT-compiled pipelines and checks a
property on the whole composition or, with \code{.at(stage)}, on the values
that reach one stage: a range check shows that a packer's inputs are in
range; a \code{lossless} test at the packer, that it is a bijection on them.
\autoref{tab:contracts} lists each format's law status and the evidence
supporting it; independently, every format's bytes are checked against
\GGML's or \Kleidi's (\S\ref{sec:eval:correct}). The tests produced that
classification, including the affine formats'
narrower domain and nvfp4's windows of \GGML's subnormal scale rounding.

\paragraph{Trust boundary.}
Each guarantee is checked by the compiler, tested, or assumed
(\autoref{tab:trust}). The compiler's checks trust Halide's simplifier and
interval analysis and the derived-component normalizer; negative tests
exercise each check (\autoref{app:checks}). Claims about compiled kernels are
empirical: they rest on the tests.

\begin{table}[t]
\caption{The trust boundary: how each guarantee is established.}\label{tab:trust}
\small
\begin{tabular}{@{}p{0.17\linewidth}p{0.79\linewidth}@{}}
\toprule
How & Guarantees \\
\midrule
{\raggedright Checked when a pipeline is built\par} &
{\raggedright Types, dimensions and component names; the chain of declared
ranges (C2); derived-component matching (symbolic) and which components may
declare a bound (\S\ref{sec:props:derived}); \code{change\_type}'s interval
proof (conservative: it may reject a safe program, with a compile error);
\code{hoist\_invariants}' invariance and empty-domain guard (built
conservatively for float sums, as the target is not yet known).\par} \\
\addlinespace
{\raggedright Tested, on sampled or exhaustive inputs\par} &
{\raggedright The round-trip law, losslessness, bounds, ranges, derived
components, and \code{closed()}, which catches a decoder that reads the
original tensor instead of its encoding (the seven grid formats skip it, as
their embedded tables count as foreign buffers; \autoref{app:laws}); bytes
against \GGML's and \Kleidi's; kernels against an f64 oracle
(\S\ref{sec:eval:correct}).\par} \\
\addlinespace
{\raggedright Assumed: user obligations\par} &
{\raggedright Supported inputs (\S\ref{sec:ident:round}); \code{sever}'s
buffers; the encoders given to \code{TrustedInverse}; derived components'
$\beta$ (also tested); \code{unsafe} accumulation; the multiples of 16 that
the scaled q4\_0 variant needs (\S\ref{app:q40variants}; checked by byte comparison).\par} \\
\bottomrule
\end{tabular}
\end{table}

\subsection{Derived components}\label{sec:props:derived}

A derived component caches a value that a kernel would otherwise recompute
from the decoded block. q8\_1 stores, with each block's integers $q$ and
scale $\hat d$, a sum $s$ approximating $\psi=\hat d\sum_j q_j$, so that a
kernel multiplying it by an affine weight block, $d_w q_w + m_w$, computes
the offset term $m_w\,\psi$ from $s$ instead of summing the activations again.
A \code{Derive} stage appends such a component, and decode ignores it.
Selecting the codec authorizes consumers to read $s$ in place of $\psi$, an
approximation whose point of application the schedule chooses
(\autoref{tab:preserve}). The compiler substitutes only where the consumer's
term matches $\psi$ as an equality over $\mathbb{R}$, with roundings opaque;
where the stored type converts exactly to the term's, so the cast adds no
error; and where the codec declares a bound $\beta$ on $|s-\psi|$, which it
can only if no later lossy stage changes what $\psi$ reads and at most one
rounds $s$. Anything else is a build error. A term $c\,\psi$ read as $c\,s$
then moves by at most $|c|\,\beta$. That $s$ lies within $\beta$ of $\psi$
is the user's tested obligation; \autoref{app:laws} gives the rule, its
normal form, the trace conditions and the proof of this bound.

\subsection{A block sum that disagrees with its block}\label{sec:props:q81}

The round-trip law (Definition~\ref{def:law}) found an inconsistency in
\GGML's q8\_1 without a reference to compare against. Decoding and
re-encoding changed \pctQEightOneDiffer of q8\_1 blocks of normally
distributed inputs, and we found two blocks with the same integers and scale
whose sums differ. The two decode to the same values, so \GGML's stored sum
is not a function of the block it describes.

The cause is the order of rounding. \GGML computes the sum from the scale
before rounding the scale to f16 for storage, so the sum keeps information
that the stored scale has lost; re-encoding the decoded block computes it from
the rounded scale instead. Our model rejects that order before any test
runs: the scale's rounding changes what the sum's definition reads, so the sum
can declare no bound, and a kernel that reads it does not build (C3,
\S\ref{sec:ident:consequences}).

Our composition rounds the scale first and derives the sum from the stored
scale and integers. The stored sum is then the decoded block's sum rounded
once, never farther from it than \GGML's, and closer on nearly every block
where the two differ (\autoref{app:checks}). Our q8\_1 passes the round-trip
law and the derived-component tests at no run-time cost. Quantization error
dwarfs the difference, so inference barely changes: the law caught a defect
of consistency, not of accuracy. \GGML does not promise the law, and either
side's kernels read the other's blocks, though imported \GGML sums carry a
weaker bound. This is our one deliberate departure from \GGML's bytes.

\section{Evaluation}\label{sec:eval}

\subsection{Methodology}\label{sec:eval:method}

We measure on an Apple M3 Max against \GGML \ggmlVersion~\cite{ggml}, which
has four paths for these kernels: plain CPU (C: hand-written Neon
\cc{vec\_dot}s, which it loops for matrix--vector (GEMV) and matrix--matrix
(GEMM) products), repack (R: interleaved weights with dot-product and i8mm
kernels), \Kleidi (K: v1.24.0~\cite{kleidiai}; q4\_0, and q8\_0 requantized
to \cc{qsi8cx}), and Metal. A \emph{case} is a weight format, an activation
format, a shape and a thread count. Shapes $K\times N$ are Llama-3-8B's
projections, $\{4096\times4096, 4096\times1024, 4096\times14336,
14336\times4096\}$, with $M=1$ (GEMV) or $M\in\{8,32,512\}$ (GEMM);
\cc{vec\_dot} has $K\in\{4096,14336\}$; CPU kernels use one or ten threads
(T1, T10; T1 at $M=512$: $4096\times1024$ only). Weights, and activations
given in a quantized format, are encoded untimed, as at model load; f32
activations are encoded inside the timed call, as by \GGML's \cc{mul\_mat}.
We report the median over 5 to 15 interleaved rounds, keeping the minimum
over re-runs of low cases; on Metal both sides wait once per 16
multiplications (\autoref{app:method} gives the protocol and other
estimators).
The metric is the speedup $\GH = t_{\GGML} / t_{\text{ours}}$; a kernel
\emph{matches} \GGML when $\GH \ge 0.95$, about twice the median half-width
of a CPU timing's 95\% confidence interval (full width \noiseCPU), and
\emph{exceeds} it when $\GH > 1$. The baseline is \GGML's fastest path other
than Accelerate (Apple's BLAS; \S\ref{sec:eval:perf} compares with it
separately), and its kind splits the cases into two strata:
\emph{format-specific}, where it is a repacked, \Kleidi or Metal kernel or a
standalone \cc{vec\_dot}, and \emph{blocked-dot fallback}, where it loops
\cc{vec\_dot} over the output. We fix one stored layout per target and
weight type, chosen on the other three shapes (\emph{held-out-shape layout}); only the diagnostics of \autoref{fig:cdf} and
\autoref{app:family} take the fastest layout per case.

\subsection{Coverage}\label{sec:eval:coverage}\label{sec:eval:matrix}

One algorithm covers every quantized type in \GGML and more: \nFormats
formats, namely all \nGGMLTypesAll quantized types in \GGML v\ggmlVersion,
\Kleidi's \cc{qsi8cx} and \cc{qai8dx}, and f16 and bf16
(\autoref{app:formats}). A table with a row per stored format instantiates
it as \nKernels CPU and Metal kernels with codecs
(\autoref{app:size}).

The same table fills the kernel matrix that \GGML fills by hand. Of the
\nPairsCPU CPU weight--activation pairs \GGML runs on this CPU, which lacks
SVE and SME, only \nPairsTiled have a tiled kernel; the others loop
\cc{vec\_dot} over the outputs or, for large f32 products, call Accelerate
(\S\ref{sec:eval:perf}). Every pair of ours is tiled, including
\nPairsNoGGML pairs for which \GGML has no kernel at all.

\subsection{Correctness}\label{sec:eval:correct}

The checker (\autoref{app:checks}) reports \nChecks results with no failures, and exhaustive sweeps
confirm the round-trip law of \nExhFormats block formats on every scale or
$(d,m)$ pair of their compiled codecs.
Every decoder and every encoder we implement produces exactly \GGML's bits,
on adversarial blocks and on blocks that cover every value of each stored
byte, wherever \GGML's result is defined. (It is undefined where \GGML's C
code converts a non-finite or out-of-range float to an integer.) There are two
exceptions: q8\_1's block sum, which we correct on purpose
(\S\ref{sec:props:q81}), and mxfp4's encoder on blocks that contain
infinities or NaNs. Likewise, our repacked layouts match \GGML's
repacking, our \cc{qsi8cx} and \cc{qai8dx} match \Kleidi's packers, and the seven grid
IQ encoders call \GGML's (\code{TrustedInverse}).
Every kernel, ours and \GGML's, is compared on every output with an f64
product of the decoded operands, within a tolerance that follows from its
precision, chain length, rounding and requantization.
Results that are not bit-exact differ in the last bits, as expected when two
kernels sum in different orders. The checks discriminate: every output-changing
mutant is rejected, some only by the bit-exact and structured checks beside
the tolerance (\autoref{tab:mutants}).

\begin{table}[t]
\caption{Marginal cost of one more format, layout or target. Ours: sloc of
application source, with no Halide or schedule changes unless stated, plus
the format-table rows and tuning knobs it adds, counted separately. \GGML:
sloc of functions named for the format or layout.}\label{tab:marginal}
\footnotesize
\setlength{\tabcolsep}{4pt}
\begin{tabular}{@{}lll@{}}
\toprule
Extension & Ours & \GGML \\
\midrule
format sharing units (q5\_1, given q5\_0) & 5 format-definition lines, 2 rows & 264 \\
K-quant sharing a quantizer (q5\_K, given q4\_K) & 15 format-definition lines, 3 rows, 3 knobs & 1297 (697 repack) \\
format with new units (mxfp4) & 68 format-definition lines, 3 rows & 469 \\
format with a new quantizer (q2\_0) & 32 format-definition lines, 3 rows & 225 \\
weight layout (q4\_0 $4{\times}8$) & 1 row & 558 + 58 shared \\
K-quant layout (q6\_K $8{\times}8$, given q4\_K's) & 8 format-definition lines, 1 row, 1 knob & 373 \\
backend for a format (q5\_1 on Metal) & 1 row & 50 + templates \\
GEMM on pre-quantized activations & 0 lines (all formats) & not provided \\
new strategy (\cc{qsi8cx}/\cc{qai8dx}) & 106 format-definition, 106 schedule lines, 3 rows & 2233 + glue \\
CPU target (x86, \S\ref{sec:eval:x86}) & \xLocSched schedule, \xLocBackend Halide lines, \xLocTable knobs & \xLocGGML \\
\bottomrule
\end{tabular}
\vspace{8pt}
\caption{Code size in lines and sloc (non-blank, non-comment lines).
Ours: whole sources. \GGML: functions named after one of our \nGGMLTypes types and
our layouts on its Arm and Metal paths, plus the \Kleidi sources it compiles for
q4\_0 and q8\_0. Bottom: our Halide additions, approximation-specific and
general (\autoref{app:size}).}\label{tab:loc}
\small
\setlength{\tabcolsep}{4pt}
\begin{tabular}{@{}lrr@{\qquad}lrr@{}}
\toprule
Ours & lines & sloc & \GGML and \Kleidi & lines & sloc \\
\midrule
format definitions\textsuperscript{a} & 2728 & \locSchemes & reference quantizers\textsuperscript{c} & 2416 & 2047 \\
algorithm & 292 & \locAlgorithm & generic CPU dot products & 1247 & 1063 \\
codec generator & 88 & 69 & Neon dot products\textsuperscript{d} & 2881 & 2221 \\
schedules & 2456 & \locSchedule & generic repack & 2041 & 1726 \\
format table\textsuperscript{b} & 377 & 260 & Neon repack\textsuperscript{d} & 3226 & 2748 \\
 & & & Metal shaders\textsuperscript{e} & 1682 & \locGGMLMetal \\
 & & & \Kleidi Neon kernels, packers\textsuperscript{f} & 7038 & 4965 \\
\midrule
total & 5941 & \locOurs & total & 20531 & \locGGML \\
\midrule
\multicolumn{3}{@{}l}{\emph{Our Halide additions}} & & & sloc \\
\multicolumn{3}{@{}l}{Halide: \code{approximate_by}, \code{Approximation}, \code{sever}} & & & \locApproxSpecific \\
\multicolumn{3}{@{}l}{total with these} & & & \locOursApprox \\
\multicolumn{3}{@{}l}{Halide: general directives, analyses, back ends} & & & \locGeneral \\
\multicolumn{3}{@{}l}{total with all our Halide additions} & & & \locOursFramework \\
\bottomrule
\end{tabular}

\smallskip
\raggedright\footnotesize
Not counted: tests (with our \locPropTesting-sloc property-testing library),
harnesses, thread pools, our x86 (\S\ref{sec:eval:x86}) and CUDA code.
\textsuperscript{a}Without 323 sloc of grid tables from \GGML.
\textsuperscript{b}With \nKnobRows rows of knobs.
\textsuperscript{c--f}\autoref{app:size}, which also lists the sources.
\end{table}

The round-trip law (\autoref{tab:contracts}) holds exactly on their domains
for the formats with closed-form encoders, except q8\_K and \cc{qai8dx},
whose scales round-trip within 1--2 ulp. It fails for the search encoders,
which re-fit their own decoded block; the grid IQ formats declare none.

\subsection{Code size}\label{sec:eval:size}

Codecs and schedules make our application \locRatio times smaller than
\GGML's (\autoref{tab:loc}), and
an extension costs a few lines (\autoref{tab:marginal}). Reusing units costs
a few lines and a table row; new units, as for mxfp4, need no schedule
changes; even a new \emph{strategy}, \Kleidi's per-row accumulation, costs
212 lines. \GGML hand-writes every extension. The schedules name no format.
Even counting all our Halide additions, including directives, analyses and
back ends that serve any Halide client, our whole implementation is smaller
than \GGML's, though we would not count improvements to a C compiler's
vectorizer or register allocator against \GGML.

\begin{table}[t]
\caption{Speedup \GH over \GGML's fastest path, with held-out-shape layouts:
cases, median \GH and share matched ($\GH\ge0.95$), by target and stratum of
\GGML's path. Accelerate: the Neon cases, with \GGML's BLAS path counted.
x86: one and twelve threads (\autoref{tab:x86}). Neon and Metal together:
format-specific, median \medHHFixed with \pctHHFixed matched; all, median
\medAllFixed with \pctMatchAllFixed matched.}\label{tab:summary}
\footnotesize
\setlength{\tabcolsep}{3pt}
\centering
\begin{tabular}{@{}l*{4}{rrr}@{}}
\toprule
& \multicolumn{9}{c}{Apple M3 Max} & \multicolumn{3}{c@{}}{Xeon w5-2455X} \\
\cmidrule(lr){2-10}\cmidrule(l){11-13}
& \multicolumn{3}{c}{Neon} & \multicolumn{3}{c}{Accelerate}
& \multicolumn{3}{c}{Metal} & \multicolumn{3}{c@{}}{x86} \\
\cmidrule(lr){2-4}\cmidrule(lr){5-7}\cmidrule(lr){8-10}\cmidrule(l){11-13}
& Cases & Med. & $\ge0.95$ & Cases & Med. & $\ge0.95$
& Cases & Med. & $\ge0.95$ & Cases & Med. & $\ge0.95$ \\
\midrule
Format-specific & \nHHCPU & \medHHCPUFixed & \pctHHCPUFixed
& \nHHCPU & \medHHCPUBlasFixed & \pctHHCPUBlasFixed
& \nHHMetal & \medHHMetal & \pctHHMetal & \xFSN & \xFSMed & \xFSPct \\
Blocked-dot & \nLoopCPU & \medLoopCPUFixed & \pctLoopCPUFixed
& \nLoopCPU & \medLoopCPUBlasFixed & \pctLoopCPUBlasFixed
& --- & --- & --- & \xBDN & \xBDMed & \xBDPct \\
All & \nCasesCPU & \medCPUFixed & \pctMatchCPUFixed
& \nCasesCPU & \medCPUBlasFixed & \pctMatchCPUBlasFixed
& \nCasesMetal & \medHHMetal & \pctHHMetal & \xCases & \xMed & \xPctMatch \\
\bottomrule
\end{tabular}
\end{table}

\subsection{Performance}\label{sec:eval:perf}

Our kernels are on par with \GGML's hand-written format-specific kernels
and much faster than its blocked-dot fallbacks (\autoref{tab:summary}).
\autoref{app:fixed} gives the distributions, bandwidth and whole-block
projections behind this section.

\paragraph{Format-specific kernels.}
Against \GGML's repacked, \Kleidi, Metal and \cc{vec\_dot} kernels, we match
in \pctHHFixed of the cases, with a median speedup of \medHHFixed{}$\times$.
The result does not come from bf16, where the baseline is weak (\GGML's Arm
bf16 \cc{vec\_dot} is a scalar loop accumulating in double, and \Kleidi's bf16
kernels need SME2): without bf16, both figures are essentially unchanged.
GEMVs whose weights exceed the caches are limited by memory bandwidth; there
our kernels reach at least as large a share of the measured bandwidth as
\GGML's.

\paragraph{Blocked-dot fallbacks.}
Where \GGML loops \cc{vec\_dot}, our kernels are tiled: the median speedup is
\medLoopGEMMFixed{}$\times$ for GEMM and \medLoopGEMVFixed{}$\times$ for GEMV,
largest for the grid IQ formats (\autoref{app:family}). On the CPU, the
advantage carries over to the projections of a whole Llama-3-8B block, in
decode and in prefill.

\paragraph{Accelerate.}
For most f32 GEMMs at $M\ge32$, \GGML can dequantize the weights for
Accelerate's \cc{sgemm}, which runs on Apple's AMX coprocessor and is usually
faster than our kernels (\autoref{app:blas}). Counting
it, the blocked-dot GEMM median falls to \medLoopGEMMBlasFixed{}$\times$, and
512-token prefill of blocked-dot formats is slower than \GGML's.
This gap concerns hardware access, not the representation: Accelerate
uses Apple's undocumented AMX, which Halide does not target. Separating codecs
from schedules does not depend on the ISA, though a new ISA can
need backend and scheduling work. Newer Apple chips expose Arm's documented
SME, which Halide targets but our M3~Max lacks.

\paragraph{Where we fall short.}
\nBelowCPU CPU cases and \nBelowMetal Metal cases stay below 95\% of
\GGML's speed (\autoref{app:fixed} lists them). Each has an identified cause
in a schedule, in Halide's code generation or in its runtime, not in the
representation. On the CPU, most are ten-thread GEMVs at the smallest shape,
too short to amortize starting and joining the threads, or q1\_0 at $M=8$,
where \GGML's repacked kernel is faster. In one \cc{qsi8cx} case both
kernels are limited by \cc{smmla} throughput, but our fixed weight layout
takes about three times as many L1 misses per weight cache line as \Kleidi's.
On Metal, GEMMs at $M=8$ are slower because Halide loads
\cc{simdgroup\_matrix} fragments only from memory, whereas \GGML builds
dequantized fragments in registers. GEMVs, mostly at the smallest
shape, pay for Halide's runtime setting up a command buffer and pipeline
state per launch, and bf16 GEMMs at $M=512$ for its Metal backend's
lack of the \cc{bfloat} type \GGML uses.

\subsection{A second CPU target: x86}\label{sec:eval:x86}

We ported the CPU schedules to x86 and timed them on an Intel Xeon w5-2455X
(Sapphire Rapids, 12 cores, with AVX-512 VNNI, BF16 and FP16, and Intel's
AMX matrix unit, unrelated to Apple's) against the same \GGML release, built
natively with its AMX and repacked kernels. \autoref{app:x86} gives the
protocol, the line counts, results by path and the gaps. The cases are
those of \S\ref{sec:eval:method}, on one and twelve threads (T1, T12). Besides
looping \cc{vec\_dot}, \GGML has three format-specific paths on x86: its AMX
back end, repacked $8{\times}8$ AVX kernels and a hand-tiled AVX-512 VNNI
GEMM for K-quant and IQ types.

\paragraph{Code.}
The port adds no format definitions or layouts and does not change the
algorithm. It adds x86-only schedules for AMX and VNNI tiles, a few knob
rows, and x86 instruction selection in Halide; Halide's
AMX lowering already existed. The AMX schedules use the shared $16{\times}4$
layout (\autoref{app:formats}), which interleaves 16 rows in 4-byte pieces, as
AMX's integer tile products take their second operand. In all, the port is
\xLocOurs sloc; \GGML's x86-specific code for the same formats is \xLocGGML
sloc, \xLocRatio times as much.

\paragraph{Performance.}
Over all \xCases x86 cases we match \GGML in \xPctMatch, with a median
speedup of \xMed{}$\times$ (\autoref{tab:summary}). We are strongest in
multithreaded GEMMs with small batches ($M=8$ and $32$, as in batched
decoding and chunked prompts): there we match nearly all of \GGML's
format-specific kernels, its tiled VNNI GEMM included, and run well ahead of
its blocked-dot fallbacks. In multithreaded prefill ($M\ge32$) we are a
median \xPreAmxMed{}$\times$ faster than \GGML's AMX kernels, more on
shapes exceeding the L3 cache, and match every case against its repacked
kernels and blocked-dot fallbacks.

\paragraph{Where we fall short on x86.}
Three gaps remain, each against a kernel \GGML wrote by hand for these
formats: 512-token prefill against the tiled VNNI GEMM for K-quant and IQ
types; single-threaded GEMMs against AMX, partly because one fixed layout
serves both thread counts; and decoding iq4\_nl and mxfp4 against \GGML's
$8{\times}8$ repacked GEMV. None of these is a limit of the
representation: the formats, algorithm and codecs are those that already
run at parity or better. What remains is tile shapes, staging and
instruction selection, the same kind of work as the port.

\subsection{Threats to validity and limitations}\label{sec:eval:scope}
We time kernels, not inference, on one M3 Max and one x86 workstation, with
knobs tuned at one shape (\autoref{app:knobs}; x86 adds a few knob rows of its own). The x86 machine
runs Linux under WSL2, a virtual machine, and its sweep follows a related
but different protocol (\autoref{app:x86}). On Metal, \GGML uses one command
buffer, Halide's runtime one per launch; synchronizing once per sample
instead, we match \GGML in \pctMatchMetalPerSample
of Metal cases (\autoref{tab:sync}).
Design limits: codecs are chosen by hand, block sizes are static, and
data-dependent sizes are outside the model.

\section{Related Work}\label{sec:related}

We propose neither quantization methods, nor formats, nor a scheduler, but
approximations selected as composable codecs whose two halves are Funcs that a
Halide schedule places. Each group below locates that boundary.

\paragraph{Low-precision compilers.}
Ladder~\cite{ladder} is closest in aim. Its \emph{tType} is a tile of $n$-bit
elements sharing metadata, such as MXFP8's scale or NF4's value map, with
functions that convert it ``losslessly'' to wider types; its tile primitives
include \emph{convert}, and a hint-based policy picks the memory level where
each conversion runs, on NVIDIA and AMD GPUs. Its library,
BitBLAS~\cite{bitblas}, configures a matmul by weight type, group size and one
of three zero-point formulas, and its weight transform takes integers already
quantized: Ladder compiles and places decoding automatically. TVM~\cite{tvm}
adopts Halide's split of algorithm and schedule and packs low-precision values
into wider types for bit-serial operators; its QNN dialect~\cite{tvm-qnn}
compiles pre-quantized INT8 models. MLC-LLM~\cite{mlc-llm} writes schemes such
as group quantization as classes whose quantize and dequantize are TVM tensor
expressions; a pass fuses the dequantization into the matmul, and default
rule-based schedules apply. MLIR's quantization
dialect~\cite{mlir,mlir-quant} makes per-tensor, per-axis or block-wise scales
and zero points parameters of a \code{quant.uniform} type.

Each of these systems models numerical formats directly: as a tile type
(Ladder), operator parameters (BitBLAS, QNN), scheme classes (MLC-LLM) or a
type (MLIR). For q4\_0, the relevant choices are the signed scale rule, f16 rounding,
offset and packing. Ladder exposes decoding through tType conversions;
MLC-LLM's cited group scheme uses a different, nonnegative scale rule.
MLIR's \code{quant.uniform} stores scales in the type; q8\_0's run-time
scales would instead need to remain values in the IR.
Our compiler extension models no format: its one abstraction is a pair of Funcs related by a
round-trip law, and q4\_0 composes generic parts, only its quantizer specific
to \GGML (\S\ref{sec:example}); a variant swaps one part (\autoref{tab:marginal}), and
the user's schedule, not a fixed policy, places each half. Ladder and MLC-LLM
offer automatic scheduling; our schedule library is written by hand. As Ladder
targets discrete GPUs and MLC-LLM does not load \GGML's formats, \GGML is our
baseline.

\paragraph{User scheduling.}
Halide's~\cite{halide-pldi13,halide-cacm18} schedules preserve values, up to
\code{rfactor}'s reassociation~\cite{halide-rfactor}; we keep this and add a
selection step, the one that may change values over~$\mathbb{R}$
(\S\ref{sec:ident:round}). Halide's casts select scalar types; its storage
directives rearrange or place whole elements. Exo~\cite{exo} is closest to our
selection step: procedures can be written over an abstract numeric type
\code{R}, and \code{set\_precision} later picks a scalar type such as
\code{f32} or \code{i8}; a packed format with per-block metadata is not a
scalar type. A Func library can express both halves; our extension adds
consumer rewriting, role annotations, derived-component matching and checked
reduction retyping (\S\ref{sec:ident:handles}).

\paragraph{Compositional representations.}
TACO~\cite{taco-formats} composes level formats, described by capabilities and
properties, into dense, CSR, COO, DIA, ELL and hash tensors. Our units
compose similarly, but transform values, not coordinates, and may lose
information. Narcissus~\cite{narcissus} derives encoders and decoders from
format specifications in Coq, guaranteed to be inverses, and
lenses~\cite{lenses} pair \emph{get} and \emph{put} under round-trip laws.
Our round-trip law (Definition~\ref{def:law}) is the generalized-inverse
condition~\cite{penrose-ginverse}; it does not require losslessness. Its role
is new: an optional contract on composable lossy codecs, checked on compiled
codecs by enumeration and tests, not mechanized proof, whose consequences
govern repeated conversion and derived components (\S\ref{sec:ident:consequences}).

\paragraph{Approximate semantics.}
EnerJ~\cite{enerj} isolates data marked approximate by type qualifiers from
precise data, and the approximation comes from storage or hardware;
Rely~\cite{rely} verifies the probability that a program computes its result
on unreliable hardware; Herbie~\cite{herbie} rewrites floating-point
expressions to reduce their error. Here the user explicitly selects
transformations of represented values. Our optional numerical contracts are
validated by tests and enumeration, rather than Rely's static reliability
analysis.

\paragraph{Quantization methods and formats.}
Integer-only inference~\cite{jacob-quant}, LLM.int8()~\cite{llm-int8},
GPTQ~\cite{gptq}, AWQ~\cite{awq}, SmoothQuant~\cite{smoothquant},
QLoRA~\cite{qlora} and SVDQuant~\cite{svdquant} choose and calibrate
representations: affine scales, 16-bit outlier features, second-order weight
updates, activation-aware channel scaling, NF4, and low-rank outlier branches.
Microscaling~\cite{rouhani-mx,ocp-mx}, FP8~\cite{fp8} and
bfloat16~\cite{bfloat16} standardize types. We implement and place
conversions, which is complementary. Our units cover \GGML's and \Kleidi's
formats, including MXFP4, NVFP4 and bfloat16 (\autoref{app:formats}).
Most of these schemes change how a representation is chosen more than what it
is, so the same constructs could express them. GPTQ and AWQ calibrate
group-wise integer weights with scales and zero points, an affine format
like q4\_1; their searches run offline, so they can enter as a
\code{TrustedInverse} encoder or through severed components. NF4 with
QLoRA's double quantization is a 16-entry codebook, like iq4\_nl's, with
quantized block scales, like the K-quants'. FP8 and the other MX element types are
small floats, like mxfp4's E2M1 values, and the MX types share mxfp4's
power-of-two block scale. Per-channel and per-token int8 scales, as in integer-only inference
and SmoothQuant, are block sizes; SmoothQuant's and AWQ's channel rescaling
is ordinary arithmetic in the algorithm. Two constructs are beyond a codec
alone: LLM.int8()'s outlier columns depend on the data, and SVDQuant's
low-rank branch adds a second product to the algorithm; we implement
neither.

\paragraph{Kernel ecosystems.}
\GGML~\cite{ggml,llamacpp} registers, per format, a quantizer, a dequantizer
and a dot product paired with an activation format, plus repacked GEMMs per
layout and architecture. \Kleidi~\cite{kleidiai} ships micro-kernels per
operation and data types, in variants per tile size and architecture feature,
with separate packing kernels. FBGEMM~\cite{fbgemm} fuses quantization into
CPU GEMMs and generates shape-specific code at run time; MARLIN~\cite{marlin}
hand-designs 4-bit-weight GPU kernels, with weights reshuffled offline.
vLLM~\cite{vllm,vllm-quant} picks a linear backend per quantization scheme,
TensorRT-LLM~\cite{tensorrt-llm} offers a set of quantization recipes, and
ExecuTorch~\cite{executorch} delegates to backend libraries.
Triton~\cite{triton} and TileLang~\cite{tilelang} make kernels programmable.
TileLang's dequantizing GEMM examples explicitly implement conversion inside
the kernel. Here, a codec yields both halves as Funcs, and a shared schedule
library places them without naming formats (\autoref{tab:marginal}).

\section{Conclusion}\label{sec:conclusion}

Making both halves of a format's round trip into computations extends Halide's
separation of concerns: a programmer specifies an algorithm, selects
approximations for its values, and schedules the resulting computations. With a fraction of the application
code of \GGML and \Kleidi, our kernels keep pace with \GGML's
format-specific kernels and outrun its blocked-dot fallbacks, and a second
CPU target cost an order of magnitude less code than \GGML's. Representations need not be
built into a kernel library: they can be user programs whose encoders and
decoders share the compiler's execution machinery.

\section*{AI Use}\label{aiuse-start}
AI coding agents wrote much of the application code in this paper (the format
definitions, schedules and test harness), the experiment and analysis scripts,
and the reduced test cases for the compiler issues we found, under the
authors' direction. The authors designed the abstractions, reviewed all code,
including the changes to Halide's core, and
checked every result reported here. An AI assistant drafted parts of the text
from the authors' design notes and logs; the authors revised it and are
responsible for all of it.
\label{aiuse-end}

\section*{Data-Availability Statement}
The artifact will be an archive of the repository at its final
commit, with a README that walks through the code and how to run the tests. It
contains the implementation, the case-study application, the raw benchmark
CSVs (Arm, Metal and x86) and the summarization scripts.

\clearpage
\bibliographystyle{ACM-Reference-Format}
\bibliography{refs}

\appendix
\makeatletter\let\usc@dischyph\@dischyph\makeatother
\counterwithin{table}{section}
\counterwithin{figure}{section}
\counterwithin{lstlisting}{section}

\FloatBarrier
\section{Contributions to Halide}\label{app:core}

\begin{table}[!hbp]
\caption{Our general-purpose changes to Halide. None mentions a format. Their
size is at the bottom of \autoref{tab:loc}.}\label{tab:core}
\small
\begin{tabular}{@{}lp{0.76\linewidth}@{}}
\toprule
Area & Change \\
\midrule
\code{Approximation} library &
\code{Func::approximate_by} replaces a Func's uses by the decoding of its encoding.
Combinators \code{Compose}, \code{Parallel}, \code{Choose}, \code{Pointwise} (with
\code{Pointwise::affine} for affine code maps), \code{Identity} and
\code{TrustedInverse}; derived components (\code{Derive}). Lossless layout units:
\code{BlockReshape} (also over index tiles), \code{StructLayout} (with
\code{port_field}), \code{BitFields} and its shorthand \code{PlanarFieldPack},
\code{SplitBits}, \code{Permute}, \code{LittleEndianScalarPack}. Named, typed components
with declared value ranges, component tags, signatures, and static checks of types and
ranges. \code{encode_funcs}/\code{decode_funcs} expose each stage's Funcs for
scheduling. \\
Types &
Struct types: records of mixed-type fields, loaded and stored whole, which
\code{StructLayout} uses for block records. \\
Pipelines &
\code{Pipeline::sever} replaces the computation of given Funcs by fresh
inputs of the same shape, so that work such as encoding static weights moves
out of the kernel. \code{Func::precompute} evaluates a constant Func at compile
time and embeds the result as a constant buffer. \\
Directives &
\code{eager_inline} inlines Funcs at schedule time, so that later directives see
the inlined expression. \code{hoist_invariants} factors loop-invariant
multiplicands out of a semiring reduction; invariance is decided modulo the
reduction variables' bounds, through shifts of bit fields, exact widenings and
inlined masks. \code{change_type} retypes a reduction after checking that the new
type cannot overflow; its \code{unsafe} form fuses a float sum of products into
multiply-adds. \code{vectorize(v, n, VectorizeIn::SWAR)} vectorizes within a
general-purpose register. \\
Queries and helpers &
\code{Func::is_precomputed}, \code{Stage::is_order_independent},
\code{Stage::value}/\code{values}. Schedule helpers \code{vectorize_all_stages},
\code{unroll_all_stages}, \code{eager_inline_all}, \code{in_callers} and
\code{clone_in_callers}. \\
Lowering &
\code{rfactor}: no split-tail guard on the merge stage; tail clamps partitioned
out of the steady state. Simplifier rules for divisions of scaled sums; a
promise's \code{likely} tag is dropped with the promise. Dense canonical forms and
alignment-aware division of multi-dimensional ramps. Bounds of \code{select}.
A store no longer clobbers a store of another type. Prefetches of struct-typed
buffers. \\
Arm backend &
i8mm: \cc{usdot}, and \cc{smmla}/\cc{ummla}/\cc{usmmla} for $2\times2$ tiles;
Armv8.6 feature detection on Apple CPUs. Bit-field shifts fused into one variable
shift. Small constant-table lookups lowered to \cc{tbl} (and to \cc{pshufb},
\cc{vpermb} and WebAssembly \cc{swizzle} on other targets). Intrinsics
\code{fixed_to_float} (one fixed-point convert) and \code{round_half_away}. \\
GPU backends &
A Metal backend for \cc{simdgroup\_matrix} tile memory. \code{gpu_lanes} warp
shuffles on Metal. Fixes to shuffle hoisting, to the liveness of shared and
\code{gpu_lanes} allocations in serial loops, and to stage-skipping flags inside
branches. CUDA: \cc{wmma} told the layout of its staged operands. \\
Runtime &
A reference-counted keep-awake count for the thread pool. \\
\bottomrule
\end{tabular}
\end{table}

\autoref{tab:core} lists the changes to Halide that the case study needed. Each is
general-purpose and independent of any format. The schedules also use
\code{split_storage} and multi-dimensional ramps~\cite{halide-repo}, recent
Halide features that we do not count as ours.

\FloatBarrier
\section{Format compositions}\label{app:formats}

The \nFormats formats (\nFormatsLayouts stored formats with layouts) compose
\nQuantizers quantizers and \nLossless lossless units, \nCoreUnits of them
from Halide's library.
\autoref{tab:formats} gives the composition of each format, in encode order,
after the \code{BlockReshape} into blocks. A semicolon separates stages of a
\code{Compose}; ``$a \,\&\, b$'' are the components of a \code{Parallel}. Every format
ends with its layout: a \code{StructLayout} of \GGML's block record, or an
\code{Interleave} into a repacked record. Units in \code{this font} are from the
core library, the others from the application; \autoref{tab:glossary}
(\autoref{app:glossary}) glosses every unit. Wherever \GGML's result is
defined, all decoders and the encoders we implement are bit-exact against \GGML's or
\Kleidi's, except q8\_1's block sum and mxfp4 on non-finite values; the grid formats use \GGML's encoders
(\S\ref{sec:eval:correct}).

\autoref{lst:q40gen} gives \code{Type0Quant}, \code{twos} and the
application's q4\_0 builder, which \autoref{lst:q40} specializes. The builder's arguments select the code
representation (\code{offset(8)} or \code{twos(4)}), the layout, the scaled
representation (\S\ref{app:q40variants}) and an indexed block reshape.
In the full \code{Type0Quant}, \code{code\_shift} multiplies the integers by
$2^{\textit{shift}}$ and divides the scale by the same power; its signature is
\code{codes\_and\_scale}, an \code{Int(8)} \emph{codes} component and an f32
\emph{scale} component, and its bound is one step, $|d|$, with a relative margin of
$10^{-4}$ for its reciprocal \code{inverse(d)}, \code{select(d != 0, 1/d, 0)}.
The margin needs $1/d$ finite, $|d|>2^{-128}$; below that, $x/d$ is
computed as an infinity or NaN. Large scales are harmless: a finite $|d|\le2^{125}$ keeps $1/d$ normal. The
type-1, q8\_0 and q8\_K quantizers share the margin and the precondition (q8\_K
on its $127/x_m$).
Bit-level layouts are units too. iq1\_m has no field for its f16 super-block
scale: \code{fp16\_bits} reinterprets the scale as 16 bits,
\code{Inverse(BitFields)} cuts them into four nibbles, and a second
\code{BitFields} stores one nibble in the top bits of each of the four 16-bit
words that hold the 3-bit sub-block scales.

\begin{lstlisting}[float=h,caption={The q4\_0 quantizer (without \code{code\_shift}), the general code representation and the q4\_0 builder, abridged from the application. \code{\_} denotes Halide's implicit variables, which make each unit independent of tensor rank.},label=lst:q40gen]
struct Type0Quant {  // GGML's "type-0" quantizer, h = 2^(bits-1)
  int bits;
  vector<Func> encode(const vector<Func> &in) const {
    Func x = in[0], d("scale"), q("codes");  Var j, b;  RDom r(0, QK);
    Tuple m = argmax(abs(x(r, b, _)));  // first strict max, as GGML's scan
    d(b, _) = strict_float(select(x(clamp(m[0], 0, QK - 1), b, _) < 0, m[1], -m[1]) / h);
    q(j, b, _) = min(cast<int8_t>(strict_float(x(j, b, _) * inverse(d(b, _)) + (h + .5f))),
                     cast<int8_t>(2 * h - 1)) - h;
    return {q, d};
  }
  vector<Func> decode(const vector<Func> &e) const { return {dequant(e)}; } // q*d
  ...                         // signature: integers in [-h, h-1], f32 scale; bound: one step
};
Approximation twos(int bits, int shift = 0) {  // GGML's repacked nibbles: offset(8) XOR 8
  int h = 1 << (bits - 1), s = 8 - bits, k = 1 << shift;
  return Pointwise{"twos",
      [=](Expr x) { return cast<uint8_t>(shift ? x >> shift : x) & cast<uint8_t>(2 * h - 1); },
      [=](Expr x) { Expr t = cast<int8_t>(x << s); return s > shift ? t >> (s - shift) : t; }}
    .with_types(Int(8), UInt(8))
    .with_ranges(ApproximationRange(-h * k, (h - 1) * k), ApproximationRange(0, 2 * h - 1))
    .with_lossless();
}
Approximation q4_0(Approximation codes, string lay, int shift = 0, bool indexed = false) {
  Approximation codec = shift ? Parallel{{"codes", codes}, {"scale", scale_by(1 << shift)}}
                              : Parallel{{"codes", codes}};
  return Compose{BlockReshape{QK, indexed}, Type0Quant{4, shift}, codec,
                 Parallel{{"codes", PlanarFieldPack{4, QK / 2}}, {"scale", fp16()}},
                 layout(lay, {{"d", Float(16)}, {"qs", UInt(8), QK / 2}}, {"qs", "d"})};
}
\end{lstlisting}

\begin{table}[!htbp]
\caption{The \nFormats formats as compositions. bpw: bits per weight of \GGML's
block layout. Layouts: memory layouts besides \GGML's block record (\emph{soa}:
planar components; $r{\times}c$: $r$ rows interleaved in $c$-byte pieces, most of them
\GGML's or \Kleidi's repackings). Quantizers are in bold.}\label{tab:formats}
\footnotesize
\setlength{\tabcolsep}{4pt}
\begin{tabular}{@{}llp{0.61\linewidth}l@{}}
\toprule
Format & bpw & Composition (encode order, after \code{BlockReshape}) & Layouts \\
\midrule
\multicolumn{4}{@{}l}{\emph{Legacy, block of 32}} \\
q4\_0 & 4.5 & \textbf{Type0}(4); integers: \code{affine} (offset 8) or \textit{twos}(4); \code{PlanarFieldPack}(4) \,\&\, \textbf{fp16} & soa, $4{\times}4$, $4{\times}8$, $8{\times}8$, $16{\times}4$ \\
q4\_1 & 5 & \textbf{Type1}(4); \code{affine} (unsigned); \code{PlanarFieldPack}(4) \,\&\, \textbf{fp16} \,\&\, \textbf{fp16} & $4{\times}8$, $16{\times}4$ \\
q5\_0 & 5.5 & \textbf{Type0}(5); integers: \code{SplitBits}(4); \code{PlanarFieldPack}(4) \,\&\, [\code{affine}; \code{PlanarFieldPack}(1), \textit{tabulated}] \,\&\, \textbf{fp16} & \\
q5\_1 & 6 & as q5\_0 with \textbf{Type1}(5), plus \textbf{fp16} for the minimum & \\
q8\_0 & 8.5 & \textbf{Q8\_0}; \textbf{fp16} & soa, $4{\times}4$, $4{\times}8$, $16{\times}4$ \\
q8\_1 & 9 & \textbf{Q8\_0}; \textbf{fp16}; \code{Derive}(block sum); \textbf{fp16} & \\
\multicolumn{4}{@{}l}{\emph{K-quants, super-block of 256}} \\
q2\_K & 2.625 & \textbf{KAffine}(2); \code{affine}; \code{PlanarFieldPack}(2) \,\&\, \code{PlanarFieldPack}(4) \,\&\, \textbf{fp16} \,\&\, \textbf{fp16} & \\
q3\_K & 3.4375 & \textbf{Q3K}; integers: \code{SplitBits}(2) into packs of 2 and 1 bits \,\&\, scales: \textit{six-bit scales} (\code{SplitBits}(4) into packs of 4 and 2) \,\&\, \textbf{fp16} & \\
q4\_K & 4.5 & \textbf{KAffine}(4); \code{affine}; \code{PlanarFieldPack}(4) \,\&\, \textit{KScales} \,\&\, \textbf{fp16} \,\&\, \textbf{fp16} & $8{\times}8$\textsuperscript{a}, $16{\times}4$ \\
q5\_K & 5.5 & \textbf{KAffine}(5); \code{SplitBits}(4) into packs of 4 and 1 bits \,\&\, \textit{KScales} \,\&\, \textbf{fp16} \,\&\, \textbf{fp16} & $8{\times}8$\textsuperscript{a}, $16{\times}4$ \\
q6\_K & 6.5625 & \textbf{Q6K}; \code{SplitBits}(4) into packs of 4 and 2 bits \,\&\, \textbf{fp16} & $8{\times}8$\textsuperscript{b} \\
q8\_K & 9.125 & \textbf{Q8K}; \code{Derive}(sums of 16) & $4{\times}8$ \\
\multicolumn{4}{@{}l}{\emph{Codebook and microscaling}} \\
iq4\_nl & 4.5 & \textbf{IQ4}(32); \textit{codebook} \,\&\, \textbf{fp16}; \code{PlanarFieldPack}(4) & $4{\times}4$ \\
iq4\_xs & 4.25 & \textbf{IQ4}(256); \textit{codebook}; \code{PlanarFieldPack}(4) \,\&\, \textit{six-bit scales} \,\&\, \textbf{fp16} & $16{\times}4$ \\
mxfp4 & 4.25 & \textbf{MXFP4}; \textit{codebook} (E2M1) \,\&\, \textit{e8m0\_half}; \code{PlanarFieldPack}(4) & $4{\times}4$ \\
nvfp4 & 4.5 & \textbf{NVFP4} (block 64); \textit{codebook} (E2M1) \,\&\, \textit{Tabulate}(\textit{ue4m3\_half}); \code{PlanarFieldPack}(4) & \\
\multicolumn{4}{@{}l}{\emph{Low-bit: ternary, binary and 2-bit}} \\
tq1\_0 & 1.6875 & \textbf{Ternary}; \code{affine}; \textit{Segments}; \textit{TritPack} per segment & \\
tq2\_0 & 2.0625 & \textbf{Ternary}; \code{affine}; \code{PlanarFieldPack}(2) & \\
q1\_0 & 1.125 & \textbf{Sign} (block 128); \code{affine} (select form) \,\&\, \textbf{fp16}; \code{PlanarFieldPack}(1) & $4{\times}1$ \\
q2\_0 & 2.25 & \textbf{Q2} (block 64); \code{affine}; \code{PlanarFieldPack}(2) & \\
\multicolumn{4}{@{}l}{\emph{Grid IQ, super-block of 256: \code{TrustedInverse}(\GGML's encoder, decoder below)}} \\
iq2\_xxs & 2.0625 & \textit{ScaledCodes}; integers: [\textit{SignSplit}; \textit{Tabulate}(\textit{codebook}; \textit{Lattice}) \,\&\, \textit{Tabulate}(\code{affine}; \textit{EvenSigns})] \,\&\, scales: \code{affine} \,\&\, scale: [\textit{scale\_by}; \textbf{fp16}]; \code{BitFields} & \\
iq2\_xs & 2.3125 & as iq2\_xxs, larger grid; scales \code{PlanarFieldPack}(4); \code{BitFields} over 16 bits & \\
iq2\_s & 2.5625 & as iq2\_xs, but the grid index \code{SplitBits} into 8 and 2 bits, raw sign bits (\code{affine}; \code{PlanarFieldPack}(1)), no \code{BitFields} & \\
iq3\_xxs & 3.0625 & as iq2\_xxs with a 3-bit grid of 4 values per point & \\
iq3\_s & 3.4375 & as iq2\_s with a 3-bit grid & \\
iq1\_s & 1.5625 & \textit{ScaledCodes}; integers: [\textit{RunShift}; \textit{Tabulate}(\textit{codebook}; \textit{Lattice}); \code{SplitBits} \,\&\, \code{affine}] \,\&\, scales: \code{affine} \,\&\, scale: [\textit{scale\_by}; \textbf{fp16}]; \code{BitFields} & \\
iq1\_m & 1.75 & as iq1\_s; the f16 scale's bits are spread over the sub-block scales (\textit{fp16\_bits}; \textit{Inverse}(\code{BitFields})) & \\
\multicolumn{4}{@{}l}{\emph{\Kleidi, one scale per row}} \\
\cc{qsi8cx} & 8\textsuperscript{c} & \textbf{Qsi8cx}; \code{Derive}(row sum) & $4{\times}8$ \\
\cc{qai8dx} & 8\textsuperscript{c} & \textbf{Qai8dx} (scale and zero point) & \\
\multicolumn{4}{@{}l}{\emph{Floating point, no block}} \\
f16 & 16 & \textbf{MiniFloat}(binary16) & \\
bf16 & 16 & \textbf{MiniFloat}(bfloat16) & \\
\bottomrule
\end{tabular}

\smallskip
\raggedright\footnotesize
\textsuperscript{a}\textit{Interleave}; the scales of 8 rows \textit{Transpose}d and re-packed by
\textit{KScales}. \textsuperscript{b}\textit{Interleave}; scales \textit{Transpose}d.
\textsuperscript{c}Plus 64 bits per row (a 32-bit float scale and a 32-bit sum or zero point).
Italic: application units and helpers (\textit{twos}, \textit{codebook},
\textit{scale\_by}, \textit{fp16\_bits}, \textit{e8m0\_half} and \textit{ue4m3\_half} are maps
written as \code{Pointwise}; \textit{six-bit scales} and \textit{tabulated} are compositions).
\end{table}

\FloatBarrier
\subsection{Variants from the same parts}\label{app:q40variants}

A format string \code{<type>[.<codes>][.<layout>]} selects the quantizer, the
integer representation and the layout independently; a colon separates a stored
format from a compute format (\code{q4\_0:f16}, \S\ref{sec:sched:metal}), and
\code{f32:q8\_0.4x4} names activations given in f32 and encoded in the kernel.
Each choice changes stages of \autoref{lst:q40} that the application's builder
(\autoref{lst:q40gen}) takes as arguments:
\begin{itemize}
\item \code{q4\_0} is \GGML's plain block: \code{offset(8)} and
\code{layout("aos")}. \code{q4\_0.soa} stores one array per field.
\code{q4\_0.i4.4x8} is \GGML's i8mm repack, \code{Interleave\{4, 8\}}.
\item The \emph{scaled} variant (\code{code\_shift = 4}) is a second code
representation with the same bytes. \code{Type0Quant} emits $16q$ and $d/16$,
which decode to the same values, since a power of two moves exactly between
the factors; \code{twos(4, 4)} encodes $16q$ as the same nibble; and
\code{scale\_by(16)} restores \GGML's scale before rounding. Its motive is
speed: a nibble's decode becomes one shift left, or a mask of the high nibble,
which leaves $16q$ in a byte, ready for \cc{sdot} (\autoref{lst:asm}). It is a
choice of format, not of schedule, because the components change meaning, though
the bytes do not. \code{check\_ranges} verifies that the integers lie in
$[-128,112]$. It does not establish divisibility by 16, which
\code{twos(4, 4)} also needs; it is assumed (\code{Type0Quant} emits
multiples of 16) and checked by byte comparison with \GGML. Ranges with arithmetic congruences~\cite{granger-congruences,cousot-ai}
would let the shift be a lossless stage of its own.
\item Activation format \code{q8\_0.4x4} reuses \code{BlockReshape},
\code{fp16()} and \code{layout("4x4", \ldots)} around \GGML's q8\_0 quantizer.
\end{itemize}

\section{Notation and Units}\label{app:glossary}

\paragraph{Format strings.}
A stored format is \code{<type>[.<codes>][.<layout>]}: the quantizer's
format, an optional integer representation (\code{i4}: two's-complement
nibbles) and an optional layout (\code{soa}: one array per field;
\code{<rows>x<chunk>}: \code{Interleave}; default: \GGML's struct, one record
per block). A colon separates a stored format from the format the kernel
computes in: \code{q4\_0:f16} stores q4\_0 and computes in f16
(\S\ref{sec:sched:metal}); \code{f32:q8\_0.4x4} takes f32 activations and
encodes them in the kernel.

\paragraph{Units.}
\autoref{tab:glossary} glosses the units of \autoref{tab:formats} (\autoref{app:formats}). All are
codecs; the bold units, the quantizers and the minifloat roundings such as
\code{fp16()}, are lossy, and every other unit is lossless on
its declared input range, except \code{twos(4, 4)}, which also needs multiples
of 16 (\S\ref{app:q40variants}).

\begin{table}[!htbp]
\caption{The units of \autoref{tab:formats}. In \code{this font}: Halide's
library; the others: the application, quantizers in
bold. Quantizers transcribe \GGML's or \Kleidi's reference encoders.}\label{tab:glossary}
\footnotesize
\begin{tabular}{@{}lp{0.30\linewidth}@{\quad}lp{0.28\linewidth}@{}}
\toprule
\multicolumn{2}{@{}l}{\emph{Combinators, reshaping, layout and bit packing}} &
\multicolumn{2}{l@{}}{\emph{Quantizers, then value maps}} \\
\cmidrule(r){1-2}\cmidrule(l){3-4}
Unit & Meaning & Unit & Meaning \\
\midrule
\code{Compose} & {\raggedright sequential composition; decode runs the stages' decodes in reverse\par} &
\textbf{Type0} & {\raggedright $d=x_m/({-h})$, $x_m$ the signed extreme, $h=2^{b-1}$; integers $x/d$ rounded half up\par} \\
\code{Parallel} & {\raggedright a codec per component, side by side (not parallel execution)\par} &
\textbf{Type1} & {\raggedright affine: $d=(\max-\min)/(2^b-1)$, and the minimum\par} \\
\code{TrustedInverse} & {\raggedright an external encoder paired with a decoder; nothing declared relates them\par} &
\textbf{Q8\_0} & {\raggedright $d=\max|x|/127$, integers $x/d$ rounded\par} \\
\textit{Inverse} & {\raggedright a unit with its encode and decode swapped\par} &
\textbf{KAffine}, \textbf{Q3K}, \textbf{Q6K} & {\raggedright K-quant search: sub-block scales (and mins) fitted, then quantized against super-block scales\par} \\
\code{Derive} & {\raggedright appends a component computed from other components; decode drops it\par} &
\textbf{Q8K} & {\raggedright $d=x_m/({-127})$ (as $1/(-127/x_m)$ in f32), integers rounded\par} \\
\textit{Segments} & {\raggedright a block cut into runs, each stored its own way\par} &
\textbf{IQ4} & {\raggedright search over trial scales; integers from a 16-value codebook\par} \\
\textit{Tabulate} & {\raggedright decodes through a table of the decode of every index (e.g.\ every byte)\par} &
\textbf{MXFP4}, \textbf{NVFP4} & {\raggedright a power-of-two (mx) or ue4m3 (nv) block scale; E2M1 values\par} \\
\code{BlockReshape} & {\raggedright $(k,\ldots)\leftrightarrow(k\bmod n,\lfloor k/n\rfloor,\ldots)$\par} &
\textbf{Ternary}, \textbf{Q2} & {\raggedright $d=\max|x|$, integers in $\{-1,0,1\}$ (\textbf{Q2}: 4 levels, code 2 only where $1/d$ overflows)\par} \\
\textit{Transpose} & {\raggedright an $a\times b$ record's elements transposed\par} &
\textbf{Sign} & {\raggedright $d$ the mean of $|x|$, integers $\pm1$\par} \\
\code{StructLayout} & {\raggedright places named components in the fields of a C struct record\par} &
\textbf{Qsi8cx}, \textbf{Qai8dx} & {\raggedright per-row int8: symmetric (\cc{qsi8cx}), or affine with a zero point (\cc{qai8dx})\par} \\
\textit{Interleave} & {\raggedright the records of $r$ rows as one, each field in pieces of $c$ bytes from the rows in turn\par} &
\textbf{MiniFloat} & {\raggedright a small float of given exponent and mantissa widths, bias, specials and overflow; \code{fp16()} is \code{MiniFloat\{5, 10, 15\}}\par} \\
\code{BitFields} & {\raggedright packs unsigned fields into words at given bit offsets and strides\par} &
\textbf{fp16} & {\raggedright \textbf{MiniFloat} for f16, rounding to nearest even\par} \\
\code{PlanarFieldPack} & {\raggedright \code{BitFields} of one $b$-bit field into bytes, byte $p$ of a group of $P$ holding elements $p$, $p+P$, \ldots\par} &
\code{affine} & {\raggedright quantized integers $c=a\,u+b\leftrightarrow$ fields $u\in[0,n)$, $|a|$ a power of two (\code{Pointwise::affine})\par} \\
\code{SplitBits} & {\raggedright an integer $\leftrightarrow$ its low $k$ bits and the rest; decode adds them\par} &
\textit{twos} & {\raggedright integers $\leftrightarrow$ their two's-complement fields\par} \\
\textit{six-bit scales} & {\raggedright 6-bit scales as low nibbles and top two bits (\code{SplitBits})\par} &
\textit{codebook} & {\raggedright integers $\leftrightarrow$ their indices in a table of values\par} \\
\textit{KScales} & {\raggedright K-quant 6-bit scales and mins $\leftrightarrow$ \GGML's 12 bytes\par} &
\textit{scale\_by} & {\raggedright $\times k$, $k$ a power of two: exact on normal floats\par} \\
\textit{TritPack} & {\raggedright up to five ternary digits per byte, base 3\par} &
\textit{e8m0\_half}, \textit{ue4m3\_half} & {\raggedright a block scale $d$ $\leftrightarrow$ the byte of $2d$, as \GGML decodes it\par} \\
\textit{EvenSigns} & {\raggedright 8 sign bits of even parity $\leftrightarrow$ their first 7\par} &
\textit{SignSplit} & {\raggedright integers $\leftrightarrow$ magnitudes and signs\par} \\
\textit{fp16\_bits} & {\raggedright an f16 $\leftrightarrow$ its 16 bits\par} &
\textit{Lattice} & {\raggedright a grid point $\leftrightarrow$ its index in \GGML's grid list\par} \\
 &  &
\textit{RunShift} & {\raggedright iq1 integers $\leftrightarrow$ grid values and a $\pm1$ shift per run\par} \\
 &  &
\textit{ScaledCodes} & {\raggedright decode of sub-block-scaled formats: integers $\times\,d\times$ sub-block scale\par} \\
\bottomrule
\end{tabular}
\end{table}

\FloatBarrier
\section{Framework size}\label{app:size}

We count our additions to Halide's \cc{src/} and \cc{tools/} directories at
the evaluated revision. We attribute each line to the commit that last wrote it
and count the sloc from the commits made for this work. Tests,
documentation, tutorials and language bindings are not counted, nor are the
x86 and CUDA back-end changes (x86: \S\ref{sec:eval:x86}); neither is the
work of other developers that the schedules use. The generated kernels
comprise \sizeCPU{}\,MiB of machine code and \sizeMetal{}\,MiB of compiled Metal libraries.
A format table of \nFormatRows rows instantiates the algorithm into
\nKernelsCPU CPU kernels and, for \nMetalFormats of the weight formats,
\nKernelsMetal Metal kernels, plus an encoder and a decoder per stored format.
Generation takes \genTimeCPU{}\,s (CPU) and \genTimeMetal{}\,s (Metal, with
Apple's compiler), at most \genTimeMax{}\,s per kernel.

The bottom of \autoref{tab:loc} counts as ours only what is specific to
approximations: \code{approximate_by}, its hooks into \code{Func} and the
generator, and the \code{Approximation} library, with its combinators, components and
core units (\locApproxLib sloc); and \code{sever} (\locSever). The rest,
\locGeneral sloc, is general-purpose: it is excluded from that
approximation-specific subtotal and included in the all-additions total,
\locOursFramework sloc. It comprises the directives,
analyses and simplifier changes (\locDirectives sloc: \code{eager_inline},
\code{hoist_invariants}, \code{change_type}, \code{precompute}, struct types,
SWAR vectorization, and others in \autoref{tab:core}), and the back-end changes
(\locBackends sloc: Arm i8mm, dot-product and table-lookup lowering, the Metal
\cc{simdgroup\_matrix} backend and warp shuffles, the CUDA \cc{wmma} fix, GPU
lowering fixes and the thread-pool keep-awake count). The property-testing library (\locPropTesting
sloc) is counted with the tests.

The rows of \autoref{tab:loc} count these sources. Ours: the 13 format
headers (\cc{schemes/*.h}), \cc{matmul.cpp}, the codec generator,
\cc{schedule.h} and the CMake format table. \GGML's: \cc{ggml-quants.c},
\cc{quants.c}, \cc{arm/quants.c}, \cc{repack.cpp}, \cc{arm/repack.cpp} and
three Metal files. The \GGML side has these limits.
(c)~The reference quantizers exclude the grid IQ encoders, which we call rather than
replace. (d)~The Neon files are counted without their SVE and SME branches.
(e)~The Metal shaders are \cc{dequantize.h}, \cc{mul\_mv.metal} and
\cc{mul\_mv\_mma.metal}, for our \nMetalFormats Metal formats except f16 and bf16;
\cc{mul\_mm.metal} only instantiates templates per format. (f)~The \Kleidi
sources are C and assembly, 32 files; \GGML also builds SVE and SME variants
(11411 lines in all) and has 3140 lines of \Kleidi glue.
Our side is \locRatio times smaller in sloc, 3.5 times in lines and 2.6
times in sloc without \Kleidi; with the approximation-specific Halide code,
\locRatioApprox times (\locRatioApproxNoKleidi without \Kleidi), and with
all our Halide additions, \locRatioFramework times.
The ratios are indicative: we count f16 and bf16; a quarter of \GGML's
CPU lines are portable fallbacks absent from ours; its reference
quantizers count, our tests do not.

\FloatBarrier
\section{Measurement details}\label{app:method}

\paragraph{Builds.}
\GGML is built natively by \ggmlCompiler{} with the flags
\begin{center}\ggmlFlags.\end{center}
llama.cpp's default
\cc{GGML\_LLAMAFILE=ON} would not change the baseline: \GGML compiles
llamafile out on CPUs with int8 matrix multiply, and an ON build produces
identical CPU code.
We compile our kernels with LLVM~22 for Armv8.6-A and for Metal, all from
one revision of the application; a build of it passes all 11 test suites and every check of \S\ref{sec:eval:correct}.

\paragraph{Samples.}
For each case the harness builds every kernel that supports it, ours and
\GGML's, and checks it (\autoref{app:checks}). It then calibrates each kernel's
repetition count $r$ so that one sample, $r$ back-to-back calls, takes at least
20\,ms, and runs three warm-up samples. It then runs 5 to 15 rounds (see Protocol below). Each round times
one sample of every kernel in a fresh random order. A sample's time is divided by
$r$, and by 16 for a \GGML graph and for our Metal calls (Metal synchronization
below). We report the median, and as an approximate 95\% confidence interval
for it the order statistics (0-based, clamped to the sample) $\lfloor n/2 -
0.98\sqrt n\rfloor$ and $\lceil n/2 + 0.98\sqrt n\rceil$, a normal
approximation; at $n=5$ this is the sample range, whose exact coverage is
$1-2^{-4}=93.75\%$.

\paragraph{Boundaries.}
Weights are quantized by \GGML from $\mathcal N(0,1)$ values, then encoded,
repacked or requantized into each kernel's layout, untimed. Activations in a
quantized format are encoded untimed, by \GGML's \cc{from\_float}. Kernels whose
activations are f32 encode them inside the timed call: ours in the pipeline,
\GGML's in \cc{mul\_mat}. Halide Metal kernels run on device buffers that hold
the inputs before timing; every 16 calls end with \cc{halide\_device\_sync}, and
the output is copied back only for checking. \GGML's backends run a graph of 16
multiplications per call; its Metal backend runs them in order
(\cc{GGML\_METAL\_CONCURRENCY\_DISABLE}), as dependent layers would.
The graph spreads \GGML's per-call costs over 16 multiplications. On CPU these
are a thread-pool start and a barrier per node, as each of our calls pays a
fan-out and a join, so the graph can only help \GGML. On Metal,
\cc{ggml\_backend\_graph\_compute} waits for the GPU at the end of each graph,
once per 16 multiplications, and so do we (Metal synchronization below).

\paragraph{Threads and memory.}
The machine has 12 performance and 4 efficiency cores and 48\,GB, and runs
macOS~26.5; \GGML is linked as shared libraries to fix its code layout.
The timing thread runs at \cc{QOS\_CLASS\_USER\_INTERACTIVE}, which macOS
schedules on performance cores; macOS offers no thread pinning. Both thread
pools spin while a sample runs and sleep between samples: \GGML's is resumed and
paused around each sample, and ours holds a keep-awake count. \cc{vec\_dot} runs
on one thread. At startup the harness touches every page of each loaded image's
constant sections and leaves 16\,KB of resident slack after each input. Without
this, a kernel that streams a buffer to within a few KB of an untouched page, or
indexes a constant table next to one, pays a failed hardware page walk per row.
This slowed some kernels on both sides by up to 25\%, depending on link layout.

\paragraph{Protocol.}
The main sweep used these rounds per case (re-runs used 15):
\begin{center}
\footnotesize
\begin{tabular}{@{}lll@{}}
\toprule
Column & Rounds & Paddings \\
\midrule
\cc{vec\_dot} & 15 & 4 (median over paddings) \\
CPU GEMV, T1 and T10 & 15 & 1 \\
CPU GEMM $M=8$ (T1, T10), $M=32$ (T10) & 9 & 1 \\
CPU GEMM $M=32$ (T1) & 5 & 1 \\
CPU GEMM $M=512$ (T1 at $4096\times1024$ only; T10) & 5 & 1 \\
Metal, all $M$ & 15 & 1 \\
\bottomrule
\end{tabular}
\end{center}
Only
\cc{vec\_dot} was timed at four text-section paddings of our binary (0, 1056, 2144
and 3232 bytes); their \GH spread is 1.0\% at the median and 2.9\% at the 90th
percentile, small next to run-to-run noise at T10. All jobs ran in one night, holding a
machine-wide lock that other benchmarks also take, under \cc{caffeinate}, and
only on AC power, each after waiting until the machine was at least 75\% idle
(the median was 96\%). After each job, the system power log was scanned for
sleep, wake, background (dark) wake or battery events in the job's window; a job
with one would have been redone, and none had one.
Cases below 0.95 were re-run twice, and a third time where the two disagreed by more than 10\%; a re-run job times one case (all layouts and
providers) at 15 rounds. Every case
with repeats combines them by the min-time rule: for each provider separately,
ours and \GGML's, take the minimum time over all repeats, then form \GH from the
fastest of each side. Of the 85 re-run cases, 63 stayed below 0.95 in every
repeat, 9 were mixed and remain below 0.95, and 13 rose above it. The median
full width of the approximate 95\% confidence interval of a median is 4.5\% for CPU GEMM and
3.9\% on Metal (90th percentile: 15\% and 16\%); T10 GEMM is noisiest, because
a call returns only when all ten threads finish.

\autoref{tab:sensitivity} recomputes the match rates with three ways of combining
repeats: the first run of each case only (the sweep; on Metal, the first
matched repeat), the minimum time per
kernel over repeats (reported), and the median time per kernel.
\begin{table}[!htbp]
\caption{Match rates against \GGML's fastest non-BLAS path (last block:
with BLAS), by how repeats are combined, under both layout policies: per-case fastest layout (\autoref{tab:cpu},
\autoref{tab:metal}) and the held-out-shape layout of the abstract and
\autoref{tab:summary} (\autoref{app:fixed}). Cases with $\GH\ge0.95$ per set,
the median \GH over all cases and the number below 0.95. Metal has one stored
layout per type, so its column is the same under both policies; Metal per
sample: the previous protocol (\autoref{tab:sync}).}\label{tab:sensitivity}
\footnotesize
\setlength{\tabcolsep}{4pt}
\begin{tabular}{@{}lrrrrrrrr@{}}
\toprule
Estimator & all & CPU & Metal & format- & blocked & median & below & Metal \\
 & (\nCases) & (\nCasesCPU) & (\nCasesMetal) & specific (\nHH) & dot (\nLoopCPU) & & 0.95 & per sample \\
\midrule
\multicolumn{9}{@{}l}{per-case fastest layout:} \\
first run & 96.5\% & 98.8\% & \pctMatchMetalEstFirst & \pctHHEstFirst & \pctLoopCPUEstFirst & 1.56 & 72 & 82.9\% \\
minimum time (reported) & 96.7\% & 99.2\% & \pctMatchMetal & \pctHH & \pctLoopCPU & 1.57 & 67 & 84.9\% \\
median time & 96.7\% & 99.1\% & \pctMatchMetalEstMedian & \pctHHEstMedian & \pctLoopCPUEstMedian & 1.56 & 67 & 85.5\% \\
\midrule
\multicolumn{9}{@{}l}{held-out-shape layout (the headline):} \\
first run & \pctMatchAllFixedEstFirst & \pctMatchCPUFixedEstFirst & \pctMatchMetalEstFirst & \pctHHFixedEstFirst & \pctLoopCPUFixedEstFirst & 1.53 & \nBelowAllFixedEstFirst & \\
minimum time (reported) & \pctMatchAllFixed & \pctMatchCPUFixed & \pctMatchMetal & \pctHHFixed & \pctLoopCPUFixed & \medAllFixed & \nBelowAllFixed & \\
median time & \pctMatchAllFixedEstMedian & \pctMatchCPUFixedEstMedian & \pctMatchMetalEstMedian & \pctHHFixedEstMedian & \pctLoopCPUFixedEstMedian & 1.53 & \nBelowAllFixedEstMedian & \\
\midrule
\multicolumn{9}{@{}l}{per-case fastest layout, with BLAS in \GGML's fastest path:} \\
first run & 86.9\% & 87.0\% & & & & 1.31 & 268 & \\
minimum time (reported) & \pctMatchAllBlas & \pctMatchCPUBlas & & & & \medAllBlas & 264 & \\
median time & 87.1\% & 87.1\% & & & & 1.31 & 264 & \\
\bottomrule
\end{tabular}
\end{table}

\paragraph{Metal synchronization.}
\GGML runs a graph of 16 multiplications to 16 output tensors and waits once
per graph. We time ours the same way: 16 calls to 16 distinct outputs, then one
device synchronization, divided by 16 (two spaced repeats of every Metal case, a
third where they differed by over 10\%; min-time rule). One difference remains:
\GGML encodes the 16 nodes into one command buffer, while Halide's
Metal runtime creates a command buffer with a serial encoder, and a pipeline
state, per launch: at least 17 command buffers per group. This run-time overhead, not generated code, sets a floor of
29--30\,\textmu s per call at the smallest GEMV ($4096\times1024$) for every
format so timed, against \GGML's 21--30. A previous per-sample protocol, one synchronization per $\ge$20\,ms
sample on our side, gave \pctMatchMetalPerSample of Metal cases (median
\medMetalPerSample; \nBelowMetalPerSample below 0.95) and \pctMatchAllPerSample
of all; it hid the floor and penalized us at $M=8$, where distinct outputs let
consecutive calls overlap (\autoref{tab:sync}). \GGML's own times, measured
the same way in both runs, differ between them by a median of 1--6\% per
column. Our Metal libraries embed precompiled \cc{metallib}s, which we time after
the first load; compared with run-time-compiled source, they
cut a cold first load from \firstLoadSrcLo--\firstLoadSrcHi{}\,s to
\firstLoadLibLo--\firstLoadLibHi{}\,ms and libraries from \sizeMetalLibsSrc to
\sizeMetalLibsLib{}\,MB.
\begin{table}[!htbp]
\caption{Metal \GH under the previous per-sample protocol and the matched
protocol of \autoref{tab:metal}, with our fastest layout per case: min--max
(median), cases below 0.95.}\label{tab:sync}
\footnotesize
\setlength{\tabcolsep}{4pt}
\begin{tabular}{@{}lrlrlr@{}}
\toprule
Column & cases & per-sample & $<$0.95 & matched & $<$0.95 \\
\midrule
GEMV & 72 & 0.86--1.72 (1.17) & 8 & 0.71--1.63 (1.00) & 28 \\
\quad $4096\times1024$ & 18 & 1.02--1.32 (1.22) & 0 & 0.71--1.01 (0.78) & 16 \\
$M=8$ & 72 & 0.49--3.48 (1.01) & 30 & 0.68--3.76 (1.23) & 22 \\
$M=32$ & 124 & 0.91--2.60 (1.13) & 9 & 1.01--2.34 (1.30) & 0 \\
$M=512$ & 124 & 0.79--1.16 (1.00) & 12 & 0.78--1.21 (1.10) & 3 \\
\bottomrule
\end{tabular}
\end{table}

\FloatBarrier
\section{The BLAS path}\label{app:blas}

\GGML's BLAS backend accepts a multiplication when the activations are f32 and
$M$, $N$ and $K$ are all at least 32; it dequantizes the whole weight matrix to
f32 (with up to $T$ worker threads) and calls \cc{cblas\_sgemm}. It supports all
of our quantized and 16-bit \GGML weight types, and we timed it for all of them. Accelerate takes no thread count
from \GGML: at T1 the process used about \blasCores cores while timing
(\cc{top}), and capping Accelerate at one thread made the call 1.35 times
slower; at T1 and $M=512$ our median is \blasMedMfivetwelveTone of its speed. We timed it against our fastest f32-activation kernel for each type at
$M\in\{32,512\}$, all four shapes, T1 and T10 (\nBlasCases cases; 9 and 5
rounds; two spaced repeats, min-time rule), with the protocol of
\autoref{app:method}. \autoref{tab:blas} gives $G_{\text{BLAS}}/H$.
\begin{table}[!htbp]
\caption{Our kernels against \GGML's BLAS path on f32 activations,
$G_{\text{BLAS}}/H$, with our fastest layout per case; cells as in
\autoref{tab:cpu}.}\label{tab:blas}
\footnotesize
\setlength{\tabcolsep}{3pt}
\begin{tabular}{@{}lcccc@{}}
\toprule
Weights & $M{=}32$ T1 & $M{=}32$ T10 & $M{=}512$ T1 & $M{=}512$ T10 \\
\midrule
legacy & 0.50--1.23 (0.61)$^{16}$ & 1.99--6.17 (3.22) & 0.08--0.19 (0.10)$^{20}$ & 0.60--1.53 (0.81)$^{12}$ \\
K-quants & 0.49--1.48 (0.61)$^{16}$ & 1.93--4.51 (3.08) & 0.08--0.18 (0.10)$^{20}$ & 0.56--1.12 (0.73)$^{16}$ \\
grid IQ & 0.27--0.68 (0.54)$^{28}$ & 1.17--2.31 (1.64) & 0.05--0.09 (0.06)$^{28}$ & 0.34--0.58 (0.43)$^{28}$ \\
codebook & 0.76--2.06 (1.33)$^{8}$ & 1.79--5.73 (3.33) & 0.08--0.24 (0.15)$^{16}$ & 0.43--1.43 (0.88)$^{8}$ \\
low-bit & 0.62--1.10 (0.75)$^{12}$ & 2.18--4.94 (3.54) & 0.09--0.15 (0.12)$^{16}$ & 0.69--1.25 (0.93)$^{10}$ \\
f16, bf16 & 0.15--0.69 (0.38)$^{8}$ & 1.00--2.52 (1.67) & 0.03--0.09 (0.06)$^{8}$ & 0.30--0.59 (0.44)$^{8}$ \\
\midrule
all (108 each) & 0.63 & 2.49 & 0.09 & 0.68 \\
\bottomrule
\end{tabular}
\end{table}
Of the \nCasesCPU CPU table cases, BLAS applies to \nBlasApplies and is faster
than every other \GGML path in \nBlasFaster. Counting it, with per-case
layouts, we match \GGML in \pctMatchCPUBlas of CPU cases (median \medCPUBlas)
and \pctMatchAllBlas of all cases (median \medAllBlas); with held-out-shape
layout selection (\autoref{app:fixed}), in \pctMatchCPUBlasFixed and \pctMatchAllBlasFixed.
In the BLAS-eligible GEMMs alone, both policies give \pctMatchBlasEligible (median
\medBlasEligible). Stock llama.cpp places a CPU weight in a repacked buffer when it can,
and BLAS cannot read those, so it sends only \nBlasRouted of our types to BLAS
(q4\_0, q8\_0, q4\_K, q5\_K, q6\_K, iq4\_nl, mxfp4 and q1\_0 go to the repacked
kernels); with that routing and per-case layouts we match in \pctMatchCPUBlasRouted of CPU cases
(\pctMatchAllBlasRouted of all). On a Mac, llama.cpp offloads all layers to
Metal by default.
In the \nLoopCPU blocked-dot cases, where \GGML's fastest other path loops \cc{vec\_dot},
BLAS applies only at $M\ge32$: the GEMV median stays \medLoopGEMV (\pctLoopGEMV
match). With \GGML's best of blocked-dot and BLAS, the GEMM median falls from
\medLoopGEMM to \medLoopGEMMBlas (\pctLoopGEMMBlas match; all \nLoopCPU:
\medLoopBlas, \pctLoopBlas); in the \nLoopBlasElig GEMMs BLAS can run, we
match in \pctLoopBlasElig (median \medLoopBlasElig). These are per-case
layouts; with held-out-shape selection, the GEMV median is \medLoopGEMVFixed and the
GEMM median falls from \medLoopGEMMFixed to \medLoopGEMMBlasFixed.

\FloatBarrier
\section{Fixed layouts and bandwidth}\label{app:fixed}

\paragraph{Held-out-shape layout selection.}
The tables take, per case, the fastest of a type's stored layouts (AoS, SoA or a
repacked tile; \emph{per-case layout}), but a deployment fixes the layout when it converts the model.
Every layout was timed in every case of its type, so we
re-score the same samples with one layout per target and weight type within
each held-out fold, fixed across activations, $M$, shapes and threads. We select the layout by
leave-one-shape-out selection: for each shape, the layout with the best geomean \GH on the other three. Every
fold picks the same layout, except for \cc{qsi8cx} in one: on the CPU, q4\_0
and q8\_0 4$\times$4, K-quants 8$\times$8, iq4\_nl and mxfp4 4$\times$4,
q1\_0 4$\times$1; on Metal, AoS. One global choice in hindsight, the layout
with the best geomean over all four shapes, gives \pctHHHind on
format-specific kernels and \pctMatchAllHind overall. Run-time choices that keep the stored bytes
stay per case: our activation encoding and, on Metal, the pipeline by $M$, as
\GGML picks its kernel per call. The tables likewise give \GGML its best of
plain, repacked and \Kleidi paths; its stock rule instead fixes the layout at
load (repacked for the types that have one, at every $M$), which differs from
that best in \nGGMLOffRule of 1595 CPU cases. The fastest layout changes
across $M$ and threads for \nLayoutSwitch of \nLayoutGroups (type,
activation, shape) groups, but rarely by enough to cross 0.95: fixing it
moves \nFixedLost cases below. \GGML's own layout for each type on this CPU
(4$\times$8 for q4\_0 and q8\_0) gives \pctMatchAllRule (\pctHHCPURule on
format-specific CPU kernels). Below, cells give the share of cases with $\GH\ge0.95$
(median \GH); in the last column, \GGML's layout is fixed too, by its stock rule.
\begin{center}
\small
\begin{tabular}{@{}lrlll@{}}
\toprule
cases & $n$ & per case & held-out & + \GGML's rule \\
\midrule
format-specific, CPU & 311 & 97.7\% (1.13) & 96.8\% (1.10) & 96.8\% (1.12) \\
format-specific, Metal & 392 & 86.5\% (1.13) & 86.5\% (1.13) & 86.5\% (1.13) \\
blocked-dot fallback, GEMV & 372 & 98.1\% (1.18) & 97.6\% (1.17) & 97.8\% (1.17) \\
blocked-dot fallback, GEMM & 966 & 100\% (2.36) & 100\% (2.36) & 100\% (2.36) \\
both strata & 2041 & 96.7\% (1.57) & 96.5\% (1.53) & 96.5\% (1.53) \\
both strata, CPU with Accelerate & 2041 & 87.1\% (1.31) & 86.8\% (1.30) & -- \\
\bottomrule
\end{tabular}
\end{center}

\begin{figure}[!htbp]
\centering
\includegraphics[width=0.5\linewidth]{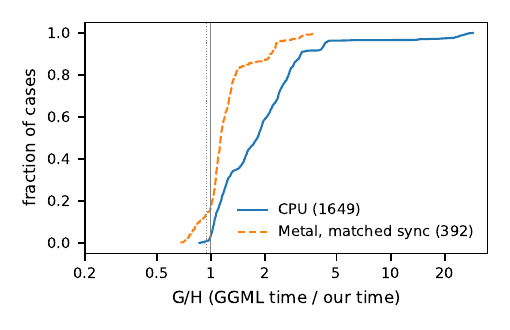}
\Description{Two step curves, CPU and Metal, of the fraction of cases with speed ratio at most x, on a log axis from 0.2 to 30. The Metal curve rises steeply around 1; the CPU curve lies to the right, mostly between 1 and 5, with a tail to about 28.}
\caption{Diagnostic, per-case layouts: cumulative distribution
of \GH over both strata (Metal: matched synchronization). Dotted: 0.95;
solid: parity. Medians: CPU \medCPU, Metal \medMetal.}\label{fig:cdf}
\end{figure}

\paragraph{Distribution and bf16.}
Over both strata, our kernels exceed \GGML's speed ($\GH>1$) in
\pctExceedAllFixed of the cases. \autoref{fig:cdf} shows the distribution
with per-case layouts: Metal rises steeply just above parity, and the CPU
lies mostly between 1 and 5. Without bf16, whose \GGML baseline is weak
(\S\ref{sec:eval:perf}), the format-specific median is
\medHHFixedNoBF{}$\times$ and we match in \pctHHFixedNoBF of these cases
(all cases: \pctMatchAllFixedNoBF, median \medAllFixedNoBF{}$\times$).

\paragraph{Cases below 0.95.}
With per-case layouts, \nBelowCPU CPU cases fall below 0.95: seven
T10 GEMVs at $4096\times1024$ (15--32\,\textmu s, too short to amortize fan-out), six q1\_0 GEMMs
against \GGML's repack, and a \cc{qsi8cx} GEMM. Held-out-shape
layouts add \nFixedLost: three \cc{qsi8cx} cases at
$4096\times14336$ and two f16 GEMVs. In all, \nBelowCPUFixed CPU cases fall below 0.95 with held-out-shape
layouts. On Metal, \nBelowMetalSmall cases at $M=8$ (\metalMeightLo--\metalMeightHi)
are limited by \cc{simdgroup\_matrix} loads from memory, the
\nBelowMetalSmallGEMV $4096\times1024$ GEMVs by the runtime's per-launch
overhead (\autoref{app:method}), and bf16 at $M=512$ (0.78--0.87) by the
missing \cc{bfloat} type.

\paragraph{Bandwidth.}
For GEMV, we divide the bytes a kernel must move (stored weights, activations
and output) by its time, and compare with a read benchmark at the same
footprint (\cc{stream\_read}: Neon loads on 10 threads; a Metal read kernel).
It sustains about \bwPeakCPU{}\,GB/s on the CPU and \bwPeakMetal on Metal at
128\,MB and beyond (specification: \bwSpec), more at smaller footprints. Our
timings are warm, so weights up to 32\,MB come from the caches. For the 18 CPU
and 12 Metal GEMVs whose weights exceed them (48--118\,MB), each case's
fraction divides its bandwidth by the benchmark's at its footprint (the larger
of the two nearest measured sizes: \bwDenCPU{}\,GB/s on 10 CPU threads,
\bwDenMetal on Metal), and we report medians of these per-case fractions: ours
\bwFracCPUOurs on the CPU (median \bwCPUOurs{}\,GB/s, above the \bwPeakCPU{}\,GB/s
at 128\,MB because the benchmark too reads faster at these footprints;
\GGML \bwFracCPUGGML, \bwCPUGGML) and \bwFracMetalOurs on Metal (\bwMetalOurs; \GGML
\bwFracMetalGGML, \bwMetalGGML). Over all GEMVs, the median fraction is \bwFracAllCPUOurs on 10 CPU threads
(\GGML \bwFracAllCPUGGML) and \bwFracAllMetalOurs on Metal (\GGML
\bwFracAllMetalGGML): smaller shapes fall well short of the caches' bandwidth,
and the grid and low-bit formats are bound by decoding (6\% and 9\% on the CPU).

\paragraph{One transformer block.}
To weigh the shapes as a model does, we sum each format's times over the
projections of one Llama-3-8B block: $q$ and $o$ ($4096\times4096$), $k$ and
$v$ ($4096\times1024$, grouped-query attention), gate and up
($4096\times14336$) and down ($14336\times4096$), all in one weight format.
The ratio is \GGML's total time over ours, $\sum_s G_s/\sum_s H_s$, which no
threshold affects; all 32 blocks are alike, so it is also the model's
projection ratio. Activations are f32, as llama.cpp passes them, so both
sides' times include converting them; decode is $M=1$, prefill $M=512$, and
each shape uses its own fold's held-out-shape layout, so only qsi8cx's block
mixes two layouts. Attention itself, norms, RoPE,
the embedding and the output head are not included. Cells give the median
ratio over formats and the formats with a ratio of at least 0.95;
format-specific formats are those for which \GGML's fastest path is
format-specific on some shape, blocked-dot formats the others.
\begin{center}
\small
\begin{tabular}{@{}llll@{}}
\toprule
regime & format-specific & blocked-dot & all \\
\midrule
CPU decode, T10 & \tokDecHHMed{} (9/9) & \tokDecFbMed{} (19/19) & \tokDecMed{} (28/28) \\
CPU decode, T1 & \tokDecToneHHMed{} (8/9) & 1.13 (19/19) & \tokDecToneMed{} (27/28) \\
CPU prefill, T10 & \tokPreHHMed{} (9/9) & \tokPreFbMed{} (19/19) & \tokPreMed{} (28/28) \\
\quad with Accelerate & \tokPreBlasHHMed{} (\nTokPreBlasHHMatch/9) & \tokPreBlasFbMed{} (\nTokPreBlasFbMatch/19) & \tokPreBlasMed{} (\nTokPreBlasMatch/28) \\
Metal decode & \tokMetalDecMed{} (\nTokMetalDecMatch/16) & --- & \tokMetalDecMed{} (\nTokMetalDecMatch/16) \\
Metal prefill & \tokMetalPreMed{} (\nTokMetalPreMatch/16) & --- & \tokMetalPreMed{} (\nTokMetalPreMatch/16) \\
\bottomrule
\end{tabular}
\end{center}

\FloatBarrier
\section{Correctness checks}\label{app:checks}

\paragraph{Breakdown.}
At the evaluated revision the checker reports \nChecks results. \nChecksCodec are
codec results: 330 encoder and 330 decoder comparisons of our stored formats, 100
decoder comparisons on crafted blocks, and 20 comparisons of our transcriptions
of \GGML's repacking and \Kleidi's packers with the originals.
\nChecksKernel are kernel results: \nChecksKernelOurs of ours (\nChecksKernelOursVecDot
\cc{vec\_dot}, \nChecksKernelOursMulMat GEMV and GEMM) and \nChecksKernelGGML of \GGML's and \Kleidi's
paths, held to the same tolerance. The kernel shapes are small ($K\times N\times
M$ of $256\times64\times1$, $512\times96\times5$, $1024\times256\times40$;
\cc{vec\_dot} at $K\in\{256,4096\}$) so that edges and tails are exercised; the
codec rows run at up to $2^{20}$ values. Kernels are compared on
\nChecksOutputs outputs in all. \nChecksBitExact results are bit-exact: the
codecs, and \nChecksGGMLDot f16-accumulating kernel rows that also match
\GGML's \cc{vec\_dot}. The other \nChecksTolOnly differ in the last bits and
lie within the tolerance.

\paragraph{Large-scale codec runs.}
A further run compares the \nExactFormats unrepacked non-grid formats on
\nExactBlocks blocks (Gaussian, skewed, ties on half-integer and codebook grids,
sub-block scales across the float range, offsets and constants, tiny, huge,
infinite and NaN values), and the seven grid IQ formats on \nExactGridBlocks
blocks (without those \GGML's search aborts on). Every decoder is bit-exact, and so
is every encoder (the grid ones are \GGML's own) except mxfp4's on blocks
with infinities or NaNs and the legacy ones on \nLegacyUBDiff of \nLegacyBlocks
blocks where \GGML's result is undefined (a non-finite or out-of-range value
converted to an integer). The legacy
encoders use strict float. Under Halide's default float semantics, which fuse
$x\cdot d^{-1}+c$ and divide by a constant through its reciprocal, \nLegacyNonStrict
blocks differed where \GGML's result is defined (fused multiply-adds,
reciprocal division, an underflowing scale test, a zero minimum's sign): the
compiler's float freedom is an approximation, which a schedule must not take
silently.

\paragraph{Kernel tolerance.}
Let $W$ and $A$ be the decoded operands of one output, $\delta w_k$ and
$\delta a_k$ bounds on what any requantization inside the kernel changes in them
(for an 8-bit absmax block whose f16 scale is normal, $\max_\text{block}|x|/127\cdot(1+2^{-7})$; for f16
and bf16, half an ulp), $c$ the declared accumulation chain length, $u$ the
accumulator's unit roundoff, $u_o$ the operand rounding, and
$\gamma_n(u)=nu/(1-nu)$. The check requires
\[
|y - y_{64}| \le \sum_k \bigl(|W_k|\,\delta a_k + |A_k|\,\delta w_k + \delta w_k\,\delta a_k\bigr) + g\,S + \gamma_K(2^{-53})\sum_k |W_k|\,|A_k|,
\]
\[
S = \sum_k \bigl((|W_k|+\delta w_k)(|A_k|+\delta a_k) + 2\,\overline W_k\,|A_k|\bigr),
\]
\[
g = \gamma_{c+4}(u) + \gamma_{\lceil\log_2(K/c)\rceil+1}(2^{-24}) + 2u_o + u_o^2,
\]
where $y_{64}$ is the dot product computed in f64, in any order (the last
term bounds its own rounding), and $\overline W_k$ the largest $|W|$ in
$k$'s weight block. The term $2\,\overline W_k|A_k|$ covers formats
that subtract a block minimum; $g$ covers $c$ accumulations with up to four
operations per term, an f32 tree over the $K/c$ partial sums, and operand
rounding. These roundings multiply: each term carries a factor
$\prod_i(1+\theta_i)$ with $|\theta_i|\le a_i$ and $\sum_i a_i=g$, and
$\prod_i(1+a_i)-1\le e^{g}-1\le g/(1-g)$ for $g<1$. Every check has $g<0.08$
(f16 accumulation over $K=4096$ is the largest); $\gamma_n(u)$ is infinite when
$nu\ge1$, which would make the check vacuous. The rigorous coefficient is thus
$g/(1-g)$; the check uses $g$, which is stricter: a kernel that passes meets
the rigorous bound, and none failed. With q8\_1 activations a further $\gamma_2(2^{-11})\sum_k
\overline W_k|A_k|$ covers its block sum, while the sum is within f16's range
(a block of 32 values of 2048 overflows it). Our sum is never farther
from the decoded sum than \GGML's; at a tie, where the decoded sum is an f16
midpoint, ours is the even neighbour: for $d=1$ and $\sum_jq_j=2049$, \GGML
stores 2050 and we store 2048 (\S\ref{sec:props:q81}). A kernel's error ratio to this
bound measures the bound, not what it would detect: with cancelling terms, a
wrong output can lie inside it (with f16 accumulation over $K=4096$, an output
of 0 where the exact dot product is 20). The tolerance is therefore
complemented by exact checks: every encoder and decoder is compared bitwise
with \GGML's (\S\ref{sec:eval:correct}), and kernels that accumulate in f16,
whose tolerance is the widest, must also equal \GGML's f16 \cc{vec\_dot} bit
for bit on every output.

\paragraph{Do the checks discriminate?}
Because the bound is worst-case, we ran \nMutKinds mutants (a dropped block,
shifted and tail rows, a tail column, a flipped sign, an omitted offset,
swapped scales, a zeroed output, and a bug confined to one interior row) of one
kernel per arithmetic family (\nMutFamilies), with the correct kernel run on
mutated inputs. Inputs were the harness's data and exact structured data:
cancelling $\pm1$ activations, affine offsets, distinct neighbouring rows, the
test shapes' output tails, and $K=4096$; on structured data every output's
exact value is an f32, and the structured test (S) requires every output to
equal it. Of \nMutNonEquiv non-equivalent mutant runs, the tolerance (T)
alone rejects \nMutTRejects, and T, G or S together \nMutAnyRejects
(\autoref{tab:mutants}). Every mutant it passes
is rejected by bit-equality, either with \GGML's f16 \cc{vec\_dot} (G; with
f16 accumulation, where a zeroed output passes the bound at
\nMutZeroRatioSOne of it) or by S. T and G compare every output with the f64
oracle and \GGML. An earlier checker sampled 64 rows $\times$ 4 columns, and
on random data \nMutMissedSampled runs of a one-row bug passed it; checking
every output, T rejects all \nMutMissedSampled. The tolerance also misses, under cancellation with f32
accumulation, a tail column that changes each result by $2/28$ of its value;
S rejects it.

\begin{table}[!htbp]
\caption{Mutants of one kernel per arithmetic family (\nMutFamilies), run on the harness's data and on exact
structured inputs. Counts are of non-equivalent runs (\nMutEquiv of the \nMutRuns runs leave every output
unchanged), summed over the families a mutant applies to (fam.); a cell is rejected/applicable. T: the
derived tolerance; G: bit-equality with \GGML's \cc{vec\_dot} (f16 accumulation only); T and G check
every output. S: the exact-result test, on the structured-input runs only; any: T, G or S. Interior row:
a bug in one row that an output sample of 64 rows would skip. The f32 columns sum six
families (q4\_0, q4\_1, q4\_K, iq4\_nl, iq2\_xxs and f16 with f32 accumulation); the last row is the
total over all \nMutNonEquiv runs.}
\label{tab:mutants}
\small
\setlength{\tabcolsep}{4pt}
\begin{tabular}{@{}lrrrrrrrrrrr@{}}
\toprule
& \multicolumn{4}{c}{f32 accumulation} & \multicolumn{4}{c}{f16 accumulation} & \multicolumn{3}{c}{all} \\
\cmidrule(lr){2-5}\cmidrule(lr){6-9}\cmidrule(l){10-12}
Mutant & fam. & runs & T & S & runs & T & G & S & runs & T & any \\
\midrule
drop last block & 6 & 72 & 72/72 & 48/48 & 12 & 4/12 & 12/12 & 8/8 & 84 & 76/84 & 84/84 \\
row shift & 6 & 72 & 72/72 & 48/48 & 12 & 9/12 & 12/12 & 8/8 & 84 & 81/84 & 84/84 \\
tail row ($N{-}1$) & 6 & 72 & 72/72 & 48/48 & 12 & 7/12 & 12/12 & 8/8 & 84 & 79/84 & 84/84 \\
tail column ($M{-}1$) & 6 & 54 & 48/54 & 36/36 & 9 & 4/9 & 9/9 & 6/6 & 63 & 52/63 & 63/63 \\
sign flip & 6 & 72 & 72/72 & 48/48 & 12 & 6/12 & 12/12 & 8/8 & 84 & 78/84 & 84/84 \\
omit offset/min & 3 & 36 & 36/36 & 24/24 & --- & --- & --- & --- & 36 & 36/36 & 36/36 \\
swap scales & 5 & 40 & 40/40 & 20/20 & --- & --- & --- & --- & 40 & 40/40 & 40/40 \\
zero output & 6 & 72 & 72/72 & 48/48 & 12 & 9/12 & 12/12 & 8/8 & 84 & 81/84 & 84/84 \\
interior row & 6 & 36 & 36/36 & 24/24 & 6 & 4/6 & 6/6 & 4/4 & 42 & 40/42 & 42/42 \\
\midrule
total & &  526 & 520/526 & 344/344 & 75 & 43/75 & 75/75 & 50/50 & 601 & 563/601 & 601/601 \\
\bottomrule
\end{tabular}
\end{table}

The bound assumes normal-range data: a block whose f16 scale is
subnormal loses relative precision in it, one whose scale rounds to zero
decodes to zeros (an error of $\max_\text{block}|x|$), and an underflowing
product adds an absolute error that $g\,S$ omits (Lemma~\ref{lem:kernel}'s
$K\eta$, as in \code{Accumulate}'s bound, \S\ref{app:accumulate}).
The tests' inputs are normal-range; these are preconditions on supported
inputs, which no compiler check enforces.

\paragraph{Negative tests.}
Halide's tests show each build-time check of \S\ref{sec:props:laws}
rejecting a violation: a broken chain of ranges; a term that differs from its
component's definition by a rounding, a scale, a block or a summation range; a
bf16--f16 widening; an overflowing retyped sum; and a hoist over a
data-dependent domain. Two normalizer rules, the float-cast predicate and an
invariant sum's extent, were hardened with such tests.

\paragraph{Property tests.}
\autoref{tab:props} lists the property tests. A trial draws one input of the
given size; distributions mix normal values at several scales, one-signed
blocks, outliers, constants and zeros, per block or sub-block, so that the
searches reach their corner cases. Seeds are fixed (trial $i$ uses seed $i$).
In all, they add \nPropChecks checks over \nPropTrials trials, and
\nExhaustive exhaustive sweeps.

\begin{table}[!htbp]
\caption{Property tests of the application. ``Checks'' counts format $\times$ distribution $\times$ property instances; each passed all
of its trials.}\label{tab:props}
\small
\begin{tabular}{@{}p{0.2\linewidth}p{0.42\linewidth}rrr@{}}
\toprule
Formats & Properties & Checks & Trials & Values/trial \\
\midrule
q4\_0, q4\_1, q5\_0, \mbox{q5\_1}, q8\_0, q8\_1 & bound, ranges, closure & 180 & 1440 & 32768 \\
q8\_1; 5 interleaved layouts & law, derived block sum, closure; the law & 13 & 44 & 32768+ \\
mxfp4, nvfp4 & bound, ranges, closure, law & 71 & 562 & 32768+ \\
iq4\_nl, iq4\_xs & bound, ranges, closure & 12 & 36 & 65536 \\
q8\_K, q2\_K--q6\_K & bound, ranges, closure & 54 & 162 & 65536 \\
tq1\_0, tq2\_0, q1\_0, q2\_0 & bound, ranges, closure; q1\_0 also the law & 29 & 84 & 65536 \\
\cc{qsi8cx}, \cc{qai8dx} (2 layouts) & bound, ranges, closure, law; \cc{qsi8cx} also derived row sum & 146 & 1154 & 2048 \\
grid IQ (7) & ranges of the full scheme & 7 & 14 & 16384 \\
grid and sign tables (8) & lossless, with and without tabulation & 14 & 14 & all indices \\
minifloat units & bound, closure, law & 42 & 336 & 4096 \\
\midrule
fp16, bf16, e8m0 & bitwise against Halide's and \GGML's conversions & \multicolumn{3}{l}{exhaustive: 5 sweeps} \\
12 minifloat units & encoder against a reference, and the law, on every f32; every code round-trips & \multicolumn{3}{l}{exhaustive: 36 sweeps} \\
\nExhFormats block formats & the law on every scale or $(d,m)$ pair & \multicolumn{3}{l}{exhaustive, compiled codecs} \\
\bottomrule
\end{tabular}
\end{table}

\begin{table}[!htbp]
\caption{Each format's round-trip law (Definition~\ref{def:law}) and its
evidence: sampled, edge and adversarial blocks (tested), or every f32, scale
or $(d,m)$ on the compiled codecs, extended to all blocks by
\autoref{app:checks}'s reduction lemmas (exhaustive). Every format but grid IQ
also declares a tested error bound and passes the input-dependency check
(\S\ref{sec:props}); grid IQ declares no bound and skips the check. q8\_1: ours (\GGML's fails,
\S\ref{sec:props:q81}). Layouts keep their format's status; windows and
counterexamples: \autoref{app:checks}.}\label{tab:contracts}
\footnotesize
\setlength{\tabcolsep}{3pt}
\begin{tabular}{@{}llp{0.5\linewidth}@{\hspace{10pt}}l@{}}
\toprule
Formats & Law & Where, or why not & Evidence \\
\midrule
f16, bf16 & exact & every f32 & exhaustive \\
q4\_0, q5\_0, q8\_0, q8\_1 & on domain & f16 scale finite, nonzero & exhaustive \\
q4\_1, q5\_1 & on domain & finite f16 $d$, $m$; not where f32 $m{+}qd$ absorbs integers & exhaustive \\
tq1\_0, tq2\_0, q2\_0 & on domain & f16 scale finite, nonzero & exhaustive \\
q1\_0 & on domain & not where a negative value's scale underflows & exhaustive \\
mxfp4 & on domain & $2^{-125}\lesssim\max|x|\lesssim2^{128}$ & exhaustive \\
nvfp4 & on domain & all but three narrow windows of $\max|x|$ & tested \\
\cc{qsi8cx} & no failure & no counterexample in sampled rows & tested \\
q8\_K, \cc{qai8dx} & 1--2 ulp & integers exact (q8\_K: up to sign on a tie); scale off by 1--2 ulp, repairable & tested \\
q2\_K--q6\_K & fails & the search re-fits the block & counterexample \\
iq4\_nl, iq4\_xs & fails & each of 16 trial scales is re-fit & counterexample \\
grid IQ (7) & none & \GGML's search (\code{TrustedInverse}); only declared ranges and agreement with \GGML & -- \\
\bottomrule
\end{tabular}
\end{table}

\paragraph{Where the law holds.}
The law holds exactly for \nEqExact formats on their domains (for
\cc{qsi8cx}, no counterexample was found), up to 1--2 ulp of the scale for
the \nEqUlp closed-form encoders q8\_K and \cc{qai8dx}, and not for the
\nEqRefit search encoders; the \nEqNone grid IQ formats declare none.
A \emph{closed-form} quantizer computes a block's fields from the block and
rounds each value to nearest; it satisfies the law where decoding is exact in
f32 and reproduces those fields. A type-0 decode $q\,d$ with an f16 $d$ and
$|q|\le2^{13}$ is exact in f32, so q4\_0 satisfies the law wherever its scale
neither underflows nor overflows. An affine decode, $m+q\,d$, can absorb code
differences ($m=4096$, $d\approx6.5\cdot10^{-5}$ turns a code 4 into 7), so
q4\_1 and q5\_1 satisfy it only where $m$ and a nonzero $d$ are finite and
each $m+q\,d$ is exact or
$d\ge\nExhUlpQFourOne{}\,\mathrm{ulp}(\max_q|m+q\,d|)$ (q5\_1: \nExhUlpQFiveOne): the law is a property of a codec
under a concrete arithmetic. For the same reason their declared bound, like the
K-quants' affine ones, is half a step plus $2^{-23}(|m|+q_{\max}|d|)$, the f32
rounding of $q\,d+m$, with $q_{\max}$ the top code. Outside q4\_0's domain (\S\ref{sec:example:compose}),
$\{2^{-23}\times32\}$ and $\{2^{-23},2^{-24}\times31\}$ decode alike, as their
scales underflow to zero. Elsewhere the classification is empirical
(\autoref{tab:contracts}). q8\_K and \Kleidi's qai8dx miss the law only
through scale arithmetic; integers and zero point round-trip. q8\_K stores
$d=1/(-127/m)$, twice rounded, signed by the block's signed maximum $m$, so a
$\pm127$ tie can flip it; qai8dx stores $1/(255/r)$ for the range $r$, and its
decoded endpoints are rounded products, so the scale moves by 1--2 ulp. As
for q8\_1, repairs are cheap: for q8\_K, a positive $d=\max|x|/127$, as
\cc{qsi8cx} computes (integers equal up to sign); for qai8dx, a scale rounded
down to 16 significant bits, making every decoded endpoint exact, and a zero
point from the minimum (worst-case error 0.507 of a step, not 0.500). The law
then holds on all \nLawQEightK q8\_K blocks and \nLawQai qai8dx rows we test
(deployed encoders: \pctLawQEightKFail and \pctLawQaiFail fail); we keep the
deployed encoders for byte compatibility. The drift the repairs remove is at
most 2 ulp of the scale, or q8\_K's sign on a tie, which leaves the decoded
values unchanged; only the qai8dx repair changes the error bound, as above.
Search-based encoders re-optimize their fit on requantizing, so the law fails
on ordinary data.

\paragraph{Truncation.}
With a fixed scale and exact division, a truncating encoder satisfies the
law, since $\lfloor q\,d/d\rfloor=q$; a truncating q8\_0 fails it only through
its reciprocal multiply or its f16 scale. On \nLawBlocks blocks of 32 normal values, a q8\_0-style encoder satisfies the
law on every block. With truncation instead of rounding, it fails on
\pctTruncRecipBlocks of blocks (\pctTruncRecipCodes of integers) when integers are
computed as $x\cdot(1/d)$, as in \GGML, where $q\,d\cdot(1/d)$ can fall just
below $q$; and on \pctTruncDivBlocks with exact division by the f16 scale,
wherever rounding to f16 raises the scale and the block maximum's code
truncates to 126.

\paragraph{The round-trip law.}
\autoref{tab:props} counts the property tests. We also added a test of the
law for every stored format; its results
are in \autoref{tab:contracts}. For each format it encodes, decodes and re-encodes
sampled inputs (the distributions above, plus blocks that put the scale at the
edges of its type: subnormal, underflowing to 0, overflowing), and tallies the
blocks whose bytes change by the class of their encoded scale: zero, underflow,
subnormal, normal or non-finite. An adversarial generator adds blocks with a
large offset and a small spread: minima at or just off $\pm2^k$ for
$k\in[-14,15]$ (a few f32 ulps away, or just below a binade so that the f16
rounding crosses it) and spreads of 1 to $2^{16}$ f32 ulps (35,840 blocks for
the legacy formats). It found that q4\_1 and q5\_1 fail with a normal scale when
rounding $m+qd$ to f32 absorbs code steps: in the block $[4095.9990234375,
4095.999267578125, 4096, \ldots]$, integers 0, 4, 15 re-encode as 0, 7, 15. Of the
adversarial blocks, q4\_1 changes 0 of 14,886 that decode exactly, 1186 of 13,954
that decode inexactly with $d<512$ ulps and 0 of 403 with $d\ge512$ ulps (q5\_1:
0, 1605 and 0); a changed block has at most $d=104.5$ ulps (q5\_1: 54.5).
``Tested'' means no block in the format's domain changed in these distributions.
Where the law fails, the test asserts that blocks of normal scale
change, and a hand-shrunk minimal block (below) fails on its
own; mxfp4 and nvfp4 also sweep blocks of one value across their windows.

\paragraph{q8\_1's block sum.}
\GGML computes q8\_1's sum from the unrounded scale, $s=\fpsixteen(d_{32}\sum_j
q_j)$, and stores the scale as $\fpsixteen(d_{32})$; the law test
(\code{idempotent\_requantize(0)}) fails on \pctQEightOneDiffer of blocks of
normally distributed inputs, and two blocks with the same integers and scale
have sums one f16 ulp apart. In \GGML's stage order the scale's \code{fp16()}
follows the \code{Derive}, which condition (i) of \autoref{app:laws} rejects.
Ours derives $s=\fpsixteen(\fpsixteen(d)\sum_j q_j)$. The decoded sum
$\fpsixteen(d)\sum_jq_j$ is exact in f32 (11 significand bits times
$|\sum_jq_j|\le 4064<2^{12}$), so within f16's range our $s$ is it correctly
rounded, at most half an f16 ulp away; \GGML's rounds a sum computed with the
unrounded scale, which can itself differ from the decoded sum, so it can be up
to about one f16 ulp away. The two differ on \pctQEightOneDiffer of blocks,
exactly where \GGML's fails the law; ours is strictly closer on
\pctQEightOneStrict, and the other \pctQEightOneTie are f16-midpoint ties,
where ours is the even neighbour. In a q4\_1$\cdot$q8\_1 kernel, the stored
$s$'s rounding is the largest arithmetic error against an f64 dot of the
decoded operands; the consistent $s$ makes it smaller by a factor of
\oracleRatioLo--\oracleRatioHi{} (the median, over rows, of the per-row
ratio). Against unquantized operands, \GGML's is only
\oracleQuantLo--\oracleQuantHi of the error. Imported \GGML records, whose $s$
may lie about one f16 ulp from $\psi$, carry the weaker bound
$\gamma_2(2^{-11})$ of the kernel tolerance above.

\paragraph{Exceptions in \autoref{tab:contracts}.}
q4\_1 and q5\_1's domain (finite f16 $d$, $m$, with the conditions above)
is a superset of the reachable blocks, and its thresholds are not tight;
the law fails where rounding $m+qd$ to f32 absorbs integers. q1\_0 is exact
unless the scale underflows on a negative value ($-0\ge0$). mxfp4 is exact
for $2^{-125}\lesssim\max|x|\lesssim2^{128}$ (exactly: \emph{Exhaustive
check} below); outside, its exponent wraps or overflows. nvfp4 is exact
unless $\max|x|\in[18,20]/1024$, $\max|x|=30/1024$, or
$\max|x|\gtrsim2^{34}$ (scale saturated). For the search encoders, a
minimal counterexample, with all other values 0: q3\_K, $\{1,-1.1\}$;
iq4\_nl and iq4\_xs, which try 16 trial scales and re-fit each, three
nonzero values.

Blocks per format: legacy 5632 sampled plus the adversarial set; tq1\_0 and
tq2\_0 1024 super-blocks, q1\_0 2048 blocks, q2\_0 4096; mxfp4 65,536 plus an 836-block
sweep, nvfp4 81,920 plus 3937; each K-quant 768 super-blocks, q8\_K 2304
blocks; iq4\_nl and iq4\_xs 2048 and 256 blocks; each grid format 64
super-blocks; \Kleidi formats 1472 rows of 256; minifloats every f32. Only the
minifloat check and the nvfp4 sweep feed NaN or infinity, and only the minifloat
check and the mxfp4 sweep feed subnormal f32s; the sampled blocks are finite,
and non-finite or subnormal f16 scales reached from them are tallied separately.
NaN equals NaN. Float components compare as values, so a $\pm0$ scale is not a
change, but quantized integers compare exactly (q1\_0's sign flip is caught), and
minifloats compare bitwise.

\paragraph{q4\_0's domain and bound.}
f16 rounds $|d|\le2^{-25}$ to 0 and $|d|\ge65520$ to infinity, so with
$|x_m|=8|d|$ the domain of \S\ref{sec:example:compose} is $x_m=0$ or
$2^{-22}<|x_m|<524160$, subnormal f16 scales included. \code{Type0Quant}
declares one step, $|x-d\,q|\le|d|$, with a relative margin of $10^{-4}$
for its f32 reciprocal, which needs $1/d$ finite ($|d|>2^{-128}$, true on
q4\_0's domain); \code{fp16()} declares half an f16 ulp, or $2^{-25}$ below
the normal range. Proposition~\ref{prop:bound}(b) at the \code{Parallel}
holding \code{fp16()}, then (a) with $S^d_1=|q|\le8$, gives
$|x-\hat d\,q|\le|x-d\,q|+|d-\hat d|\,|q|$ for $\hat d=\fpsixteen(d)$
finite. For the unrounded
$|d|\ge2^{-14}$, $|d-\hat d|\le2^{-11}|d|$, giving the bound
$|d|(1+10^{-4}+2^{-8})$; for smaller $|d|$ (including those that round up to
$2^{-14}$), $|d-\hat d|\le2^{-25}$, and the bound is $|d|(1+10^{-4})+2^{-22}$.
Both hold in f32, as $\hat d\,q$ is exact there ($\hat d\in$ f16, $|q|\le8$).

\paragraph{Exhaustive check.}
For \nExhFormats closed-form formats we check the law on every encoding in
the format's domain, calling the compiled codecs. Each encoder reduces a block to its
scale fields (a maximum, minimum or sum) and then computes each code from its
own value and those fields (one lemma per encoder family:
\autoref{tab:exh-lemmas}). A decoded block holds its extreme code, so
re-encoding recovers the fields from the stored ones alone, and each new code
depends only on those fields and its own old code. It therefore suffices to
enumerate every stored scale, with arrangements that put every code at every
position and keep the extreme present: all \nExhScales f16 scales for q4\_0
and q5\_0, \nExhBlocksQEight blocks each for q8\_0 and q8\_1 (whose block sum
is compared on a second round trip), every positive f16 scale for tq1\_0, tq2\_0,
q2\_0 and q1\_0, and every e8m0 exponent from 0 to 252 for mxfp4. No block fails
with a finite nonzero scale the encoder can store (mxfp4: on its domain
$2^{-125}(1-11\cdot2^{-22})\le\max|x|<2^{128}(1-11\cdot2^{-22})$; outside it,
the wrapped exponents and $E=253$ fail). Each lemma's premise, the extreme
integers it needs, is a function of one f32 and was checked for every f32.
For q4\_1 and q5\_1 we
enumerate all \nExhPairsFull $(d,m)$ pairs (\nExhBlocksQFourOne and
\nExhBlocksQFiveOne blocks) without checking which an encoder can produce, so
the failing set over-approximates the reachable one. No pair on the domain
fails; failures need an
inexact $m+qd$ and $d<\nExhUlpQFourOne$ ulp (q5\_1: \nExhUlpQFiveOne; the
largest failing ratios are \nExhMaxFailUlpQFourOne and \nExhMaxFailUlpQFiveOne).
The adversarial tests' largest failure, 104.5 ulp (q5\_1: 54.5), is below
these because the enumeration includes pairs no f32 block produces: with
$|m|\gg d$, the f32 spacing near $m$ limits which $(d,m)$ are jointly
reachable; we do not characterize the reachable set.
\begin{table}[t]\small
\caption{Reduction lemmas for the exhaustive law check. On any decoded block, the encoder's reduction depends only on the stored fields, provided the listed codes are present; each code is then re-encoded from the fields and its own old code. Each premise is a function of one f32 (the extreme, or $R=\mathrm{fl}(\max-\min)$); swept over every f32, it holds wherever the stored scale is finite and nonzero, and for mxfp4 on its domain, $\lfloor\log_2\max|x|\rfloor\in[-125,127]$, outside which the law fails. $h=2^{b-1}$, $t=2^b-1$, $y=\dec(e)$, and a rotation puts every code at every position.}
\label{tab:exh-lemmas}
\begin{tabular}{@{}p{.1\linewidth}p{.19\linewidth}p{.32\linewidth}p{.28\linewidth}@{}}
\toprule
family & reduction & fixed by the stored fields because & tested: $\enc(\dec(e))=e$ for \\ \midrule
\raggedright q4_0, q5_0 & \raggedright signed max; $d=\max/(-h)$ & \raggedright code $0$ (value $-h$) present; $y=(c-h)\,d_{16}$ exact, so $\max'=-h\,d_{16}$ & \raggedright all $2^{16}$ $d_{16}$ $\times$ $2h$ rotations of $0..2h{-}1$ \tabularnewline
\raggedright q8_0 & \raggedright $d=\max|x|/127$ & \raggedright $\pm127$ present; $\max|y|=127\,d_{16}$ exact & \raggedright all $d_{16}$ $\times$ 1020 rotations of $[-127,127]$ with a $\pm127$ \tabularnewline
\raggedright q8_1 & \raggedright q8_0; $s=\mathrm{fp16}(d_{16}\sum_j c_j)$ & \raggedright as q8_0; $s$ is a function of $(d_{16},c)$, both restored for every $c$ & \raggedright as q8_0, $s$ compared on a second round trip \tabularnewline
\raggedright q4_1, q5_1 & \raggedright $m=\min$, $d=(\max-\min)/t$ & \raggedright codes $0,t$ present; $y=\mathrm{fl}(c\,d_{16}+m_{16})$ is monotone: $\min'=m_{16}$, $\max'=\mathrm{fl}(t\,d_{16}+m_{16})$ & \raggedright all $d_{16}>0$ $\times$ all finite $m_{16}$ $\times$ $2^b$ rotations of $0..t$; $d=0$ \tabularnewline
\raggedright tq1_0, tq2_0, q2_0 & \raggedright $d=\max|x|$ & \raggedright $\pm1$ present; $\max|y|=d_{16}$ & \raggedright all $d_{16}>0$ $\times$ 3 rotations of $\{-1,0,1\}$ \tabularnewline
\raggedright q1_0 & \raggedright $d=\sum_j|x_j|/128$ & \raggedright none: $|y_j|=d_{16}$ for every code, and the sum is exact & \raggedright all $d_{16}>0$ $\times$ both alternations of $\pm1$ \tabularnewline
\raggedright mxfp4 & \raggedright $E=\lfloor\log_2\max|x|\rfloor+125$ & \raggedright $|kv|\in\{8,12\}$ present; both give $\lfloor\log_2\max|y|\rfloor=E-125$ & \raggedright $E\le252$ $\times$ 28 rotations (extreme 12 or 8) \tabularnewline
\bottomrule
\end{tabular}
\end{table}

\paragraph{Proofs of the lemmas.}
Each row of \autoref{tab:exh-lemmas} claims that on $y=\dec(e)$, with the
premise integers present, the reduction is a function of the stored fields $F$,
and each new code is $G(F,c_j)$. Three facts serve every row.
(i)~Each $y_j$ is exact and finite: a code of at most 7 bits times an 11-bit
$d_{16}$ (type 1: then one rounding with $m_{16}$), below $128\cdot65504<2^{23}$
in magnitude (mxfp4: $12\cdot2^{124}$), far below $2^{128}$.
(ii)~The encoders' first strict maximum picks a position, but the reductions
use only its value, and all elements tied at the extreme have the same value,
sign included where the reduction keeps it; so which one is first does not matter.
(iii)~Each code depends on its own value and the reduced fields only, so a
rotation family that puts every code at every position covers every code
vector with the same fields.
\par\noindent\emph{q4\_0, q5\_0.} Integers lie in $[0,2h-1]$, so $|y_j|=|c_j-h|\,|d_{16}|\le h|d_{16}|$,
with equality only for code 0, whose value is $-h\,d_{16}$; hence the signed
maximum is $-h\,d_{16}$ and $d'=-h\,d_{16}/(-h)=d_{16}$, exactly (a power-of-two
division in f32's normal range).
\par\noindent\emph{q8\_0, q8\_1.} Integers satisfy $|c_j|\le127$, with equality for the premise code, so
$\max|y|=127\,|d_{16}|$ and $d'=\mathrm{fl}(127\,|d_{16}|/127)=|d_{16}|$, which
is $d_{16}$ as the encoder stores $d\ge0$; q8\_1's
$s'=\mathrm{fp16}(d'_{16}\sum_jc'_j)$ is a function of $(d'_{16},c')$, which
q8\_0's result restores for every code vector.
\par\noindent\emph{q4\_1, q5\_1.} $c\,d_{16}$ is exact and increasing in $c$ for $d_{16}>0$,
and rounding is monotone, so $y(c)=\mathrm{fl}(c\,d_{16}+m_{16})$ is
nondecreasing in $c$: with integers $0$ and $t$ present, $\min'=y(0)=m_{16}$ and
$\max'=y(t)$, functions of $(d_{16},m_{16})$; by (i) neither overflows.
\par\noindent\emph{tq1\_0, tq2\_0, q2\_0.} $|y_j|\le d_{16}$, attained by the premise code $\pm1$,
so $\max|y|=d_{16}$ (q2\_0's code 2 needs $1/\max|x|$ to overflow, so $d_{16}=0$).
\par\noindent\emph{q1\_0.} $|y_j|=d_{16}$ for every $j$, and every partial sum $k\,d_{16}$,
$k\le128$, has at most 19 significant bits, so the in-order sum is
$128\,d_{16}$ exactly and $d'=d_{16}$ for every code vector.
\par\noindent\emph{mxfp4.} $y_j=kv_j\,2^{E-128}$ is exact; the premise code makes
$\max|y|$ either $8\cdot2^{E-128}=2^{E-125}$ or $12\cdot2^{E-128}=1.5\cdot2^{E-125}$,
both with $\lfloor\log_2\rfloor=E-125$, and the encoder's computed logarithm
agrees on them (the premise sweep covers every f32), so $E'=E$.
\par In each row, whether $G(F,\cdot)$ is the identity, and whether $F$ is
restored, is what the enumeration checks on the compiled codecs. The lemmas
are proved of the source and the enumeration runs the binary; joining them
assumes the compiler keeps the source's dependencies.

For q8\_K and \cc{qai8dx}, which hold up to the scale's last bits, a scalar sweep
over every f32 block maximum in $[1,2)$ confirms the cause: requantizing changes
q8\_K's scale for 7.2\% of them, by 1 ulp, and \cc{qsi8cx}'s for none; with
the repaired q8\_K scale $d=\max|x|/127$, $\mathrm{fl}(\mathrm{fl}(127d)/127)=d$
for all of them (see above). For the grid IQ
formats, whose encoders are \GGML's, the only other sampled check of the full
scheme is that decoded values stay in their declared ranges; their exactness
against \GGML is the codec check.

\FloatBarrier
\section{Schedule knobs}\label{app:knobs}

The format table's \nKnobRows knob rows set \nKnobSettings values of \nKnobs
parameters, for \nKnobFormats formats, all on the CPU. Every other format and
every Metal kernel uses the defaults. \emph{interleave} (18 settings) sets how
many blocks one loop iteration covers; one where a second block's loads leave the
load instruction's immediate range, as with q8\_K activations. \emph{sublanes} (6)
spreads sub-blocks across vector lanes so that their scales take one multiply-add
per four sub-blocks. \emph{slp} (6) disables LLVM's SLP vectorizer where it packs
scalar grid-index arithmetic into vectors (iq1\_s, iq1\_m). \emph{row\_pairs} (5)
turns off computing two weight rows at once where that spills. \emph{fold\_offsets}
(4) chooses where a constant code offset is folded into the activations' block
sums, as \GGML's \cc{vec\_dot} does. \emph{gpr\_scales} (4) decodes six-bit
sub-block scales in general-purpose registers (q4\_K, q5\_K), as \GGML does.
\emph{gemm\_index\_words} (3) stages iq3\_s's grid-index bytes a word at a time in
the GEMM tiles. We chose each value by A/B timing on the evaluation machine; the
format table's comments record the measured differences (for example, 1.02 to
2.7 times for \emph{interleave}).

\paragraph{Ablation.} We rebuilt our kernels with every knob at its default
(the \nKnobRows rows touch 42 CPU libraries of \nKnobFormats formats; no Metal library) and timed both builds back to
back on \nKnobCases cases: the \nKnobFormats formats on their paired and on f32
activations, for \cc{vec\_dot}, GEMV at T1 and T10, and GEMM at $M=32$ and 512
at T10, all at $4096\times4096$ (at least two repeats; minimum time per kernel).
The default build passes the checker. \autoref{tab:knobs} summarizes the
result. Defaults already match \GGML in \nKnobDefaultMatch of \nKnobCases cases
(all \nKnobCases tuned). Large gains come from four formats: q3\_K (1.9--2.3
times in every column; the default offset fold doubles its loop's instruction
count), iq1\_s and iq1\_m (1.03--1.39, from the LLVM flag), and iq2\_xs
(1.15--1.18, where two q8\_K blocks per iteration spill). Seven formats (iq2\_xxs,
iq3\_xxs, iq3\_s, iq4\_xs, q2\_K, q5\_K, tq2\_0) match in every column with the
defaults. No knob makes a case slower (lowest speedup of tuned over default: 0.98, within noise).
The values were chosen on this machine and the ablation covers one of the four
shapes.

\begin{table}[!htbp]
\caption{Knob ablation on the \nKnobCases cases of the \nKnobFormats knob formats:
cases where our default build matches \GGML ($\GH\ge0.95$), and the median
speedup of the tuned build over the default one.}\label{tab:knobs}
\footnotesize
\begin{tabular}{lrrr}
\toprule
Kernel & cases & default matches & speedup of tuned over default (median) \\
\midrule
\cc{vec\_dot} & 14 & 8 & 1.06 \\
GEMV (T1, T10) & 56 & 41 & 1.02 \\
GEMM ($M=32$, 512) & 56 & 54 & 1.01 \\
\midrule
all & \nKnobCases & \nKnobDefaultMatch & \medKnobGain \\
\bottomrule
\end{tabular}
\end{table}

\FloatBarrier
\section{Speed ratios by format family}\label{app:family}

Tables~\ref{tab:cpu} and~\ref{tab:metal} break \S\ref{sec:eval:perf}'s
results down by weight family, activation format and operation. They are
diagnostics: each case takes our fastest stored layout (per-case layout),
whereas the summaries of \autoref{tab:summary} fix one layout per target and
weight type within each held-out fold (\autoref{app:fixed}). Both tables cover both strata of \GGML's
fastest non-BLAS path; \autoref{app:blas} gives the CPU baseline including
Accelerate.

\begin{table}[!htbp]
\caption{CPU speedups against \GGML's fastest non-BLAS path, with per-case
layouts (a diagnostic; held-out-shape summaries: \autoref{tab:summary}): \GH (\GGML time\,/\,our time; $>1$: ours faster). Cells: min--max (median) over a row's formats and $K\times N$
shapes; superscript: the number of cases
below 0.95 (sensitivity: \autoref{tab:sensitivity}). Activations: paired, the format of \GGML's \cc{vec\_dot} for the
weights (q8\_0, q8\_1, q8\_K, f16 or bf16). Fastest \GGML paths, by frequency: C plain CPU, R repack,
K \Kleidi.}\label{tab:cpu}
\footnotesize
\setlength{\tabcolsep}{4pt}
\begin{tabular*}{\linewidth}{@{\extracolsep{\fill}}lllccc@{}}
\toprule
\multicolumn{6}{@{}l}{\emph{(a) Single row: \cc{vec\_dot} and GEMV}}\\
Weights & Activations & \GGML & \cc{vec\_dot} & GEMV T1 & GEMV T10 \\
\midrule
legacy & paired & C & 0.99--1.04 (1.01) & 1.00--2.13 (1.74) & 0.95--2.65 (1.49) \\
 & f32 & C, R, K & n/a & 0.99--2.13 (1.08) & 1.01--2.28 (1.11) \\
K-quants & paired & C & 1.01--1.24 (1.07) & 1.02--1.45 (1.15) & 1.01--2.04 (1.21) \\
 & f32 & R, C & n/a & 1.02--1.25 (1.10) & 0.88--1.75 (1.13)$^{3}$ \\
grid IQ & paired & C & 0.99--1.29 (1.19) & 0.99--1.30 (1.15) & 0.97--1.35 (1.13) \\
 & f32 & C & n/a & 0.99--1.30 (1.15) & 0.86--1.32 (1.10)$^{3}$ \\
codebook & paired & C & 1.02--1.31 (1.08) & 1.18--1.75 (1.51) & 1.20--1.98 (1.65) \\
 & f32 & R, C & n/a & 1.02--1.56 (1.17) & 1.07--1.75 (1.22) \\
low-bit & paired & C & 0.98--1.08 (1.07) & 1.01--1.35 (1.15) & 1.05--1.44 (1.21) \\
 & f32 & C, R & n/a & 1.02--1.20 (1.08) & 0.88--1.21 (1.07)$^{1}$ \\
\cc{qsi8cx} & f32 & K & n/a & 0.98--1.84 (1.45) & 1.07--1.79 (1.44) \\
f16 & both & C & median \medFHVecDot & \multicolumn{2}{c}{median \medFHGEMV} \\
bf16 & both & C & median \medBFVecDot & \multicolumn{2}{c}{medians \rangeBFGEMV} \\
\midrule[\heavyrulewidth]
\multicolumn{6}{@{}l}{\emph{(b) GEMM, also over $M\in\{8,32,512\}$}}\\
Weights & Activations & \GGML & & GEMM T1 & GEMM T10 \\
\midrule
legacy & paired & C & & 1.75--2.85 (2.12) & 1.73--3.19 (2.15) \\
 & f32 & C, R, K & & 0.99--2.84 (1.89) & 0.98--3.49 (1.82) \\
K-quants & paired & C & & 1.89--2.42 (2.20) & 1.80--2.63 (2.18) \\
 & f32 & R, C & & 0.97--2.42 (1.22) & 0.97--2.46 (1.32) \\
grid IQ & paired & C & & 1.81--4.38 (2.92) & 1.86--4.57 (2.95) \\
 & f32 & C & & 1.81--4.55 (2.89) & 1.00--4.43 (2.65) \\
codebook & paired & C & & 1.45--1.78 (1.55) & 1.44--1.81 (1.60) \\
 & f32 & R, C & & 1.14--1.62 (1.39) & 0.97--1.67 (1.35) \\
low-bit & paired & C & & 1.43--3.23 (2.53) & 1.40--3.40 (2.53) \\
 & f32 & C, R & & 0.92--2.77 (1.91)$^{4}$ & 0.87--2.85 (1.80)$^{2}$ \\
\cc{qsi8cx} & f32 & K & & 0.92--2.63 (1.00)$^{1}$ & 0.96--2.48 (1.04) \\
f16 & both & C & & \multicolumn{2}{c}{median \medFHGEMM} \\
bf16 & both & C & & \multicolumn{2}{c}{medians \rangeBFGEMM} \\
\bottomrule
\end{tabular*}

\smallskip
\begin{minipage}{\linewidth}
\emph{Rows:} legacy (q4\_0--q8\_1),
K-quants (q2\_K--q6\_K), grid IQ, codebook (iq4\_nl, iq4\_xs, mxfp4, nvfp4;
Metal: no nvfp4), low-bit (tq1\_0, tq2\_0, q1\_0, q2\_0; Metal: q2\_0 only).
\emph{f16, bf16:} a cell covers both activation formats and, if two columns
wide, both thread counts; ``medians'' gives the range of the medians per
activation format and thread count.
\end{minipage}
\end{table}

\begin{table}[!htbp]
\caption{Metal speedups \GH against \GGML's Metal backend, with per-case
layouts (a diagnostic), both sides waiting once per 16 multiplications. Cells:
min--max (median) over a row's formats and $K\times N$ shapes; superscript:
the number of cases below 0.95. Rows as in \autoref{tab:cpu}. n/a: \GGML has
no kernel.}\label{tab:metal}
\footnotesize
\setlength{\tabcolsep}{3pt}
\begin{tabular}{@{}llcccc@{}}
\toprule
Weights & Activations & GEMV & $M{=}8$ & $M{=}32$ & $M{=}512$ \\
\midrule
legacy & f32 & 0.72--1.63 (1.04)$^{6}$ & 0.74--1.25 (0.96)$^{9}$ & 1.06--2.29 (1.33) & 1.01--1.21 (1.12) \\
 & f16 & n/a & n/a & 1.07--2.34 (1.32) & 1.02--1.20 (1.09) \\
K-quants & f32 & 0.76--1.38 (0.93)$^{11}$ & 0.68--3.76 (0.87)$^{12}$ & 1.02--2.30 (1.27) & 1.05--1.20 (1.13) \\
 & f16 & n/a & n/a & 1.04--2.34 (1.28) & 1.04--1.19 (1.12) \\
codebook & f32 & 0.81--1.34 (1.23)$^{3}$ & 1.19--3.12 (2.49) & 1.02--2.29 (1.28) & 1.01--1.14 (1.07) \\
 & f16 & n/a & n/a & 1.06--2.29 (1.30) & 1.03--1.13 (1.06) \\
low-bit & f32 & 0.82--0.93 (0.89)$^{4}$ & 1.63--3.67 (3.38) & 1.12--2.08 (1.29) & 1.07--1.20 (1.13) \\
 & f16 & n/a & n/a & 1.16--2.11 (1.32) & 1.09--1.19 (1.11) \\
f16, bf16 & paired & 0.72--1.23 (1.00)$^{2}$ & 1.31--2.24 (1.79) & 1.19--2.13 (1.38) & 0.96--1.17 (1.01) \\
 & f32 & 0.71--1.21 (1.00)$^{2}$ & 0.76--1.92 (1.15)$^{1}$ & 1.01--1.96 (1.33) & 0.78--1.15 (0.96)$^{3}$ \\
\bottomrule
\end{tabular}
\end{table}

\section{x86 measurement}\label{app:x86}

\paragraph{Machine and build.}
The x86 results of \S\ref{sec:eval:x86} come from an Intel Xeon w5-2455X
(Sapphire Rapids; 12 cores, 24 hardware threads; AVX-512 with VNNI, BF16 and
FP16, and AMX; 24\,MiB L2 and 30\,MiB L3). It runs Ubuntu~24.04 under WSL2
(Linux~6.18). Our kernels are compiled by Halide on LLVM~22 for the host's
features. \GGML is the release of \S\ref{sec:eval:method}, built for the
native ISA with its AMX back end and CPU repacking enabled, and without the
llamafile SGEMM. The sweep ran overnight on an otherwise idle machine, with
frequency scaling off (the operating system's high-performance mode). The
harness, the checks and the cases are those of \autoref{app:method}, except that
CPU kernels use one or twelve threads (T1, T12).

\paragraph{Protocol.}
The x86 sweep used a related but different protocol from the M3 Max's. A
sample takes at least 30\,ms, and a case runs up to 25 interleaved rounds. It
stops after 5 rounds once the 95\% confidence interval of every \GH lies
wholly below 0.92 or wholly above 1.03, that is, once each kernel is clearly
behind or clearly ahead. Each of the 323 cases below 0.97 was re-run once at 25
rounds without early stopping; we keep each kernel's minimum time over the two
runs (the min-time rule of \autoref{app:method}). Every run's sampled check
passed. The held-out-shape layout rule is that of \autoref{app:fixed}.

\paragraph{Strata.}
We assign each case to the \GGML kernel that was fastest for it. Cases where
it is an AMX kernel, a repacked kernel or the tiled GEMM
(\cc{ggml-cpu/tiled}: AVX-512 VNNI, $M\ge8$) are \emph{format-specific}.
Those where \GGML loops \cc{vec\_dot} are \emph{blocked-dot}, as on the M3
Max (\S\ref{sec:eval:method}). The tiled GEMM is not AMX; it covers iq1\_s,
iq1\_m, iq2\_xxs, iq2\_xs, iq2\_s, iq3\_xxs, iq3\_s, iq4\_xs and q2\_K--q6\_K.
\GGML's AMX back end covers q4\_0, q4\_1, q8\_0, q4\_K, q5\_K, q6\_K, iq4\_xs
and f16 (the last with AVX-512), and its repacked $8{\times}8$ kernels cover
iq4\_nl, mxfp4, q2\_K and q4\_K.

\paragraph{Code counts.}
Ours: the x86 ranges of the schedule file and of the format table
(\autoref{app:size}), plus Halide back-end commits for x86 (category X of
\autoref{app:core}, less its CUDA commits). The upper bound adds every
application line last written since the port began, by \cc{git blame}, which
includes some CUDA and Arm edits, and the general Halide commits made since
then. \GGML: its functions in \cc{arch/x86} named for our formats, counted as
in \autoref{tab:loc}, and its \cc{amx/} and \cc{tiled/} directories in full.
Ours comprises \xLocSched sloc of x86-only schedules (AMX and VNNI tiles),
\xLocTable knob rows in the format table, and \xLocBackend sloc of Halide
instruction selection: VNNI for signed and three-factor byte dot products,
\cc{psignb}, \cc{pmulhuw} and small table lookups. \GGML's \xLocGGML sloc
comprise \xLocGGMLDots of AVX dot products and repacked kernels,
\xLocGGMLAmx of AMX kernels and \xLocGGMLTiled of the tiled GEMM: \xLocRatio
times ours, and still \xLocRatioMin times our upper bound (\xLocOursMax sloc).

\paragraph{Results by path.}
\autoref{tab:x86} breaks the x86 results of \autoref{tab:summary} down by
\GGML's path. In multithreaded GEMMs with small batches ($M=8$ and $32$), we
match \GGML's format-specific kernels in \xSmallFSPct of \xSmallFSN cases,
with a median of \xSmallFSMed{}$\times$; against its tiled VNNI GEMM we match
in \xSmallTlPct (median \xSmallTlMed{}$\times$), and against its blocked-dot
fallbacks the median is \xSmallBDMed{}$\times$. In multithreaded prefill
($M\ge32$), the median against the AMX kernels is \xPreAmxMed{}$\times$
(\xPreAmxBigMed{}$\times$ on the shapes that exceed the L3 cache,
\xPreAmxFiveMed{}$\times$ at $M=512$), against the repacked kernels
\xPreRpMed{}$\times$ and against the blocked-dot fallbacks
\xPreBDMed{}$\times$.
\begin{table}[!htbp]
\caption{x86: speedup \GH over \GGML's fastest non-BLAS CPU path, by that
path, with held-out-shape layouts: cases, median \GH and cases matched
($\GH\ge0.95$), on twelve and one threads. \cc{vec\_dot} is timed on one
thread only; ``both strata'' covers matrix products.}\label{tab:x86}
\footnotesize
\setlength{\tabcolsep}{3.5pt}
\centering
\begin{tabular}{@{}llrrr@{\qquad}rrr@{}}
\toprule
& & \multicolumn{3}{c}{T12} & \multicolumn{3}{c@{}}{T1} \\
\cmidrule(lr){3-5}\cmidrule(l){6-8}
Stratum & \GGML's path & Cases & Median & $\ge0.95$ & Cases & Median & $\ge0.95$ \\
\midrule
format-specific & AMX & \xAmxTwelveN & \xAmxTwelveMed & \xAmxTwelvePct & \xAmxOneN & \xAmxOneMed & \xAmxOnePct \\
 & repacked & \xRpTwelveN & \xRpTwelveMed & \xRpTwelvePct & \xRpOneN & \xRpOneMed & \xRpOnePct \\
 & tiled VNNI & \xTlTwelveN & \xTlTwelveMed & \xTlTwelvePct & \xTlOneN & \xTlOneMed & \xTlOnePct \\
 & all GEMM/GEMV & \xFSTwelveN & \xFSTwelveMed & \xFSTwelvePct & \xFSOneN & \xFSOneMed & \xFSOnePct \\
 & \cc{vec\_dot} & -- & -- & -- & \xVdN & \xVdMed & \xVdPct \\
blocked-dot & & \xBDTwelveN & \xBDTwelveMed & \xBDTwelvePct & \xBDOneN & \xBDOneMed & \xBDOnePct \\
\midrule
both strata & & \xTwelveN & \xTwelveMed & \xTwelvePct & \xOneN & \xOneMed & \xOnePct \\
\bottomrule
\end{tabular}
\end{table}

\paragraph{Results by batch size.}
\autoref{tab:x86m} breaks \autoref{tab:x86} down by $M$.
\begin{table}[!htbp]
\caption{x86 speedups \GH by $M$, held-out-shape layouts: cases, median and
cases matched ($\ge0.95$). T1 at $M=512$: $4096\times1024$ only.}\label{tab:x86m}
\footnotesize
\centering
\begin{tabular}{@{}llrrrrrr@{}}
\toprule
& & \multicolumn{3}{c}{format-specific} & \multicolumn{3}{c@{}}{blocked-dot} \\
\cmidrule(lr){3-5}\cmidrule(l){6-8}
Threads & $M$ & Cases & Median & $\ge0.95$ & Cases & Median & $\ge0.95$ \\
\midrule
T12 & 1   &  28 & 1.10 & 53.6\% & 188 & 1.08 & 86.2\% \\
    & 8   & 128 & 1.33 & 92.2\% &  88 & 2.51 & 93.2\% \\
    & 32  & 128 & 1.23 & 93.8\% &  88 & 3.42 & 100.0\% \\
    & 512 & 128 & 0.97 & 55.5\% &  88 & 5.08 & 100.0\% \\
\midrule
T1  & 1   &  35 & 0.94 & 48.6\% & 181 & 1.09 & 87.8\% \\
    & 8   & 128 & 1.23 & 76.6\% &  88 & 2.73 & 92.0\% \\
    & 32  & 128 & 1.10 & 77.3\% &  88 & 4.13 & 100.0\% \\
    & 512 &  32 & 0.83 & 15.6\% &  22 & 4.97 & 100.0\% \\
\bottomrule
\end{tabular}
\end{table}

\paragraph{Cases below 0.95.}
The three gaps of \S\ref{sec:eval:x86} have medians of \xFiveTlMed{}$\times$
(512-token prefill against the tiled GEMM at twelve threads;
\xFiveTlMedPerCase{}$\times$ with per-case layouts) and \xAmxOneMed{}$\times$
(one thread against AMX), and range over \xGemvRpLo--\xGemvRpHi{}$\times$
(iq4\_nl and mxfp4 GEMVs against the repacked kernels). Per-case layouts
raise single-threaded format-specific matches from \xFSOnePct to
\xFSOnePctPerCase. Of the \xCases cases, \xBelow are below 0.95. At twelve threads, 54 of the 100
$M=512$ cases against the tiled GEMM are below, mostly K-quants in the
$16{\times}4$ layout and the IQ types (lowest: q5\_K, 0.78). At both thread
counts, 14 iq4\_nl and mxfp4 GEMVs against the repacked kernels are below
(\xGemvRpLo--\xGemvRpHi; lowest at $4096\times1024$ on twelve threads).
On one thread, 46 of the 91 AMX cases are below; q4\_K and q5\_K in the
$16{\times}4$ layout fall as low as 0.68 at $M\ge8$ (11 of their 15 cases
below 0.95 are below 0.80). \GGML repacks these
weights into AMX's operand layout, whereas our $16{\times}4$ layout is
decoded in place: each sub-block's scale then needs its own tile product and
rescale. Also on one thread, 24 of the 25 $M=512$ cases against the tiled GEMM
are below (q4\_K, q5\_K and iq4\_xs: 0.66--0.71). The 14 \cc{vec\_dot}
cases below lie between 0.86 and 0.945 (iq1\_s, iq1\_m, q4\_0, q4\_1, q5\_1,
q1\_0, q2\_K, q6\_K and iq2\_xxs). Another 61 are blocked-dot GEMVs and
$M=8$ GEMMs (median 0.92; lowest: bf16 at twelve threads, 0.67). The
remaining 60 are spread over the other format-specific cells, with at most 12
for any \GGML path, thread count and $M$.

\FloatBarrier
\section{Directive Contracts}\label{app:directives}

\autoref{tab:preserve} states what each directive preserves, under the rule
of \S\ref{sec:ident:round}: a schedule preserves the specification's value
over $\mathbb{R}$ and may change only the f32 program's rounding, overflow,
underflow and NaN behavior; whatever can change the
value over $\mathbb{R}$ is an approximation with a contract.

\paragraph{Supported inputs.}
On supported inputs, the scheduled result is an f32 evaluation of the same
real expression, possibly reassociated or factored, so its error may exceed
one rounding (Lemma~\ref{lem:kernel}). The rule covers inlined decoders too.
Supported inputs have finite data, nonzero normal-range scales and no
overflowing partial sum. These are user obligations (\autoref{tab:preserve}),
except where \code{change\_type} proves that no sum overflows; \code{sever}'s
obligation is the user's too.

\paragraph{Strict-float updates.}
Encoders compute with \code{strict\_float} (\S\ref{sec:ident:round}), as
rounding changes threshold decisions, hence integers (\autoref{app:checks}),
and
directives that would reorder such an update are rejected: \code{rfactor} and
\code{atomic()} reject strict-float updates as non-associative, and
parallelizing, vectorizing or reordering their RVars is rejected too. Only
\code{atomic()}'s explicit override, a user assertion like unsafe
\code{change\_type}, bypasses this.

\begin{table}[!htbp]
\caption{What each directive preserves, and the obligations behind it.
``The f32 program'' is preserved up to the behaviours Halide's default,
non-strict float semantics admit, such as fusion and reassociation; bit for
bit only under \code{strict\_float}.}\label{tab:preserve}
\footnotesize
\begin{tabular}{@{}p{0.25\linewidth}p{0.35\linewidth}p{0.33\linewidth}@{}}
\toprule
Directive & Preserves & Obligation (who) \\
\midrule
Halide's other directives, \code{eager\_inline} & the f32 program & none \\
\code{sever} & the f32 program & the substituted buffers hold the cut computation's values, or an encoding meeting the codec's contracts (user; unchecked) \\
\code{rfactor}, \code{hoist\_invariants} & value over $\mathbb{R}$; f32 rounding, overflow, underflow and NaNs may change & bound-based hoists guarded by \code{require()}; write-backs not preserving the identity guarded by a nonempty domain, for float sums only under strict floats (compiler) \\
\code{change\_type}, safe & value over $\mathbb{R}$, computed exactly; rounds once back & exact-integer terms, no overflow (interval proof; run-time check for symbolic extents) \\
\code{change\_type}, unsafe (\code{Accumulate} only) & nothing: an approximation & \code{Accumulate}'s bound (\S\ref{app:accumulate}); no overflow, normal range (user, unchecked) \\
\code{approximate\_by(codec)} & nothing: values become $\rho(x)$ & the codec's declared contracts (user; tested) \\
\code{approximate\_by(port, ...)} & nothing: an approximation selected with the codec, within $|c|\,\beta$ (\S\ref{sec:props:derived}); the schedule places it & term equals $\psi$ over $\mathbb{R}$ (compiler); component within $\beta$ of $\psi$ (user; tested) \\
\bottomrule
\end{tabular}
\end{table}

\begin{figure}[!htbp]
\footnotesize
\[
\begin{array}{@{}l@{\quad}l@{}}
\textsc{Hoist} & f(\bar v)\oplus\!\!= s\otimes t \;\leadsto\; g(\bar v)=1_\oplus;\ \ g(\bar v)\oplus\!\!= t;\ \ f(\bar v)=f(\bar v)\oplus(s\otimes g(\bar v))\\
\text{\emph{over} }\mathbb{R} & \text{if } s \text{ is free of } R \text{ and } (\oplus,\otimes)\in\{(+,\times),(\min,+),(\max,+),(\vee,\wedge),(\wedge,\vee)\}\\
& \text{write-back only where } R\neq\emptyset, \text{ unless } s\otimes 1_\oplus=1_\oplus;\ \min,\max \text{ only where } {+} \text{ cannot wrap}\\[4pt]
\textsc{Field} & \textstyle\sum_R t\bigl[(x\gg s)\,\&\,m\bigr] \;\leadsto\; \bigl(\sum_R t\bigl[(x\gg (s-\delta))\,\&\,(m\ll \delta)\bigr]\bigr)\gg \delta\\
\text{\emph{exact}} & \delta=\mathrm{select}(0\le s\le k,\,s,\,0), \text{ if } s \text{ is free of } R \text{ and } |R| \text{ is constant}\\
& \text{and } k \text{ is the largest shift with } [0,|R|]\cdot\mathcal{I}\bigl(t[(x\gg(s-k))\,\&\,(m\ll k)]\bigr)\subseteq L_{\mathrm{type}(t)}\\[4pt]
\textsc{Split} & \mathrm{cast}_W(a+b) \;\leadsto\; \mathrm{cast}_W(a)+\mathrm{cast}_W(b)\\
\text{\emph{exact}} & \text{if } W \text{ represents } T=\mathrm{type}(a+b) \text{ and } \mathcal{I}(a)+\mathcal{I}(b)\subseteq T\\[2pt]
\hline\\[-8pt]
\textsc{Widen} & \mathrm{cast}_F(a)\odot\mathrm{cast}_F(b) \;\leadsto\; \mathrm{cast}_F(\mathrm{widening}_\odot(a,b))\\
\text{\emph{exact}} & \text{if } a,b:T \text{ integer},\ F \text{ represents } T \text{ and } \odot\in\{+,-,\times\}\\[4pt]
\textsc{Retype} & f(\bar v)=i;\ f(\bar v)\oplus\!\!=x \;\leadsto\; f'(\bar v)=\mathrm{cast}_t(i);\ f'(\bar v)\oplus\!\!=x';\ f=\mathrm{cast}(f')\\
\text{\emph{over} }\mathbb{R} & \text{if } \mathcal{I}(i),\,\mathcal{I}(x)\subseteq L_t\text{, float leaves of } i,x \text{ are whole-number constants or int casts,}\\
& \text{and every float }{+},{-},{\times}\text{ in them stays within its type's exact integers}\\
& \text{and, for } \oplus\in\{+,-\},\ \mathcal{I}(i)\pm[0,|R|]\cdot\mathcal{I}(x)\subseteq L_t \ \text{(symbolic } |R|\text{: checked at run time)}
\end{array}
\]
\smallskip
Under each name: what the rewrite preserves, exactly or as values over
$\mathbb{R}$. $R$: the reduction domain; $\mathcal{I}(e)$: Halide's interval
for $e$; $t[f]$: an integer term linear and homogeneous in its bit fields $f$
(it has no field-free constant term, such as an offset code's $-8$, which
must be split off first), each field in $[0,m\,2^k]$ when read at shift $k$, so every
term is a multiple of $2^\delta$ and ${\gg}\,\delta$ is exact; $L_t$: the
integers $t$ represents exactly; $1_\oplus$: the identity of $\oplus$; $x'$:
$x$ after \textsc{Widen}, cast to $t$.
\caption{The rewrites of \code{hoist\_invariants} (above the line) and
\code{change\_type} (below). \textsc{Hoist} moves invariant factors out of a
reduction; \textsc{Field} and \textsc{Split} prepare terms for it, reading bit
fields in place and splitting a cast sum; \textsc{Widen} and \textsc{Retype}
change the accumulator's type. Other combiners are rejected. \textsc{Hoist}'s
nonemptiness test, \textsc{Field}'s $\mathrm{select}$ and \textsc{Retype}'s
symbolic-$|R|$ check are generated into the code and fold away for constants;
the other premises are checked when the directive is applied.}\label{fig:rules}
\end{figure}

\paragraph{\code{hoist\_invariants()}.}
On the update of a reduction, it applies one of the distributive laws
$\sum_k s\,x_k=s\sum_k x_k$, $\min_k(c+x_k)=c+\min_k x_k$ (and $\max$),
$\bigvee_k(p\wedge x_k)=p\wedge\bigvee_k x_k$ (and its dual). A factor is
hoistable if it does not depend on the reduced reduction variables; it may sit
at any depth of the term's associative--commutative chain, including inside a
widening multiply or a cast of one. Integer and boolean subexpressions are
simplified with the reduction variables' bounds and with constant
\code{bound()}s on pure variables; a hoist that relies on such a bound is
guarded by a \code{require()}. In an integer sum, a bit field
\code{(x >> s) \& m} at an invariant position $s$ is read in place and the
shift is applied once to the sum, where no intermediate value wraps
(\textsc{Field}, \autoref{fig:rules}). It is exact because masking after a
shorter shift reads the same field scaled, $(x\gg(s-\delta))\,\&\,(m\ll\delta)
=((x\gg s)\,\&\,m)\cdot2^\delta$ for $0\le\delta\le s$, and a term linear
and homogeneous in its fields scales with them, $t[2^\delta f]=2^\delta
t[f]$; the new sum is thus $2^\delta$ times the old, and ${\gg}\,\delta$
divides it exactly. The symbolic $\delta=\mathrm{select}(0\le s\le k,s,0)$ is
$s$ where $s$ is a valid shift no larger than $k$, and 0 elsewhere, where the
rewrite is the identity. The interval premise bounds the sum read at shift
$k$; a field read at $\delta\le k$ lies in $[0,m\,2^\delta]\subseteq[0,m\,2^k]$,
and $t$ is linear, so the sum at the run-time $\delta$ does not wrap either. A term
list with several outer terms is split into one intermediate per term; the
grouped form splits it by the Funcs the terms call. With neither a factor nor
several terms it is an error. The intermediates start at the identity
$1_\oplus$, which they keep on an empty domain; where the write-back does not
map it to itself ($\min$ or $\max$ with an additive factor: $c+1_{\min}$ may
wrap), it applies only where every hoisted reduction variable's extent is
positive and the domain's predicate holds at the corner toward which it is
monotone, which is exact. The guard folds to true for constant extents (all
our kernels are unchanged); a predicate not monotone in every reduction
variable, such as a data-dependent one, is rejected on every target, float
sums included: the directive runs before the target is known, so for a float
reduction it always builds the closed-form guard and lowering selects it only
under \code{StrictFloat} (conservative, not unsound). Integer sums and $\lor$/$\land$ are never guarded. Over floats,
Halide's default semantics assume no infinities or NaNs, so $s\times0=0$ and a
sum's write-back needs no guard; where infinities and NaNs are observable (the
\code{StrictFloat} target feature, or \code{strict\_float()} in the update or
the factor), $s\times0$ may be NaN, and float reductions get the same
nonemptiness guard. Over integers and booleans the rewrite is exact;
over floats it is an identity of real arithmetic that changes rounding, under
the same freedom as \code{rfactor} (\S\ref{sec:ident:round}).

\paragraph{\code{change\_type(t, unsafe = false)}.}
It applies to a single-output Func whose updates are $f=f\oplus x$, with $x$
free of $f$ and $\oplus$ one of $+$, $-$ (accumulator on the left), $\min$,
$\max$, $\wedge$, $\vee$; products are not accepted. Each leaf of $x$ is
retyped: a float sum, difference or product of two casts from one integer type
that the float type represents becomes the cast of the widening integer
operation, which is exact: as the casts of $a$ and $b$ are exact, both sides
are $F$'s rounding of the same exact $a\odot b$; for an integer $t$, a float promotion of an integer
is peeled; any other leaf is cast to $t$, losslessly where its producers'
value bounds allow. A float-typed term gets an integer interval only if it is
built from whole-number constants and integer-to-float casts by $+$, $-$,
$\times$, $\min$, $\max$ and $\mathrm{select}$, and every float $+$, $-$,
$\times$ in it stays within the float type's exact integers ($\pm2^{24}$ for
f32), checked before widening, so the integer term computes the float term's
value; any other float term, such as a quotient, would be a truncating cast,
which the proof rejects. A float-to-integer conversion truncates toward
zero, and its interval is computed that way; a constant leaf stays an
explicit cast, so an unrepresentable constant reaches the proof rather than
being folded. Every update is retyped, and safety is proved by interval
analysis, carrying the accumulator's interval through the updates in order.
The initial value need not be zero, but its proven interval must fit $t$ (a
float constant must be finite and an exact int64; conversion endpoints are
compared without undefined host conversions). Every partial sum must fit too: the proof
bounds it by the initial interval plus the term's interval times the
cardinality of the whole reduction domain, a run-time precondition if that is
symbolic. Other combiners are rejected. The integer accumulation is exact,
so it may differ from a float accumulation whose partial sums round past
$2^{24}$; only f32 rounding changes, within the \S\ref{sec:ident:round}
contract. For a float $t$ the range is its exactly representable
integers ($\pm2048$ for f16). A failed proof is an error. The intervals
over-approximate, so a safe program may be rejected: a compile error, never
a miscompile. \code{unsafe}
skips the proof: like \code{bound()}, it is an unchecked assertion by the
caller, made only by \code{Accumulate}, whose contract is its rounding bound. The original Func becomes an
inlined cast back to its type. The new reduction is exact; it equals the old
one wherever that one evaluates every product and partial sum exactly (in f32,
below $2^{24}$), and the cast back rounds once at most.

\subsection{By-element dot products}\label{app:byelem}

For the running example, \code{blocked\_rows} builds the integer dot
like \code{block\_dot}, but without inlining the activation's integer
decoders, then stages them:
\begin{lstlisting}
Stage cu = codes.update();                       // codes: the integer dot
cu.split(rx, rp, rx, chunk).split(rx, rq, ri, 4);
Func quads = cu.rfactor(rq, q);                  // four integers at a time
for (Func a : act_codes) a.in(quads).compute_at(LoopLevel(quads, u));  // a lane
\end{lstlisting}
It splits the in-block reduction variable \code{rx} into the
record's pieces and groups of four integers, and \code{rfactor}s it, so each group is
one \cc{sdot} lane. The activation's four integers for one group are staged as one
32-bit lane of a register (\code{a.in(quads)}), and each 16-byte weight load
covers the same four positions of four rows. This is \Kleidi's \cc{dotprod}
kernel: \cc{sdot} by element, the weight vector against one lane of the
activation register. Because the weight integers are two's-complement nibbles,
their decode is a shift left and an arithmetic shift right; with the scaled
variant (\S\ref{sec:example:variants}) it is only a shift left, or a mask of
the high nibble, and the factor of 16 is folded into the scale
(\autoref{lst:asm}).

\begin{lstlisting}[float=h,language={},morecomment={[l]{//}},caption={Part of the running example's generated inner loop (disassembly of the compiled kernel in standard Arm syntax, abridged). One shift or mask decodes 16 bytes of weight integers from four rows; each \cc{sdot} multiplies them by four activation integers of one row.},label=lst:asm]
ldur  q0, [x26, #0x8]             // 16 bytes of a 4x4 weight record
shl   v5.16b, v0.16b, #4          // low nibbles: 16q, decoded
ldur  q4, [x25, #-0x3c]           // activation row 0: 4 integers in lane 0
sdot  v25.4s, v5.16b, v4.4b[0]    // 4 weight rows x 4 integers
...
and   v24.16b, v24.16b, v0.16b    // high nibbles (v0 now 0xf0): 16q, decoded
sdot  v28.4s, v24.16b, v17.4b[0]
\end{lstlisting}

\subsection{Accumulation bound}\label{app:accumulate}
\code{Accumulate}'s contract is the standard worst case over every order of
summation~\cite{higham}: a sum of $n$ terms lies within
$((1+u)^n-1)\sum_j|t_j|+n(1+u)^n\eta$ of the exact sum, with $u$ the unit
roundoff and $\eta$ the accumulator's smallest subnormal, on supported inputs
(\S\ref{sec:ident:round}).

\FloatBarrier
\section{The Properties}\label{app:laws}

\paragraph{Composed bounds.}
This section states the composition result of \S\ref{sec:ident:contracts}
and proves it with the other propositions. A decoder's \emph{sensitivity} to a
component bounds how far a change in that component moves a decoded value. The
bound of a \code{Compose} is the outer stage's bound plus each inner stage's
bound scaled by the outer decoder's sensitivity; a \code{Parallel} keeps its
children's bounds component by component. For our units the sensitivities are
simple (Lemma~\ref{lem:sens}): the only lossy inner stages are \code{fp16()}
roundings of scales, minimums and sums, and every decoder is affine in those
(sensitivity $|q|$ for a scale, 1 for an offset). q4\_0's bound
(\S\ref{sec:example:compose}) is an instance.

\begin{proposition}[Composed bounds]\label{prop:bound}
(a) Let $c=\mathrm{Compose}(c_1,c_2)$ with bounds $\beta_1,\beta_2$, and let
$e=\enc_1(x)$. Suppose $\dec_1$ has sensitivities $S^p_1$ to the components $p$
that $c_2$ encodes: whenever $|e'_p-e_p|\le\beta_{2,p}(e)$ for every $p$,
\[|\dec_1(e')_i-\dec_1(e)_i|\le\textstyle\sum_p S^p_{1,i}\max_j|e'_{p,j}-e_{p,j}|,\]
over the elements $j$ of $p$ that element $i$ reads. Then
\[|x_i-\rho(x)_i|\le\beta_1(x)_i+\textstyle\sum_p S^p_{1,i}\max_j\beta_{2,p}(e)_j.\]
(b) $\mathrm{Parallel}(c_1,\ldots,c_k)$, each $c_p$ on its own components with
bound $\beta_p$, has the per-component bound $(\beta_1,\ldots,\beta_k)$.
\end{proposition}

\begin{corollary}\label{cor:tree}
A tree of \code{Compose} and \code{Parallel} over units with bounds has a
bound, obtained bottom-up by (b) at each \code{Parallel} and (a) at each
\code{Compose}, given each decoder's sensitivities.
\end{corollary}

\noindent For $\dec(q,d)=q\,d$ with both components perturbed, $S^q=|d|$ and
$S^d=|q|+\beta_q$. A lossless $c_2$ ($\beta_2=0$), such as a packer or a
layout, leaves $\beta_1$ unchanged; a lossless $c_1$ leaves $c_2$'s bound
scaled by $S_1$. Over floats, $\rho(x)=\mathrm{fl}(\dec_1(\rho_2(e)))$ adds
$\dec_1$'s rounding unless it is exact, as for q4\_0
(\S\ref{sec:example:compose}).

\paragraph{Proofs.}
We write composition as juxtaposition.
Encodings are compared up to $\approx$ (Definition~\ref{def:law}), which
$\dec$ respects, so $\approx$-equal encodings decode alike; we write $=$ for
it. Without that requirement the propositions fail: with $\enc(0)=+0$,
$\enc(1)=-0$ and a decoder that reads the sign bit, $\dec(+0)=1$ and
$\dec(-0)=0$, the law holds up to $\pm0$, yet $\rho(0)=1$ and $\rho(\rho(0))=0$.
Our decoders read float components only through arithmetic; the one stage that
stores a float's bits as an integer component (iq1\_m's scale) is in a format that
declares no law.
\emph{Proposition~\ref{prop:split}.} (a)
$\rho\rho=\dec\,(\enc\,\dec\,\enc)=\dec\,\enc=\rho$, and by induction
$\rho^n=\rho$ and $\enc\,\rho^n=\enc\,\dec\,\enc=\enc$. (b) For
$e=\enc(x)$, $\enc(\dec(e))=\enc\,\dec\,\enc(x)=e$. (c) $\psi\varphi$ is
the identity on $\enc(D)$, so $\dec\,\psi\varphi\,\enc=\rho$, which uses no
law; with the law, and as $\varphi$ respects $\approx$,
$\varphi\,\enc\,\dec\,\psi\varphi\,\enc=\varphi\,\enc\,\dec\,\enc=\varphi\,\enc$.
The composite decoder $\dec\,\psi$ respects $\approx$ on
$\varphi(\enc(D))$, as Definition~\ref{def:law} requires, because $\psi$
does there and $\dec$ does.
\emph{Proposition~\ref{prop:bound}.} (a) $\rho(x)=\dec_1(\rho_2(e))$ and $\rho_1(x)=\dec_1(e)$, so
$x-\rho(x)=(x-\dec_1(e))+(\dec_1(e)-\dec_1(\rho_2(e)))$. The first term is
bounded by $\beta_1$; since $|\rho_2(e)_p-e_p|\le\beta_{2,p}(e)$ for every
component $p$, the sensitivity hypothesis bounds the second by
$\sum_pS^p_1\max_j\beta_{2,p}(e)_j$. (b) The round trip of a
\code{Parallel} is the product of its children's, each acting on its own components
(a component no child encodes has round trip the identity), so component $p$'s error is
bounded by $\beta_p$.
\emph{Corollary~\ref{cor:tree}.} By structural induction: a unit's bound is
declared; a \code{Parallel}'s follows from its children's by (b); a
\code{Compose} is a nest of binary ones, and each node's bound follows by (a)
from its outer stage's bound and its inner stages', which bound the outer
stage's encoded components. Over $\mathbb{R}$; each $\dec$'s f32 rounding is an
added term unless exact (\S\ref{sec:ident:contracts}).

\begin{lemma}[Decoder sensitivities]\label{lem:sens}
In every format of \autoref{tab:formats}, a lossy stage that follows a unit
in a \code{Compose} is \code{fp16()} on a scale, minimum or block-sum component,
and the unit's decoder is affine in that component with the coefficient of
\autoref{tab:sens}. Every other component a decoder reads has only lossless later
stages ($\beta_2=0$), so its sensitivity is not needed.
\end{lemma}
\begin{proof}
By case analysis of the units, not checked by the compiler. In each
\code{Compose} of \autoref{tab:formats} the lossy stages after a quantizer
are its \code{fp16()} stages; the integer representations, packers, layouts,
\code{e8m0\_half}, \code{ue4m3\_half} and \code{Tabulate} after it are
declared lossless. The decoders of \autoref{tab:sens} multiply the scale component
by a value fixed by the other components (a code, a table entry, a sub-block scale
times a code) and add the minimum component times a fixed value, with no clamp or
select on the scale or minimum, so they are affine in those components. The
non-affine decodes (minifloat and e8m0 values, grid and code tables) read
integers or exponent components, which only lossless stages follow. A block sum is
dropped by decode; its error enters through \code{Derive}'s bound
(\S\ref{sec:props:derived}), not a sensitivity.
\end{proof}

\begin{table}[!htbp]
\caption{Decoder sensitivities to the components that a later \code{fp16()} rounds
(Lemma~\ref{lem:sens}); $c$ is the code's value (a signed code, a table
entry or $\pm1$), $s_j$ and $m_j$ a sub-block's scale and minimum integers.}\label{tab:sens}
\small
\begin{tabular}{@{}llll@{}}
\toprule
Quantizer & formats & decoded value & sensitivities \\
\midrule
\textbf{Type0}, \textbf{Q8\_0}, \textbf{Sign}, & q4\_0, q5\_0, q8\_0, q8\_1, & $d\,c$ & $S^d=|c|$ \\
\quad \textbf{Ternary}, \textbf{Q2}, \textbf{IQ4}(32) & q1\_0, tq1\_0, tq2\_0, q2\_0, iq4\_nl & & \\
\textbf{Type1} & q4\_1, q5\_1 & $d\,c+m$ & $S^d=|c|$, $S^m=1$ \\
\textbf{KAffine} & q2\_K, q4\_K, q5\_K & $d\,s_j\,c-m\,m_j$ & $S^d=|s_jc|$, $S^m=|m_j|$ \\
\textbf{Q3K}, \textbf{Q6K}, \textbf{IQ4}(256) & q3\_K, q6\_K, iq4\_xs & $d\,s_j\,c$ & $S^d=|s_jc|$ \\
\textit{ScaledCodes} & grid IQ (7) & $d\,s_j\,c$ & $S^d=|s_jc|$ (no bound declared) \\
\textbf{MXFP4}, \textbf{NVFP4}, \textbf{Q8K}, & mxfp4, nvfp4, q8\_K, & --- & no lossy later stage \\
\quad \Kleidi's, f16, bf16 & qsi8cx, qai8dx, f16, bf16 & & \\
\bottomrule
\end{tabular}
\end{table}
\emph{Proposition~\ref{prop:split}(d).} On $D$, the law reads $\enc\circ\rho=\enc$, so
$\enc(x)=\enc(\rho(x))=\enc(\rho(y))=\enc(y)$.

\begin{lemma}[Kernel error]\label{lem:kernel}
Assume the \S\ref{sec:ident:round} contract: every directive other than
\code{approximate\_by} and derived-component substitution preserves the value over
$\mathbb{R}$ (\code{Accumulate} changes only an accumulator's float type, so
its rounding enters $\varepsilon$). Let a kernel compute $y=\sum_k x_k z_k$
from weights $w=\rho_W(x)$ and activations $a^0=\rho_A(z)$, with
$|x_k-w_k|\le\beta^W_k$ and $|z_k-a^0_k|\le\beta^A_k$, requantizing the
activations inside the kernel as $a=\rho_R(a^0)$, $|a^0_k-a_k|\le\beta^R_k$
(f32 activations to q8\_0, or a compute format such as f16; $\rho_R=\mathrm{id}$
if none; a weight-side requantization adds to $\beta^W$ likewise). Let it read substituted components
$s_p$ in terms $c_p\psi_p$, with $|s_p-\psi_p|\le\beta_p$. On supported
inputs (\S\ref{sec:ident:round}) its result $\tilde y$ satisfies
\[
|\tilde y-y|\le\textstyle\sum_k\bigl(|x_k|(\beta^A_k+\beta^R_k)+|a_k|\beta^W_k\bigr)
+\sum_p|c_p|\beta_p+\varepsilon,\qquad
\varepsilon=\frac{g\,S+K\eta}{1-g}\ \ (g<1).
\]
Here $g$ and $S$ are those of the kernel tolerance (\autoref{app:checks}),
with $u$ the unit roundoff of the narrowest accumulator ($2^{-11}$ for f16,
$2^{-24}$ for f32), the schedule's chain length $c$, its f32 tree over the
$K/c$ partial sums (depth $\lceil\log_2(K/c)\rceil+1$), the operand rounding,
at most four roundings per term (as in all our kernels) and the operands'
magnitudes; $\eta$ is half that accumulator's smallest subnormal ($2^{-25}$
for f16, $2^{-150}$ for f32), for each term's final product, provided no
earlier rounding in a term underflows (otherwise scale $\eta$ by the later
factors).
\end{lemma}
\begin{proof}
Let $y_1=\sum_k w_k a_k$ and $y_2$ be $y_1$ with each $c_p\psi_p$ read as
$c_p s_p$. Then $|y-y_1|$ is bounded by the first sum, since
$x_kz_k-w_ka_k=x_k(z_k-a_k)+a_k(x_k-w_k)$ and $|z_k-a_k|\le\beta^A_k+\beta^R_k$;
$|y_1-y_2|\le\sum_p|c_p|\beta_p$ by the components' bounds. By the assumed
contract, every directive preserves the value over $\mathbb{R}$, so $\tilde y$
is a float evaluation of $y_2$. Each rounding multiplies by $(1+\theta)$,
$|\theta|$ at most its unit roundoff, except that a product rounded into
the subnormal range adds an absolute error of at most $\eta$ (an addition that
underflows is exact). Multiplying out as in \autoref{app:checks},
$|\tilde y-y_2|\le\varepsilon$ (the check uses the stricter $g\,S$).
\end{proof}
\noindent The checks of \S\ref{sec:eval:correct} compare against
$\sum_kw_ka^0_k$ in f64, the operands as stored; their tolerance covers the
$\beta^R$ terms (\autoref{app:checks}'s $\delta w$, $\delta a$) and $g\,S$. It
omits $K\eta$, at most $2^{-13}$ (f16 accumulation, $K=4096$), negligible
against $g\,S$ on the tests' normal-range data. The absolute term is needed:
with identity codecs, 32 products of $2^{-75}$ sum to $2^{-145}$ but compute 0.

\begin{table}[!htbp]
\caption{The checks of \S\ref{sec:props}. \emph{Build}: checked whenever a
pipeline is built. \emph{Sampled}: a property of the testing library, run on
inputs drawn from a distribution; each applies to a whole composition or, with
\code{.at(stage)}, to one stage. \emph{Test}: a check of the definitions, run
by the test harness.}\label{tab:laws}
\small
\begin{tabular}{@{}lp{0.47\linewidth}l@{}}
\toprule
Check & Law & When \\
\midrule
types, dimensions, component names & a stage's components are what the next stage expects & build \\
\code{check\_ranges} & each lossless stage's declared input range contains the
declared output range of the stage before it (C2) & build \\
derived-component match & a term equals a component's definition over $\mathbb{R}$
(\S\ref{sec:props:derived}) & build \\
\midrule
\code{idempotent\_requantize(tol)} & $\enc(\dec(e))=e$ for $e=\enc(x)$,
the round-trip law; integer components exactly, float components within \code{tol} & sampled \\
\code{lossless()} & $\dec(\enc(x))=x$, bit for bit & sampled \\
\code{within\_declared\_bound()} & $|x_i-\rho(x)_i|\le\beta(x,\enc(x))_i$ & sampled \\
\code{outputs\_within\_declared\_ranges()} & every encoded value lies in its
component's declared range & sampled \\
\code{derived\_ports\_hold()} & a derived component equals its definition, or is
within its declared bound & sampled \\
\code{closed()} & each stage's encode and decode reach their own inputs, and
no other Func, buffer or image of the round trip (a check of the call graph, run by the test harness)\textsuperscript{a} & test \\
\bottomrule
\end{tabular}

\smallskip
\raggedright\footnotesize
\textsuperscript{a}It counts a constant table a stage embeds (\code{Tabulate}, the grid
formats' tables) as a foreign buffer, so the grid formats skip it (\S\ref{sec:props}).
\end{table}

The \code{closed()} check walks each stage's encode and
decode and fails if an output reaches a Func, buffer or image parameter of the
round trip other than the stage's own inputs, or reaches none of them; scalar
parameters and Funcs the stage defines are allowed. We run it as a test.

\paragraph{Tests.}
The tutorial checks the running example's parts this way:
\begin{lstlisting}
check_ranges(q4_0).empty();                                    // static
check_property(q4_0, lossless().at(q.pack), normal, Float(32), {N});
check_property(q4_0, within_declared_bound().at(q.quantize), normal, Float(32), {N});
\end{lstlisting}
The application's tests check the declared ranges, each quantizer's bound at
its stage, and \code{closed()} (except for the grid formats, whose embedded
tables it counts as foreign inputs; accepting a constant buffer that a stage
itself embeds would remove this exception), on mixtures of distributions chosen to reach
the quantizers' corner cases: binades of every size, blocks of one sign, blocks
dominated by one value, constants and zeros.

\paragraph{Derived components.}
\code{Derive\{"sum", \{"integers", "scale"\}, block\_sum\}} appends q8\_1's
sum component (\S\ref{sec:props:derived}); \code{sever} lists it, as decode
drops it. A representation means its decoded tensor $\rho(x)$ together with
its derived components, each within a declared bound $\beta$ of its definition
$\psi$ on the decoded values. \code{approximate\_by(derived, consumers)}
applies the substitution in each consumer by one rule:
\[
\textsc{Derive}\quad \tau \;\leadsto\; \mathrm{cast}(s(\bar a))
\quad\text{if } N(\tau)=N(\psi[\bar a]),\ s \text{ declares } \beta,\ \text{and } \mathrm{type}(\tau) \text{ represents } \mathrm{type}(s),
\]
where $\bar a$ are affine indices $c\,v+w$ unified with the component's, and
$N$ is the polynomial normal form below, in which components, roundings and
integer division are opaque. A component declares $\beta$ only if the codec's
trace after its \code{Derive} meets two conditions, both checked: (i) no lossy
stage reaches a component that $\psi$ reads; (ii) at most one lossy stage
transforms $s$, and it transforms $s$ alone and declares $\beta$. A mismatch or
a failed condition is an error. Lemma~\ref{lem:kernel} sums a kernel's error
against the specification: decode and in-kernel requantization error through
the product, $|c|\,\beta$ per substituted component, and rounding at the
narrowest accumulator's precision, as the tolerance of \autoref{app:checks}
does.

\paragraph{The derived-component matcher.}
\code{approximate\_by(port, consumers)} normalizes the term and the
component's definition. Calls to the components are atoms. Other pure Funcs
are expanded, and sum reductions become sums over renamed reduction variables.
$+$, $-$ and $\times$, on floats and on signed integers of 32 bits or more,
are normalized as polynomial arithmetic: products are distributed, like terms
are collected, and invariant factors are pulled out of sums. A float division
is interpreted only by a nonzero constant. Every other division and modulus,
including every integer division, is an opaque atom over normalized children,
so $(p/2)\cdot 2$ never matches $p$. Arithmetic on other integer types
(narrower or unsigned) is opaque, except that a sum reduction accumulating in
such a type $t$ is expanded if every reduction extent is a constant and
$\mathrm{range}(t)\supseteq\mathrm{range}(\mathit{init})+[0,n]\cdot\mathrm{range}(\mathit{body})$,
with $n$ the product of the extents (each clamped at 0). Ranges are exact
integer intervals, unbounded where a subterm may overflow. Every partial sum
is the initial value plus $k\le n$ body values, so none wraps; otherwise the
reduction stays an atom.
Signed-integer overflow is not modeled, as Halide treats it as undefined. A
cast that is exact on its operand's range is dropped; every rounding is an
opaque node that matches only itself, so rounding is never taken for the
identity. A match thus establishes that the term equals $\psi$ over
$\mathbb{R}$ (over $\mathbb{Z}$ for the wide integers and such sums), with roundings and
integer division and modulus as uninterpreted functions. The gap between that
and the f32 program is the component's declared bound, the user's obligation. The matcher
does not shift indices or re-split reduction domains, so the term must have the
shape of the definition. On a mismatch, \code{approximate\_by} is an error that
prints both normal forms. The replacement carries the component's declared bound,
and it is an error if the component declares none; an option also checks the
relation at run time where the component is read.

The schedule (\code{derived\_sums} in \cc{kernels/schedule.h}) substitutes
last. It inlines the decoders (\code{eager\_inline}), splits off the
code-free terms with the grouped \code{hoist\_invariants}, \code{rfactor}s
those terms per block (and per activation sub-block), and hoists the
activation's scale, so that the term is a sum of integers over the component's block.
It then tests the match (\code{match\_derived\_port}) before calling
\code{approximate\_by}; if no component matches it computes the sum, or reports an
error where the kernel needs the component.

\end{document}